\documentclass[12pt]{article}

\usepackage{amsmath}
\usepackage{graphicx,psfrag,epsf}
\usepackage{enumerate}
\usepackage{natbib}
\usepackage{url} % not \dfrac{num}{den}crucial - just used below for the URL 
\usepackage{tabularx}
\usepackage{natbib} % for the bibliography
\usepackage{arydshln}
\usepackage{txfonts}
\usepackage[colorinlistoftodos]{todonotes}
\usepackage[colorlinks=true, allcolors=blue]{hyperref}
\usepackage{bm} 
\usepackage{bbm}
\usepackage{mathrsfs}
\usepackage{mathtools}
\usepackage{dsfont}
\usepackage{algorithm2e,setspace}
\usepackage{float} 
\usepackage{textcomp}

\usepackage{fancyhdr}

\usepackage{subcaption}
\usepackage{tikz} 

\usepackage{tkz-tab}
\usepackage[font=footnotesize]{caption}
\usepackage{diagbox}
\usepackage{booktabs}
\hypersetup{
	citecolor=blue,linkcolor=blue}

\makeatletter
\newcommand{\distas}[1]{\mathbin{\overset{#1}{\kern\z@\sim}}}%
\newsavebox{\mybox}\newsavebox{\mysim}
\newcommand{\distras}[1]{%
	\savebox{\mybox}{\hbox{\kern3pt$\scriptstyle#1$\kern3pt}}%
	\savebox{\mysim}{\hbox{$\sim$}}%
	\mathbin{\overset{#1}{\kern\z@\resizebox{\wd\mybox}{\ht\mysim}{$\sim$}}}%
}

\newcommand\floor[1]{\lfloor#1\rfloor}

\DeclareMathOperator*{\argmax}{arg\,max}
\def\BState{\State\hskip-\ALG@thistlm}
\makeatother

\floatstyle{ruled}
\newfloat{algorithm}{tbp}{loa}
\providecommand{\algorithmname}{Algorithm}
\floatname{algorithm}{\protect\algorithmname}

\usetikzlibrary{backgrounds}

\begin{document}

\def\spacingset#1{\renewcommand{\baselinestretch}%
{#1}\small\normalsize} \spacingset{1}

%%%%%%%%%%%%%%%%%%%%%%%%%%%%%%%%%%%%%%%%%%%%%%%%%%%%%%%%%%%%%%%%%%%%%%%%%%%%%%
  \title{\bf  Bayesian Model Search  for Nonstationary Periodic Time Series}
  \author{Beniamino Hadj-Amar,  B{\"a}rbel Finkenst{\"a}dt, Mark Fiecas, 
  	\\Francis L{\'e}vi and  Robert Huckstepp \thanks{Beniamino Hadj-Amar (E-mail: B.Hadj-Amar@warwick.ac.uk), B{\"a}rbel Finkenst{\"a}dt Rand (E-mail: B.F.Finkenstadt@warwick.ac.uk), Department of Statistics, University of Warwick, Coventry CV4 7AL, UK. Francis L{\'e}vi (E-mail: F.Levi@warwick.ac.uk ), Warwick Medical School, University of Warwick, Coventry CV4 7AL, UK. Robert Huckstepp (E-mail: R.Huckstepp@warwick.ac.uk), School of Life Sciences, University of Warwick, Coventry CV4 7AL, UK. Mark Fiecas (E-mail: mfiecas@umn.edu), University of Minnesota, School of Public Health, Division of Biostatistics, Minneapolis, MN 55455, USA. This version of the paper is an electronic reprint of the original article published by \href{https://taylorandfrancis.com/}{Taylor \& Francis} in \href{https://www.tandfonline.com/toc/uasa20/current}{\textit{Journal of the American Statistical Association}}. }}
  	  \date{}

  \maketitle

  \thispagestyle{fancy}
% \if0\blind
% {
%   \bigskip
%   \bigskip
%   \bigskip
%   \begin{center}
%     {\LARGE\bf Bayesian Model Search  for Nonstationary Periodic Time Series}
% \end{center}
%   \medskip
% } \fi

\spacingset{1.2} % DON'T change the spacing!

\begin{abstract}
	
		%This article considers the analysis of time series that exhibit an oscillatory behaviour. The nonstationary investigation of oscillatory data that show regime shifts in periodicity, amplitude and phase is challenging as the timing and number of such changes is usually unknown.  
		
We propose a novel Bayesian methodology for analyzing  nonstationary time series  that exhibit  oscillatory behaviour.
We approximate the time series using a piecewise oscillatory model with unknown periodicities, where our  goal is to estimate the change-points while simultaneously identifying the potentially
changing periodicities in the data. Our proposed methodology is based on a trans-dimensional Markov chain Monte Carlo (MCMC)
algorithm that simultaneously updates the change-points and the periodicities relevant to any segment between them. We show that the proposed methodology successfully identifies time changing oscillatory behaviour in two applications which are relevant to e-Health and sleep research, namely the occurrence of ultradian oscillations in human skin temperature during the time of night rest, and the detection of instances of sleep apnea in  plethysmographic respiratory traces.
 \end{abstract}
\noindent % 

\newpage
\spacingset{1.45} % DON'T change the spacing!
\section{Introduction}
Identifying the periodicities present in a cyclical phenomenon allows us to gain insight into the sources of variability that drive the phenomenon.
% \textcolor{red}{Cyclic patterns } are found on all levels in nature  and the analysis of time series that show oscillatory  
% behaviour, often involving a mix of 
% periodic contributions from several oscillators, is a
% common  task in a multitude of scientific  research areas. %not least for the analysis of physiological time series signals 
% in the emerging subject of e-Health \citep{kyriacou2003multi, paradiso2005wearable,komarzynski2018}. 
For example, respiratory traces obtained from a plethysmograph used on rodents in experimental sleep apnea research 
exhibit many abrupt changes in their periodic components % in their spectral properties 
as the rat naturally
changes their breathing pattern in the course of  its  sleep-wake activities \citep{han2002periodic,nakamura2003sleep}.
Similarly, % from a visual inspection of the data, we may hypothesize that 
human temperature, as measured  by a wearable sensing device 
over several days  at relatively high temporal resolution,  may be subject to a different periodic behaviour during the night when the individual  transitions between ultradian sleep stages \citep{carskadon2005normal,komarzynski2018}.
While the theory and methods for analyzing the periodicities of time series data whenever the time series is stationary are relatively well-developed, 
the task of modelling time series that show
regime shifts in periodicity, amplitude and phase remain challenging because the timing
 of changes and the relevant periodicities are usually  unknown.

% Stationary
There have been many developments for modeling stationary oscillatory time series data. %has For the analysis of stationary time series, several authors have considered model estimation for oscillatory signals.
\citet{rife1976multiple}  and \citet{stoica1989maximum} addressed the problem of estimating the frequencies, phases and amplitudes of sinusoidal signals under the assumption of a known number of sinusoids, where inference is based on maximum likelihood frequency estimators.  These models, however, require very long time series and a large separation in the frequencies that drive the process, which will not always be the case in practice % seems not to be valid in the interesting cases where the amount of data is small and the frequencies are close to each other 
\citep{djuric1996model, andrieu1999joint}. \cite{quinn1989estimating}, \citet{yau1993maximum} and \citet{zhang1993information} tackled the problem of model selection on the number of sinusoidal signals by employing the Akaike information criterion \citep{akaike1974new} and the minimum description length principle \citep{rissanen1978modeling}. \citet{djuric1996model} showed that these procedures tend to estimate a wrong number of components when the sample size is small and the signal-to-noise ratio is low. % \citet{fuchs1988estimating} determined the number of sinusoids of a signal by investigating the eigenvector decomposition of the estimated autocorrelation matrix.  

% Bayesian
Bayesian approaches to modelling stationary oscillatory signals were explored for the first time by \citet{bretthorst1988, bretthorst1990bayesian} with applications to nuclear magnetic resonance spectroscopy. \citet{dou1995bayesian, dou1996bayesian} presented a Bayesian approach that uses a  Gibbs sampler to identify multiple frequencies that drive the signal. % multiple frequency signals 
%where they approximate the posterior conditional distribution of the frequencies by an expansion at the maximum. 
Their method required the number of frequencies to be fixed in advance, and model selection was achieved by 
choosing the most probable model based on the estimation of the parameters for all possible models. Bayesian model selection for stationary oscillatory signals based on posterior model probabilities were also investigated by \citet{djuric1996model}.
\citet{andrieu1999joint} introduced a more efficient 
reversible-jump MCMC method \citep{green1995reversible} that jointly tackles model selection and 
parameter estimation for an unknown number of stationary sinusoidal signals and avoids the 
computationally expensive numerical optimization of  \citet{dou1995bayesian, dou1996bayesian}
by sampling the 
frequencies  one-at-time via Metropolis-Hastings (M-H) steps.  To the best of our knowledge, currently there is no extension of this methodology to analyze nonstationary oscillatory signals. 

% Nonstationary
A formal statistical modeling framework for a specific class of nonstationary time series data, called {\it locally stationary} time series, was developed by \citet{dahlhaus1997fitting}.  % developed a framework for analysing the time varying spectral properties of nonstationary time series by defining the concept of locally stationary processes and evolutionary spectra.  
%\citet{young1999dynamic} presented a dynamic harmonic regression formulated within a state space framework which exploits the  properties of the Kalman filter recursive algorithm. Although the model of \citet{young1999dynamic} considers time variable parameters for amplitude and phase, it does not allow for the number of periodicities  and their values  to change over time.  
Extending this framework to a Bayesian setting, \citet{rosen2009local} proposed an approach to model the log of the time-varying spectral density
using a mixture of smoothing splines. % Their approach required 
% that  a number of partitions of the data are pre-specified where 
% the weights of the distinct log spectral density changes with time and posterior inference is carried out using reversible-jump MCMC. 
\citet{rosen2012adaptspec} improved on this by splitting the time series into an unknown but finite number of segments of variable lengths, thereby avoiding the need to pre-select partitions, and to estimate the time-varying spectral density using a fixed number of smoothing splines.  % For a given partition of the time series, the likelihood function is approximated via a product of local Whittle likelihoods \citep{whittle1957curve}. 
 For a given partition of the time series, the likelihood function is approximated via a product of local Whittle likelihoods \citep{whittle1957curve}. The methodology was developed using a Bayesian framework and  is based on the assumption that, conditional on the position and number of partitions, the time series are  
piecewise stationary, and the underlying spectral density for each partition is smooth over frequencies. However, exploratory analyses of the time series in both of our case studies revealed spectral densities with very sharp peaks, often at several nearby frequencies, thus invalidating the assumption that the spectral density is smooth over frequencies. In addition, the frequency location of these sharp peaks changed over time.

% Contributions
In this article we propose a novel Bayesian methodology for modelling oscillatory data that
show regime shifts in periodicity, amplitude and phase. 
In contrast to previous work our approach does not require  pre-specifying 
the number of regimes or the order of the model within a regime.
We assume that, conditional on the position and number of change-points, 
the time series can be approximated by a piecewise changing sinusoidal regression model. The timing and number of changes
are unknown, along with the number and values of relevant periodicities   in each  segment.
We develop a reversible jump MCMC 
technique that jointly explores  the parameter space of the change-points and sub-models for all segments.  % Prior beliefs about the parameters are updated based on a Gaussian likelihood.  Hence, we avoid using the Whittle likelihood, which is only a large-sample approximation of the true underlying likelihood. We note that  \citet{punskaya2002bayesian} proposed a Bayesian method for fitting piecewise linear regression models, which leads to a similar structure of reversible-jump MCMC algorithm where the number of knots and the dimension of the linear regression models within each segments are both unknown. However,  our approach is targeted at modelling oscillatory behaviour and the estimation of  the relevant spectral frequencies poses an additional challenge.

The paper is organized as follows. Section \ref{model} and \ref{s:bayesian_inference} present the model, the prior specifications and the general structure of our Bayesian approach. Sections \ref{sampling_scheme} and \ref{simulation_study} provide a detailed explanation of our sampling scheme and  simulation studies to demonstrate the performance of our approach. %along with a comparison with some existing methods.
In Section \ref{applications} we illustrate the use of our methodology in two data-rich scenarios related to sleep, circadian rhythm and e-Health research, namely the identification of the  spectral properties of  experimental breathing traces arising in sleep apnea research,  
and the analysis  of human temperature data measured over several days by a wearable sensor. We conclude and discuss our current findings in Section \ref{conclusion}. 

% Code that implements the methodology is available at \hyperlink{https://github.com/Beniamino92/AutoNOM}{https://github.com/Beniamino92/AutoNOM}.

\section{The Model}  \label{model}
Consider a time series realization $y_1, \dots, y_n$  whose periodic behaviour may change at 
$k$ unknown time-points $ \bm{s}_{\, (k)} = (s_1, \dots, s_k)^{'}$ where $k$ is also unknown.
Assume that in each sub-interval $\text{I}_j = [s_{j-1}, s_j)$ there are $m_j$ relevant frequencies $\bm{\omega}_{\, j} = (\omega_{j, \, 1}, \dots, \omega_{j, \, m_j})^{'}$, for $j = 1, 2, \dots, k + 1$. Setting $s_0 = 1$ and $s_{k+1} = n$, we can write the following sinusoidal  model (\citealp{andrieu1999joint}) 

\begin{equation}\label{model_RJMCMC_1}
y_t =  \sum_{j=1}^{k+1} f \, \big(t, \, \bm{\beta}_{ \, j}, \,  \bm{\omega}_{\, j} \, \big)\mathbbm{1}_{[ \, t \, \in \,  I_j \, ]}   + \varepsilon_t,
\end{equation}
where \begin{equation}
\label{model_RJMCM_2}
f \, \big(t, \, \bm{\beta}_{ \, j}, \,  \bm{\omega}_{\, j} \, \big) = \alpha_j + \mu_j \, t + \sum_{l=1}^{m_j} \Bigg(  \beta_{j, \,l}^{\, (1)} \cos(2\pi\omega_{j,\,l} \, t) + \beta_{j,\, l}^{\, (2)} \sin(2\pi\omega_{j,\,l} \,  t) \Bigg),
\end{equation}
$\bm{\beta}_{\,j } = (\alpha_j,  \mu_{\, j}, \, \bm{\beta}_{j, \, 1}^{'}, \dots, \,  \bm{\beta}_{j, \, m_{j}}^{'})^{'}$,
 $\bm{\beta}_{j, \, l} = (\, \beta_{j, l}^{\, (1)}, \, \beta_{j, l}^{\, (2)} \, )^{'}$,
 $\mathbbm{1}_{[\cdot]}$ denotes the indicator function,
 and  $\mu_j$ and $\alpha_j$ may, if needed, account for a linear trend within each segment.  
 For simplicity we assume independent zero-mean Gaussian errors with regime-specific variances 
\begin{equation} \label{eq:residual_error}
\varepsilon_t \sim \mathcal{N}\, (0, \sigma^2_{j}), \quad \text{for} \, \, t \in \text{I}_j \, \, \, \text{and} \, \, \,  j = 1, \dots, k+1,  
\end{equation}
noting that, in principle, the methodology can in principle be extended to the non-Gaussian case.

The dimension of the model is given by the number of change-points $k$ and the number of frequency 
components in each regime denoted by $\bm{m}_{\, (k)} = (m_1, \dots, m_{k+1})^{'}$. Furthermore,
let $\bm{\beta}_{\, (k)} = (\, \bm{\beta}_{ \, 1}^{'}, \dots, \bm{\beta}_{ \, k+1}^{'}  )^{'}$, $\bm{\omega}_{\, (k)} = (\bm{\omega}_{ \, 1}^{'}, \dots, \bm{\omega}_{ \,k+1}^{'} )^{'}$, $\bm{\sigma}^{\, 2}_{\, (k)} = (\sigma^{2}_{ \, 1}, \dots, \sigma^{\, 2}_{\, k+1}  )^{'}
$ and $\bm{\theta}_{\, (k)} = ( \, \bm{\beta}_{\, (k)}^{'}, \, \bm{\omega}_{\, (k)}^{'}, \, \bm{\sigma}_{\, (k)}^{\, 2'} \,  )^{'}$. 
Using Equation \eqref{model_RJMCMC_1}, the likelihood of $( \, k, \, \bm{m}_{\, (k)}, \, \bm{s}_{\, (k)}, \bm{\theta}_{\, (k)} \, )$  given the data $\bm{y} = (\, y_1, \dots, y_n \, )^{'}$ is \begin{equation}
\label{log_likelik_RJMCM} 
\mathscr{L} ( \,  k, \, \bm{m}_{\, (k)}, \, \bm{s}_{\, (k)}, \, \bm{\theta}_{\, (k)}, \, | \,  \bm{y} \,) = \prod_{j=1}^{k+1} \mathscr{L} ( \, m_j, \, \bm{\theta}_{\, j} \, | \, \bm{y}_j \,  ), \quad \bm{y}_j = \big( \,  y_t:   \, t \,   \in \, \text{I}_j  \, \big), 
\end{equation}
where  \begin{equation}
\label{likelik_segment}
\mathscr{L} ( \,  m_j, \,  \bm{\theta}_{\, j} \, | \, \bm{y}_j \,  ) =  ( \, 2\pi \sigma^2_j \, )^{\,-n_j/2} \exp \Bigg[ -\dfrac{1}{2\sigma^2_j} \,  \sum_{ \, t \,   \in \, \text{I}_j}^{} \Bigg\{ \,  y_t - \bm{x}_t \, \big( \bm{\omega}_{ j} \big)^{\, '} \,  \bm{\beta}_{\, j} \,  \Bigg\}^{\, 2}\Bigg],
\end{equation}
$\bm{\theta}_{\, j} = (\, \bm{\beta}_{ \, j}^{'}, \, \bm{\omega}_{\, j}^{'}, \,  \sigma^{\, 2'}_{\, j}  \, )^{'}
$ is the vector of parameters, $n_j$  the number of observations of the j$^{th}$ segment, and the vector of basis functions $\bm{x}_t \, \big( \, \bm{\omega}_j \, \big)$ is defined as 
\begin{equation*}
\label{basis_functions}
\bm{x}_t \, \big( \bm{\omega}_{j} \big) = \big(1, \,  t, \, \cos(2\pi\omega_{j,1}t), \, \sin(2\pi\omega_{j,1}t), \dots, \cos(2\pi\omega_{j,m_j}t), \, \sin(2\pi\omega_{j,m_j}t)  \big)^{'}.
\end{equation*}
% , and the notation $\floor{s}$ means the largest integer less than or equal to $s$. 

\section{Bayesian Inference} \label{s:bayesian_inference}
Given some pre-fixed maximal numbers of change-points, $k_{\text{max}}$, and  frequencies   per regime, $m_{\text{max}}$,  inference is achieved by assuming that the true
model is unknown but comes from a finite class of models  where each model $\mathcal{M}_k$, with $k$ change-points,  is parameterized by the vector 
$$ (\, \bm{m}_{\, (k)}, \, \bm{s}_{\, (k)}, \,  \bm{\theta}_{\, (k)} \,)  \in \, \bm{\Pi}_k, \quad \bm{\Pi}_k \in  \, \bm{\Pi}. $$
Let  $ \bm{S}_k = \Big\{ \, \bm{s}_{\, (k)} \in [1, \, n]^{\, k} \, :  \mathbbm{1}_{[1 < s_1 < \dots < s_k < n]} \,  \Big\}$  and  $ \bm{\Omega}_{m_j} = (0, 0.5)^{\, m_j}$
denote, respectively,  the sample space for the locations of change-points and   the frequencies of the j$^{th}$ segment.
 The overall parameter space can be written as a finite union of subspaces \begin{equation*}
\bm{\Pi} = \bigcup_{k = 0}^{k_{\text{max}}} \Big\{ \,  k \, \Big\} \times \bm{\Pi}_k, \qquad \text{and} \, \, \, \, \bm{\Pi}_k =  \bm{S}_k \,  \times \, \prod_{j=1}^{k+1} \big\{ m_j \big\} \, \times \, \bigcup_{m_j = 1}^{m_{\text{max}}} \Big\{ {\rm I\!R}^{2 m_j + 2}  \times \bm{\Omega}_{m_j} \times {\rm I\!R}^{+}\Big\}.
\end{equation*}
Bayesian inference on $k, \, \bm{m}_{\, (k)}$, $\bm{s}_{\, (k)}$ and $\bm{\theta}_{\, (k)}$ is based on the following factorization 
of the joint posterior distribution 
\begin{equation*}
\pi \, (\, k, \, \bm{m}_{\, (k)}, \, \bm{s}_{\, (k)}, \, \bm{\theta}_{\, (k)} \,   | \, \bm{y} \, ) = \pi \, (k \, | \,  \bm{y}) \, \pi \, (\bm{m}_{\, (k)} \, | \,  k, \, \bm{y}  ) \, \pi \, ( \bm{s}_{\, (k)} \, | \, \bm{m}_{\, (k)}, \, k, \, \bm{y} ) \,  \pi \, (\bm{\theta}_{\, (k)} \, | \, \bm{s}_{\, (k)}, \,  k,\,  \bm{m}_{\, (k)}, \, \bm{y} ),
\end{equation*} where we use $\pi\, (\, \cdot \, )$ as generic notation for  probability density or  mass function, whichever is appropriate.
Sampling from it poses a multiple model selection problem, namely of the number of change-points and number of frequencies in each regime, which can be addressed  by constructing  a reversible-jump MCMC algorithm \citet{green1995reversible}.  The  algorithm in its basic structure iterates between 
 the following  two moves:

% A similar procedure is proposed in Punksaya et al. (2002); they address the problem of fitting a signal via sequence of piecewise constant % linear regression models, where number of knots and orders of the linear regression models within each segments are both unknown.

\begin{enumerate}
	\item \label{itm:seg_model}  \textbf{Segment model move:}
	Given a partition of the data at $k$ locations $\bm{s}_{\, (k)}$,   inference on the 
	parameters $\bm{m}_{\, (k)}$ and $\bm{\theta}_{\, (k)}$ is based on the conditional posterior    
	$$\pi\, (\bm{m}_{\, (k)}, \, \bm{\theta}_{\, (k)} \, | \, k, \, \bm{s}_{(k)}, \, \bm{y} ) = \prod_{j = 1}^{k+1} \pi \, (m_j, \, \bm{\theta}_{\, j} \, | \,  k, \, \bm{s}_{\, (k)}, \, \bm{y}_j) .$$
	A reversible-jump MCMC algorithm is performed in parallel on each of the $k+1$ segments, where 
	at each iteration the number of sinusoids $m_j$, the linear coefficients $\bm{\beta}_{\, j}$, the frequencies $\bm{\omega}_{\, j}$ and the residual variances $\sigma^2_{j}$ are sampled independently in each segment, for $j = 1, \dots, k+1$. Notice that at this stage the algorithm will explore subspaces of variable dimensionality  regarding  
	the number of frequencies per segment, while the change-point model  remains fixed. 
	\item \label{itm:cp_model}  \textbf{Change-point model move:}
	This step performs a reversible-jump MCMC algorithm for change-point model search where   the number $k$ and locations of change-points $\bm{s}_{\, (k)}$ are sampled, 
along with the linear coefficients,  number of frequencies and their values as well as  the residual variances for 
any segments affected by the move.
\end{enumerate}

Our prior specifications assume independent Poisson distributions for the number of break-points $k$ and  frequencies in each segment $m_j$,
conditioned on $k \leq k_{\text{max}}$ and $1 \leq m_j \leq m_{max}$, respectively.  
 Given  $k$, a  prior distribution 
for the positions of the change-points $\bm{s}_{\,(k)}$ can be chosen as in  \citet{green1995reversible} 	
\begin{equation}
\label{prior_locations_Green}
\pi \, (\bm{s}_{\, (k)} \, | \, k) = \dfrac{(2k+1)!}{(n-1)^{2k+1}} \prod_{j=0}^{k}\Big(  s_{j+1} - s_j \Big) \,  \mathbbm{1}_{[s_0 < s_1 < \dots < s_k < n]}, \qquad s_0 = 1, s_{k+1} =n. 
\end{equation} 
Conditional on $k$ and $\bm{m}_{\, (k)}$, we choose a uniform prior for the frequencies 
$ \omega_{j, \, l} \sim \text{Uniform}(0, 0.5), \quad l = 1, \dots, m_j, \, \, \, \text{and} \, \, \, j = 1, \dots, k+1.  $ 
Analogous  to a Bayesian regression  \citep{Bishop:2006:PRM:1162264}, a zero-mean isotropic Gaussian prior is assumed for the  coefficients of the j$^{th}$ segment,
$\bm{\beta}_{\, j} \sim \mathcal{\bm{N}}_{2m_j+2} (\, \bm{0}, \, \sigma^2_{\beta} \, \bm{I}\, ), \quad j = 1, \dots, k+1, $ 
where $\sigma^2_\beta$ is a pre-specified large value, and the prior on the residual variance $\sigma^2_j$ of the j$^{th}$ partition is 
$\text{Inverse-Gamma} \, \big(\frac{\nu_0}{2}, \frac{\gamma_0}{2}\big)$, where $\eta_0$ and $\nu_0$ are fixed at small values.

\section{Sampling Scheme for Nonstationary Periodic Processes}

\label{sampling_scheme}
Here we provide the sampling scheme associated with the nonstationary periodic processes that we wish to model.
An outline of the overall procedure is  as follows. Start with an initial configuration of number of change-points $k$, 
along with their locations $\bm{s}_{\,(k)}$; this yields  a partition of the data $\bm{y} = $ $(\bm{y}_1, \dots, \bm{y}_{k+1})$. 
Initialize the number of frequencies in each regime $\bm{m}_{\,(k)}$ and their values $\bm{\omega}_{\,(k)}$, 
along with the  coefficients $\bm{\beta}_{\,(k)}$ and residual variances $\bm{\sigma}^2_{\,(k)}.$  
At each iteration of the algorithm a segment model  and a change-point model move are estimated. 
A random choice with probabilities  (\ref{transition_prob}) based on the current number of parameters 
will determine whether to attempt a birth, death or a within-model move.
In particular, let $z$ denote the current number of parameters, 
i.e. change-points $k$ in the change-point model or  frequencies $m_j$ in the j$^{th}$ segment model;
then, the dimension may increase by one (\textit{birth step}) with probability $b_z$, 
decrease by one (\textit{death step}) with probability $d_z$ 
or remain unchanged (\textit{within step}) with probability $\mu_z = 1 - b_z - d_z$, where 
\begin{equation}
\label{transition_prob} 
b_z = c \, \text{min} \Bigg\{ 1, \dfrac{\pi\, (z+1)}{\pi\,(z)} \Bigg\}, \qquad d_{z+1} = c \, \text{min} \Bigg\{1, \dfrac{\pi\,(z)}{\pi\,(z+1)} \Bigg\},
\end{equation}
for some constant $c$ $\in [0, \frac{1}{2}]$, and $\pi \, (z)$ is the prior probability of the model including $z$. 
Reversibility of the Markov chain is guaranteed for move types that involve a change in dimensionality as $ b_z  \, \pi\,(z) = d_{z+1} \, \pi\,(z+1). $ 
Here we chose $c=0.4$ but other values are legitimate as long as $c$ is not larger than $0.5$, 
to assure that the sum of the probabilities  does not exceed 1 for some values of $z$.  
Naturally,  $b_{k = k_{\text{max}}} = b_{m = m_{\text{max}}} = 0$ and $d_{k = 0} = d_{m = 1} = 0$. 
The pseudocode of the overall algorithm that describes an iteration of the sampler is given in Algorithm \ref{a:algoritmo}. We next describe in more detail the specific procedures needed to update the moves.

\begin{algorithm}[h] 
	\caption{}
	\begin{enumerate}
		\item For each segment $j = 1, \dots, k+1$, perform a segment model move (Section \ref{Segment_model})
		\begin{itemize}
			\setlength\itemsep{0.1em}
			\item[] Draw $U$ $\sim$ $\text{Uniform} \, (0, 1) $
			\item[] \textbf{if} $U \leq b_{m_j}$ $ \, \rightarrow \, $ \textit{birth-step}
			\item[] \textbf{else if} $b_{m_j} \leq U \leq d_{m_j}$ $ \, \rightarrow \, $  \textit{death-step}
			\item[] \textbf{else} $ \, \rightarrow \, $ \textit{within-step}
		\end{itemize}  
		\item Perform a change-point model move (Section \ref{Break-point_model}):
		
		\begin{itemize}
			\setlength\itemsep{0.1em}
			\item[] Draw $U$ $\sim$ $\text{Uniform} \, (0, 1) $  
			\item[] \textbf{if} $U \leq b_{k}$ $ \, \rightarrow \, $ \textit{birth-step}
			\item[] \textbf{else if} $b_{k} \leq U \leq d_{k}$ $ \, \rightarrow \, $  \textit{death-step}
			\item[] \textbf{else} $ \, \rightarrow \, $ \textit{within-step}
		\end{itemize}
	\end{enumerate}
	\label{a:algoritmo}
\end{algorithm}

\subsection{Updating a Segment Model} \label{Segment_model}
Given the number of change-points $k$ and their locations $\bm{s}_{\, (k)}$, a segment model move 
is performed independently and in parallel on each of the $k+1$ partitions. 
Hence, throughout this subsection the subscript 
relating to the j$^{th}$ segment may be dropped and a segment of interest is  
denoted by $\bm{y} = (y_a, \dots, y_b)^{'}$, which contains $n$ observations. Assume that the current number of frequencies is set at $m$; then, an independent random choice is made between attempting a birth, death or within-model step, with probabilities given in  \eqref{transition_prob}. An outline of these moves is as follows (further details are provided in the Appendix).

{\bf Within-Model Move:} \label{segment_model_within} 
Conditioned on the number of frequencies $m$, we sample the vector of frequencies $\bm{\omega}$  following \cite{andrieu1999joint}, i.e. by sampling the frequencies one-at-time  using a mixture of M-H steps, with target distribution 
\begin{equation} \label{posterior_omega}
\pi \, (\bm{\omega} \, | \, \bm{\beta},  \, \sigma^2, \, m, \, \bm{y}) \propto \exp \Bigg[ -\dfrac{1}{2\sigma^2} \sum_{t \, = \, a}^{b} \Big\{ y_t - \bm{x}_t \, \big( \, \bm{\omega} \, \big)^{\, '} \,  \bm{\beta} \, \Big\}^{2}  \Bigg] \mathbbm{1}_{\big[ \, \bm{\omega} \,  \in  \, \bm{\Omega}_{m} \big] \, }.
\end{equation}
In particular, the proposal distribution is a combination of a Normal random walk centred around the current frequency  and a sample from values of the Discrete Fourier transform of $\bm{y}$.  The corresponding vector of linear parameters $\bm{\beta}$ is then updated in a M-H step, from the target posterior conditional distribution
\begin{equation} \label{posterior_beta}
\pi\, ( \, \bm{\beta} \, | \,  \bm{\omega}, \, \sigma^2, \,  m, \, \bm{y}) \propto \exp \Bigg[ -\dfrac{1}{2\sigma^2} \sum_{t \, = \, a}^{b} \Big\{ y_t - \bm{x}_t \, \big( \, \bm{\omega} \, \big)^{\, '} \,  \bm{\beta} \, \Big\}^{2} - \dfrac{1}{2\sigma^2_\beta} \, \bm{\beta}^{\, '}  \bm{\beta} \Bigg],
\end{equation}  
where the proposed values are drawn from normal approximations to their posterior conditional distribution. Finally, the residual variance $\sigma^2$ is then updated in a Gibbs step from 
\begin{equation}
\label{inverse_gamma}
\sigma^2_{| \, \bm{\omega}, \, \, \bm{\beta}} \sim \text{Inverse-Gamma} \, \Bigg( \, \dfrac{n + \nu_0}{2}, \, \dfrac{\gamma_0 + \sum_{t \, = \, a}^{b} \Big\{ \, y_t - \bm{x}_t \, \big( \, \bm{\omega} \, \big)^{\, '} \,  \bm{\beta} \, \Big\}^{2}}{2}  \Bigg).  
\end{equation}

{\bf Between-Model Moves: }  \label{segment_model_between} 
For this type of move, the number of frequencies is either proposed to increase by one (birth) or decrease by one (death).  If a birth move is proposed, we have that $m^{\,p} = m^{\,c} + 1$, where current and proposed values are denoted by the superscripts \textit{c} and \textit{p}, respectively. The proposed vector of frequencies is constructed by proposing an additional frequency to include in the current vector. Conditional on the frequencies, the corresponding vector of linear coefficients and the residual variance are sampled as in the within-model move. If a death move is proposed, we have that $m^{p} = m^{c} - 1$.  Hence, one of the current frequencies is randomly chosen to be removed. The proposed corresponding vector of linear coefficients is drawn, along with the residual variance.  For both moves, the updates are jointly accepted or rejected in a M-H step.

\subsection{Updating the Change-Point Model} \label{Break-point_model}
This part of the algorithm  identifies the number and locations of change-points. 
Suppose the number of change-points is currently set to some value $k$, 
then according to the probabilities given in  \eqref{transition_prob} 
a random decision is made between adding, removing or moving a  change-point. 
The rules for updating these types of moves are described below and more details are given in the Appendix.

{\bf Within-Model Move: } \label{cp_model_within} 
An existing change-point is proposed to be relocated  with probability $\frac{1}{k}$, obtaining say $s_j^{\, c}$. The update for the selected change-point is proposed from a mixture of a Normal random walk centred on the current change-point $s_j^{\, c}$ and a sample from a uniform distribution on the interval $[  s_{j-1}^{\, c} + \psi_s, s_{j+1}^{\, c} - \psi_s ]$. Here, we introduced $\psi_s$ as a fixed minimum time between change-points avoiding change-points being too close to each other. \citet{rosen2012adaptspec} used a similar scheme, but on a discrete-scale. The number of frequencies 
and their values are kept fixed, and, conditional on the relocation, the linear coefficients for the segments affected by the relocation are sampled. 
These updates are jointly accepted or rejected in a M-H step and the 
residual variances are  updated in a Gibbs step. 

\vspace{0.1cm}
{\bf Between-Model Moves:} \label{cp_model_between} 
For this type of move, the number of change-points may either increase (birth) or decrease (death) by one.  If a birth move is proposed, we have that $k^{\, p} = k^{\, c} + 1$. The new proposed change-point is drawn uniformly on $f \, (\bm{s}_{\,(k^{\, c})}^{\, c}, \, \psi_s)$, the support of $\bm{s}_{\,(k^{\,c})}^{\, c}$ given the constraints imposed by $\psi_s$, i.e. $f \, (\bm{s}_{\,(k^{\, c})}^{\, c}, \, \psi_s) = [1 + \psi_s, \,  s_1^{\,c} - \psi_s] \,   \cup \,  [s_1^{\,c} + \psi_s, \,  s_2^{\,c} - \psi_s] \cup \,  \dots \, \cup \, [s_{k^{\,c}}^{\,c} + \psi_s, \,  n - \psi_s]$.  The latter involves splitting an existing segment. The number of frequencies and their values in the proposed segments are selected from the current states. Two residual variances for the new proposed segments are then constructed from the current single residual variance. Finally, two new vectors of linear parameters are sampled.  If a death move is proposed, we have that $k^{\, p} = k^{\, c} - 1$. Hence, a candidate change-point to be removed is selected from the vector of existing change-points, with probability $\frac{1}{k^{\, c}}$. The latter involves merging two existing partitions. The number of frequencies and their values in the proposed segments are selected from the current states. A single residual variance is constructed from the current variances relative to the segments affected by the relocation. Finally, a new vector of linear coefficient is drawn.  For both type of moves, these updates are jointly accepted or rejected. 

\normalsize
\section{Simulation Studies } \label{simulation_study}
We carry out simulation studies to explore the performance of our method, which will be referred to as AutoNOM (Automatic Nonstationary Oscillatory Modelling). In Section \ref{illustrative_example} we illustrate the performance of our methodology when the simulated data are generated from the proposed model. Section \ref{misspecified_model} deals with scenarios when the model is misspecified relative to the generating process. Our results are compared with two state-of-the-art existing methods.

\subsection{Illustrative Example} \label{illustrative_example}
In this simulation example, we generate a time series consisting of  $n = 900$ data points from model \eqref{model_RJMCMC_1} with $k = 2$ change-points located at positions $\bm{s}_{\,(2)} = (300, 650)$, and fixed  number of frequencies per regime $\bm{m}_{\,(2)} = (3, 1, 2)$. (Further details of the
parameterization are available in Supplementary Material, Section 1.1.)  Figure \ref{fig:data_illustrative_1} (top panel) shows a realization  from this model. The prior means $\lambda_{\omega}$ and $\lambda_s$, say,  on the number of frequencies and  change-points, respectively, were set  to 2, to reflect a fair degree of prior information on their numbers. We  discuss in  Section 1.2 of the Supplementary Material that AutoNOM was relatively insensitive to these prior specifications, for this example.  The maximum number of change-points $k_{\text{max}}$ was set to 15, and the maximum number of frequencies per regime $m_{\text{max}}$ was set to 10.  
Furthermore, we fixed  $\psi_s = 20 $ and  $ \phi_{\omega}$ = 0.25 (Appendix \ref{appendix_segment_between}) for the uniform distribution for sampling the frequencies.  
The full estimation algorithm was ran for 20,000 updates, 5,000 of which are discarded as burn-in period.
The estimation took 390 seconds with a (serial) program written in Julia 0.62 on a Intel\textsuperscript{\textregistered} Core\textsuperscript{TM} i7-4790S Processor 16 GB RAM.
The results, summarized in Table \ref{table:posterior_k_m_illustrative_1} clearly show that  a model with two change-points has the highest estimated posterior probability (left panel) and that  AutoNOM correctly identifies the right number of significant frequencies in each regime (right panel).
\begin{figure}[htbp]
	\centering
	\includegraphics[width = 16.5cm, height = 6.0cm]{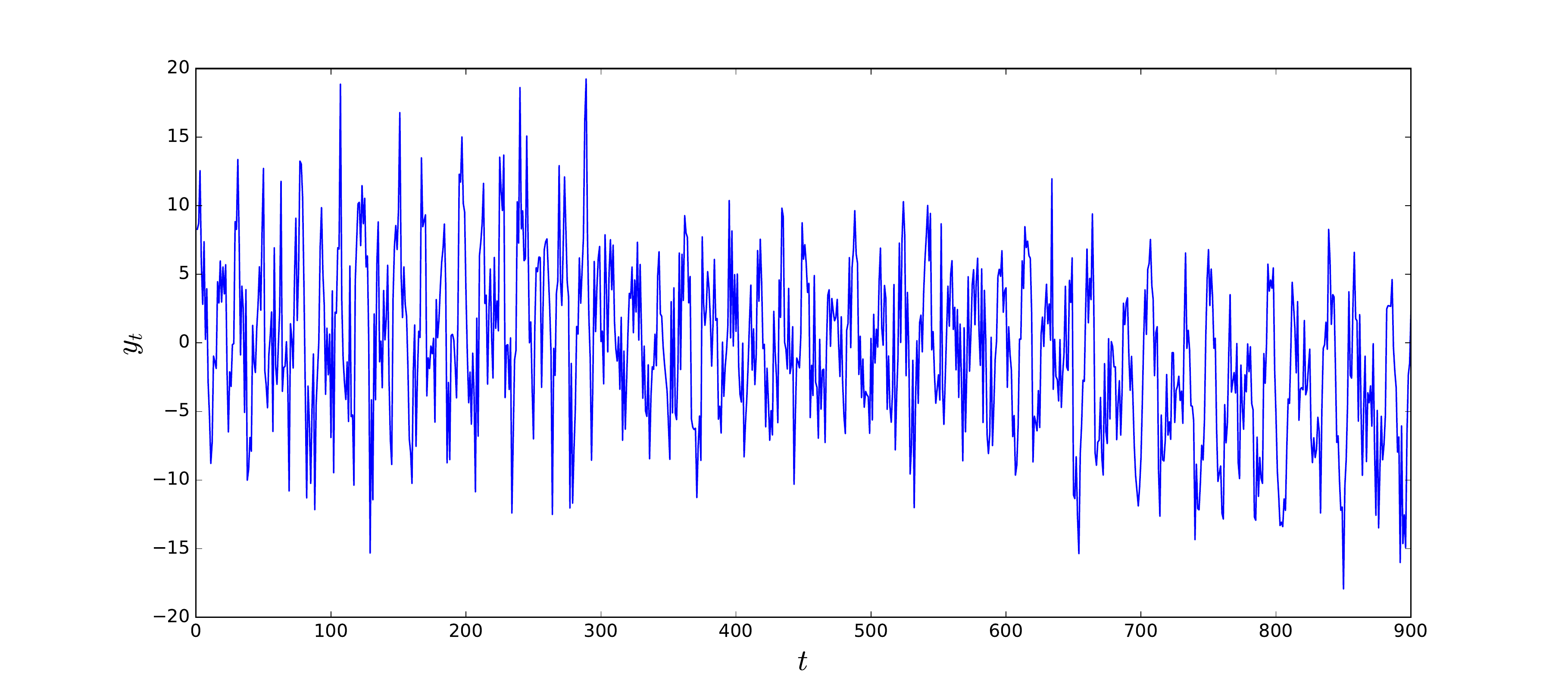}
	\includegraphics[width = 16cm, height = 4.5cm]{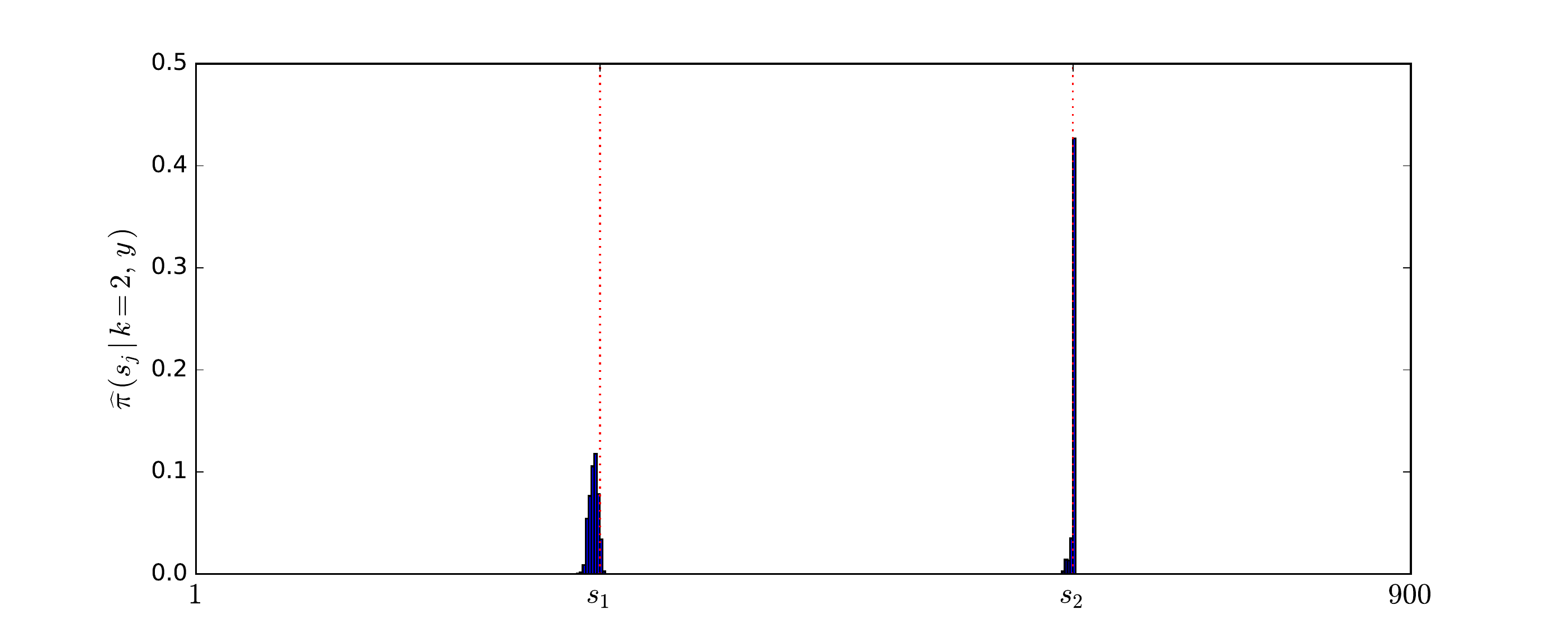}
		\includegraphics[height = 6cm, width=.30\textwidth]{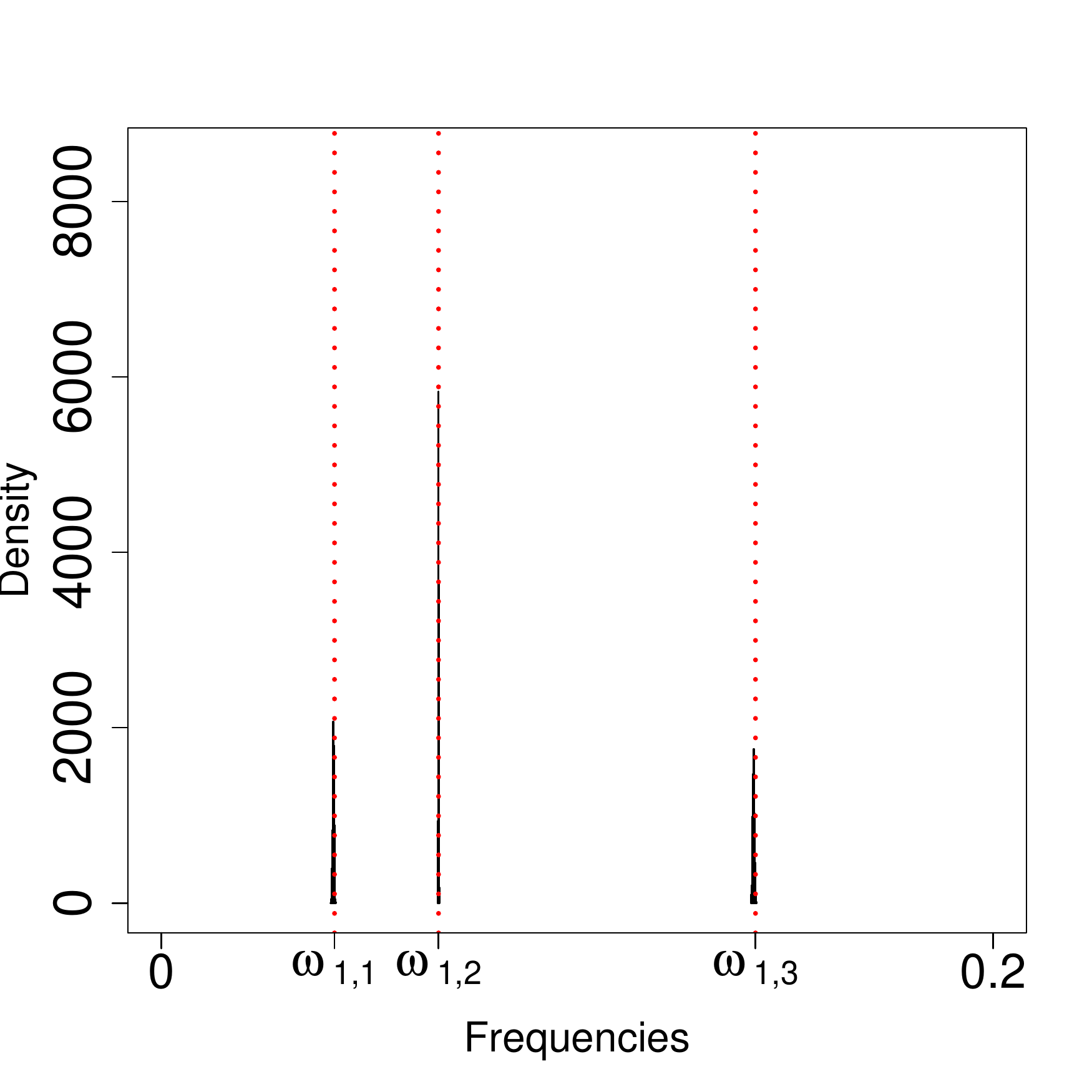}
	\includegraphics[height = 6cm,width=.30\textwidth]{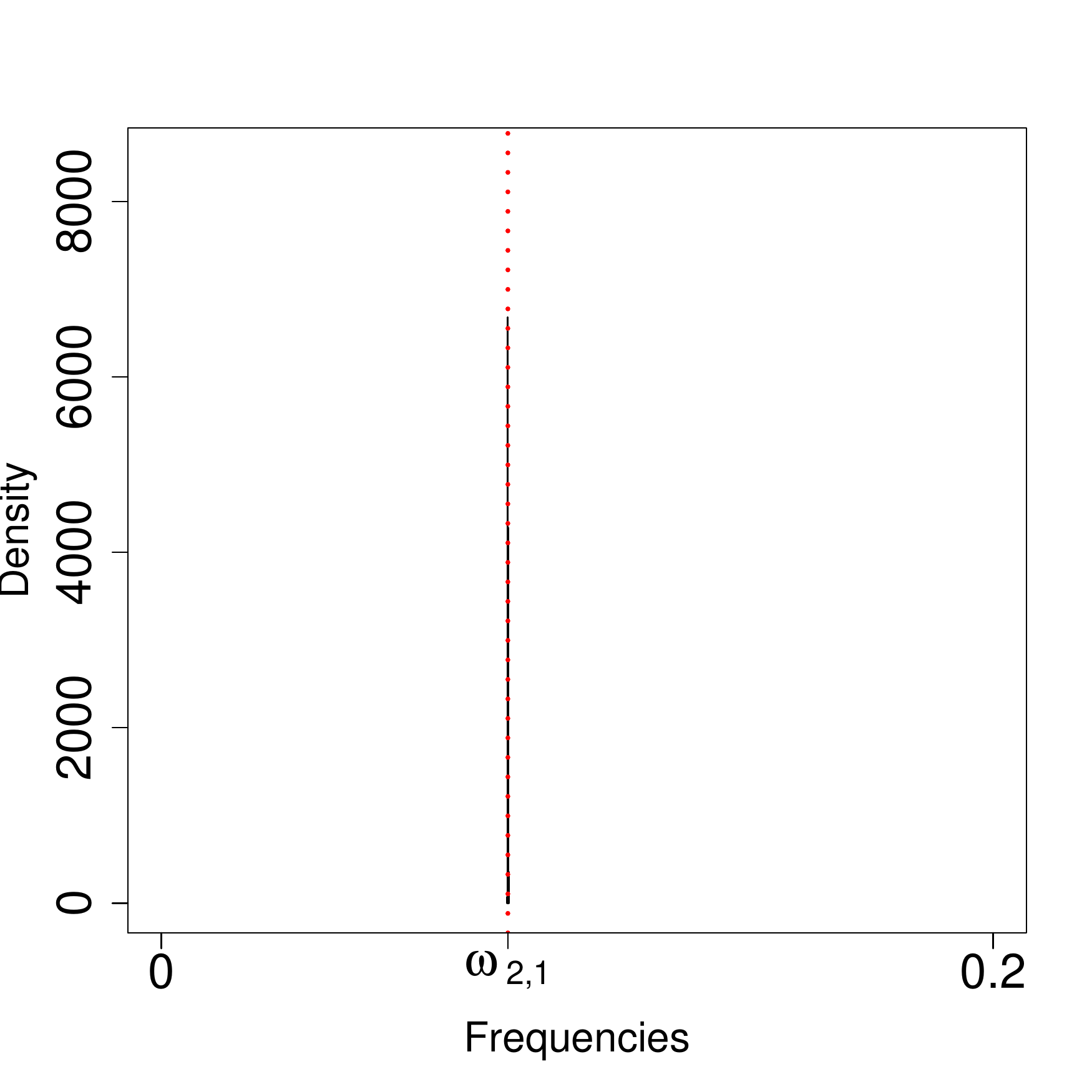}
	\includegraphics[height = 6cm,width=.30\textwidth]{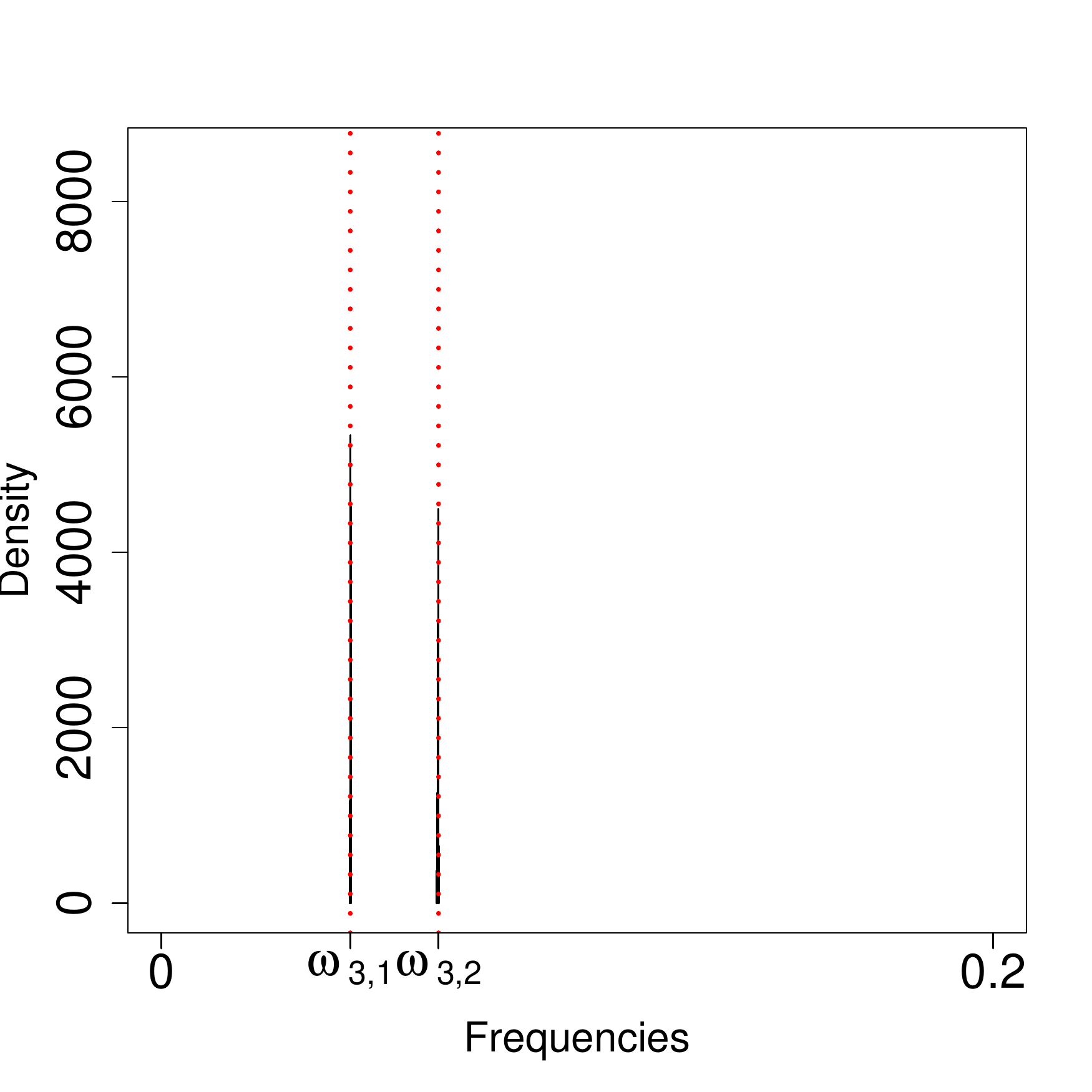}
	\caption{Illustrative example. (Top) Simulated time series.
	(Middle) Estimated posterior distribution for the location of the change-points, conditioned on $k = 2$. The  dotted vertical lines represent true location of change-points. (Bottom) Estimated posterior distribution of the frequencies for each different segment, conditioned on $k = 2$, $m_1 = 3$, $m_2 = 1$  and  $m_3 = 2$. The dotted vertical lines represent true values of the frequencies.}
	\label{fig:data_illustrative_1}
\end{figure}

%including two change-points located at positions $\bm{s}_{(2)} = (300, 650)$. The rest of the parameterization is given in Table \ref{table:parameter_illustrative_1}. 

 Figure \ref{fig:data_illustrative_1} (middle panel) shows the estimated posterior distribution for the location of the change-points, conditioned on three segments. 
 The posterior means of the change-point locations are $\hat{\mathbb{E}} \, (s_1 \, | \,k = 2, \,  \bm{y}) = 298.7$ and $\hat{\mathbb{E}} \, (s_2 \, | \,k = 2, \,  \bm{y}) = 650.1$.  Figure \ref{fig:data_illustrative_1} (bottom panel) shows that the estimated posterior distributions are an excellent match to the true frequencies. In addition, we provide details about acceptance rates in Supplementary Material, Section 2.

\begin{table}[htbp]
	\centering
	\caption{Illustrative example. (left panel) posterior probabilities for number of change-points; (right panel) posterior probabilities for number of frequencies in each regime, conditioned on $k = 2$. }
	\label{table:posterior_k_m_illustrative_1}
	\begin{tabular}{cc}
		\hline \\[-0.9em]
		\hline
		$k$ & $\hat{\pi} \, (\, k \, | \, \bm{y})$ \\[.1em] \hline
		0 &      .00\\
		1 &      .02\\
		2 &     .97 \\
		3 &     .01 \\
		4 &     .00 \\ \hline
	\end{tabular} \quad \quad
	\begin{tabular}{cccc}
		\hline \\[-0.9em]
		\hline
		$m$ & $\hat{\pi} \, (\, m_1 \, | \, k = 2,  \, \bm{y})$ & $\hat{\pi} \, (\, m_2 \, | \, k = 2,  \, \bm{y})$ & $\hat{\pi} \, (\, m_3 \, | \, k = 2,  \, \bm{y})$ \\[.1em] \hline
		1 &  .00    & .99     &  .00    \\
		2 & .00     & .01     &  .98    \\
		3 & .98      & .00     &  .02    \\
		4 & .02     & .00     &   .00   \\
		5 &  .00    & .00     &   .00  \\ \hline
	\end{tabular}
\end{table}

%Moreover, in our experiments we notice that when these hyper-parameters are small, Algorithm \ref{alg:overall_algorithm} converges quicker to the target invariant distribution.

%$$ \hat{f}_t =  \dfrac{1}{P} \sum_{i=1}^{P} \sum_{j=1}^{k^{\, (i)} + 1} \Big\{ f \, \big(t, \, \bm{\beta}_{ \, j}, \,  \bm{\omega}_{\, j} \, \big)\mathds{1}_{[ \, t \, \in \,  I_j \, ]}\Big\}  $$ are given in

%displays as it has noticeable little impact on the results we obtained. 

% Theoretically, a small value of $\lambda_\omega$ supports models with a small number of frequency components, whereas a large value of  $\lambda_\omega$ leads to models with a large number of sinusoids; a similar argument follows for the prior mean $\lambda_s$ for the number of change-points

\subsubsection{Detecting Spectral Peaks}

We simulate a time series from the same simulation model as above with the only difference  that the  residual variances were set equal to one for all segments and thus are smaller than above. The performance of AutoNOM is compared with two existing methods, namely the Bayesian adaptive spectral estimation for nonstationary time series proposed by \citet{rosen2012adaptspec}, referred to as AdaptSPEC, and the frequentist piecewise vector autoregressive method of \citet{davis2006structural}, referred to as AutoPARM.  Specifically, we explore the performances of these methodologies  in identifying the number and location of change-points, and the number and location of frequency peaks in each estimated segment. AdaptSPEC requires the user to specify in advance the number of basis function $J$ used for smoothing the 
periodogram in the segments. We run AdaptSPEC for two different specifications, namely $J = 7$ and $ J= 15$ basis functions. The model is fitted with a total of 15,000 iterations, 5,000 of which are discarded as burn-in, by using the R package provided by the authors. Posterior samples of peak frequencies are obtained by considering the modes of the spectrum per MCMC iteration.  AutoPARM is performed with default tuning parameters. We note that \citet{davis2006structural} do not discuss computation of confidence intervals for frequencies.

\begin{figure}[htbp]
	
	\centering
	\includegraphics[width=.32\textwidth]{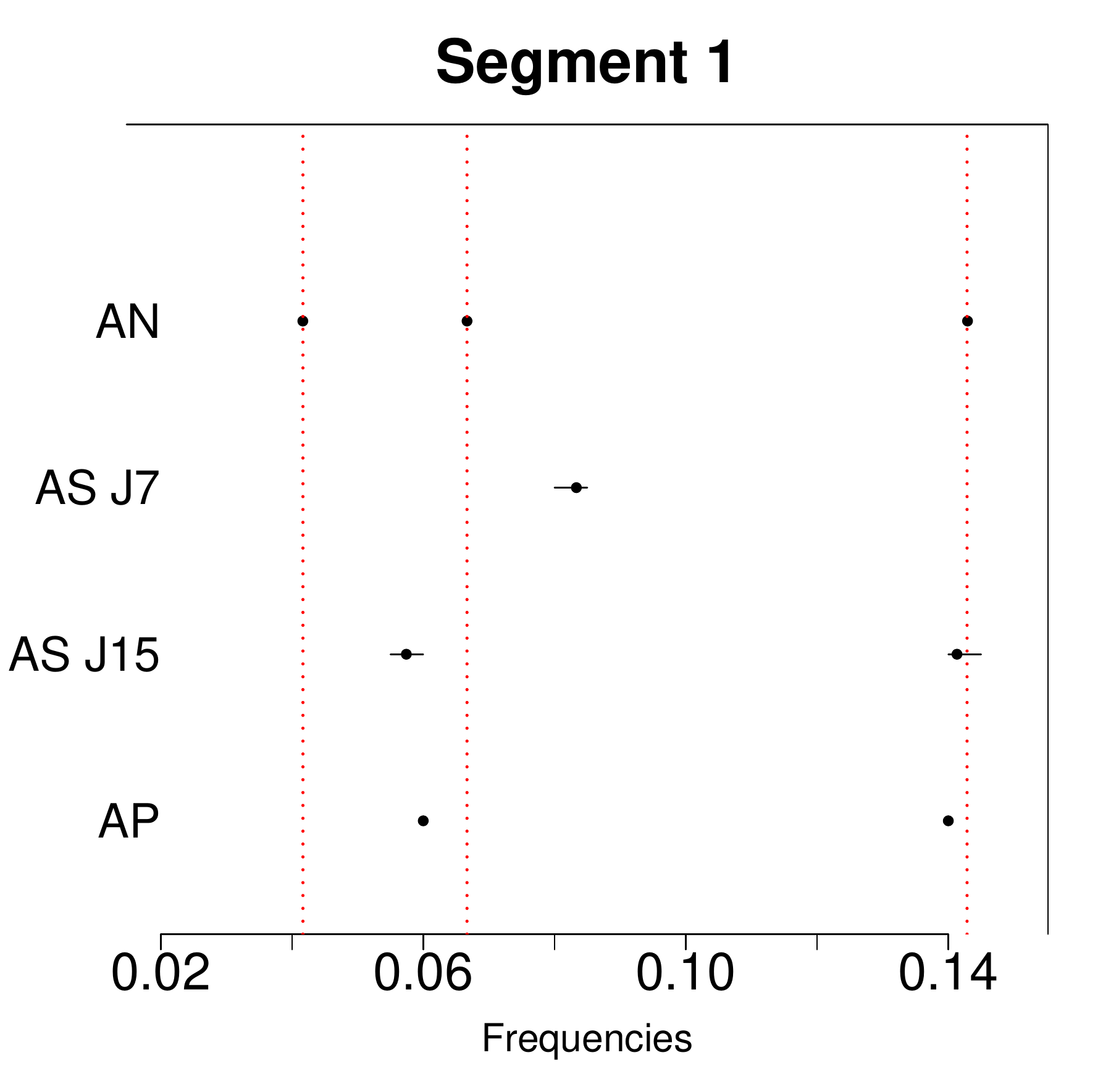}\hfill
	\includegraphics[width=.32\textwidth]{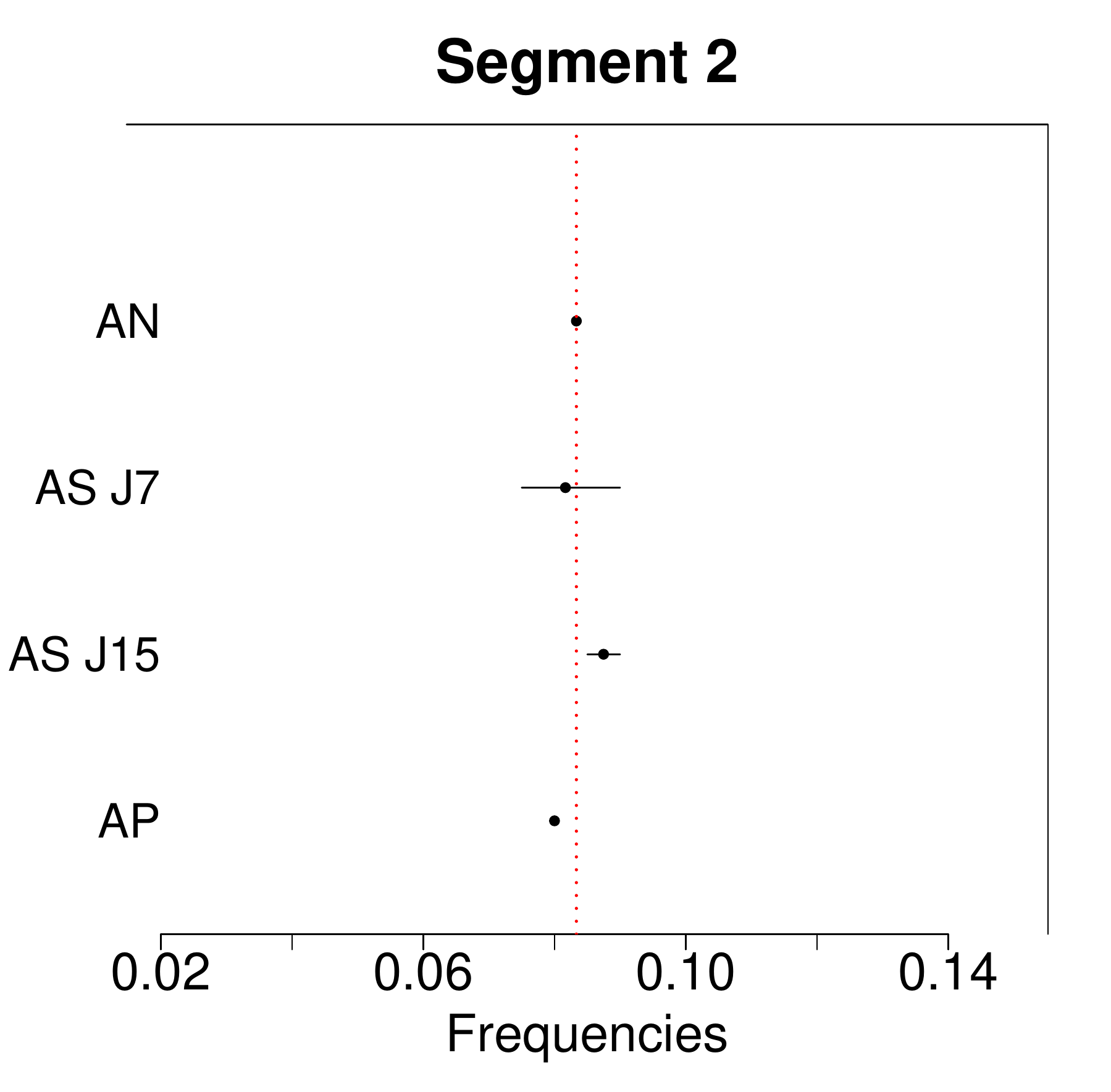}\hfill
	\includegraphics[width=.32\textwidth]{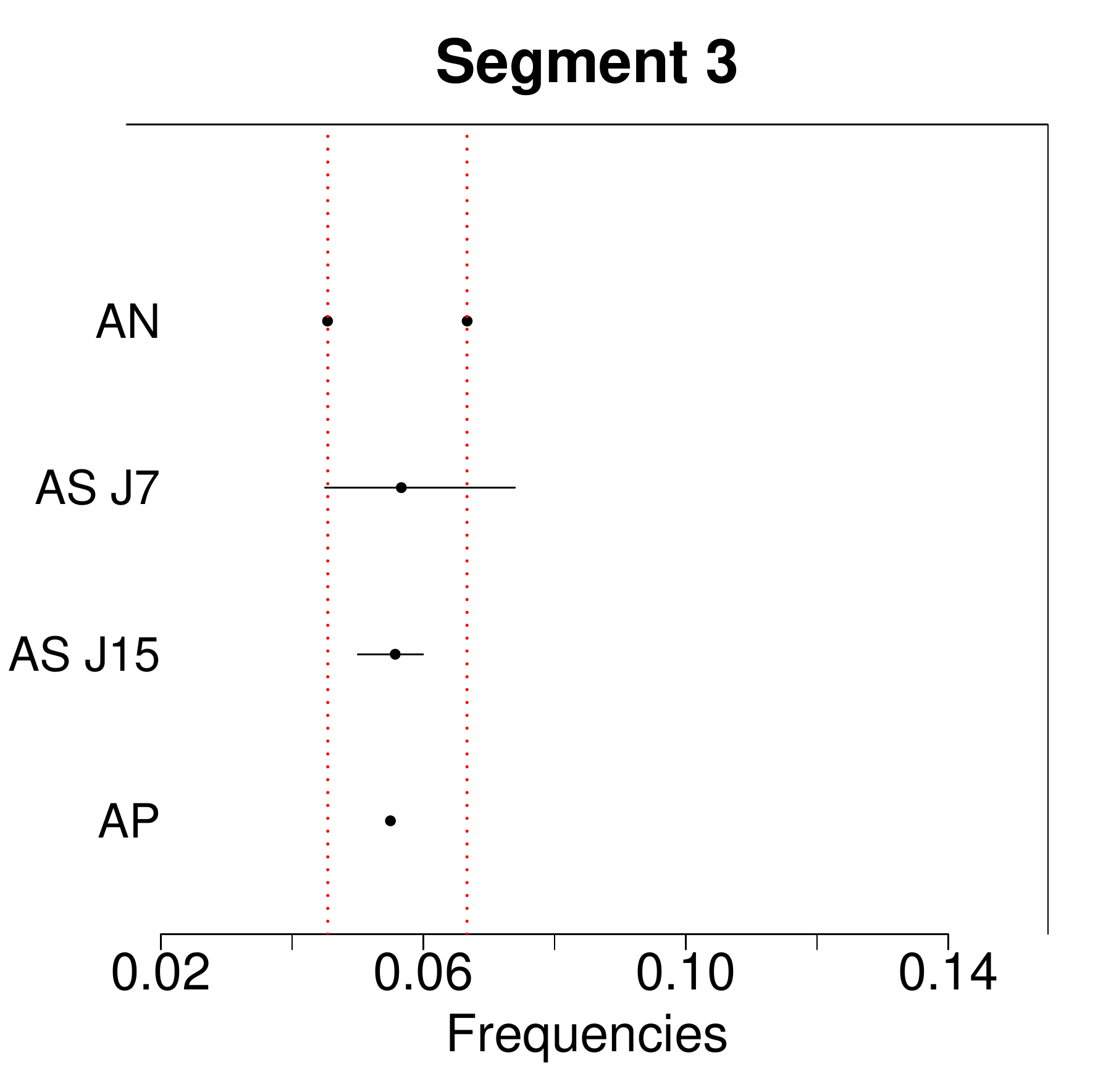}
	\caption{Illustrative example with unitary residual variances. Estimated frequency peaks for AutoNOM (AN), AdaptSPEC (AS, $J = 7, 15$) and AutoPARM (AP); 95\% credible intervals (horizontal lines) are also reported for Bayesian methods. Dotted vertical lines are true locations of the frequency peaks. }
	\label{fig:comparison_methods}
	
\end{figure}

The modal number of change-points for AdaptSPEC is 2 for both $J=7$ and $J=15$, with posterior probability $\widehat{\pi} \, ( k = 2 \, | \, \bm{y})$ of 76\% and 88\%, respectively; the modal number of change-points for AutoNOM is 2 and AutoPARM identifies 2 change-points as well. Conditioned on the modal number of change-points, Table \ref{table:comparison_methods} displays the estimated  location of changes (left panel) and frequency peaks (right panel) for the different compared methods, where we report the standard deviation for the estimate obtained from the empirical distribution of the posterior samples.  Similarly, we show in Figure \ref{fig:comparison_methods} the estimated location of the frequency peaks and their 95\% credible intervals, for each of the three identified segments; dotted vertical lines represents the true location of the frequency peaks. Results for AutoNOM are conditioned on the modal number of frequencies per regime.

\begin{table}[htbp]
	\centering 
	\caption{Illustrative example with unitary residual variances. Estimated  change-points locations (left panel) and frequency peaks (right panel) for AutoNOM (AN), AdaptSPEC (AS $J = 7, 15$) and AutoPARM (AP); posterior standard deviations are also reported for Bayesian methods. }	\begin{tabular}{lrrlrrrrrr}
		\hline \\[-0.9em]
		\hline
		\multicolumn{1}{r}{} & \multicolumn{1}{l}{$s_1$}                                  & \multicolumn{1}{l}{$s_2$}                                  &  & $\omega_{1,1}$                                                   & $\omega_{1,2}$                                                    & $\omega_{1,3}$                                                    & $\omega_{2,1}$                                                     & $\omega_{3,1}$                                                     & $\omega_{3,2}$                                                   \\ \cmidrule{2-3} \cmidrule{5-10}
		True              & 300                                                       & 650                                                       &  & .042                                                     & .067                                                     & .143                                                      & .083                                                     & .046                                                     & .067                                                      \\[0.5em]
		AN                    & \begin{tabular}[c]{@{}r@{}}300.87\\  \footnotesize (.97)\end{tabular}  & \begin{tabular}[c]{@{}r@{}}650.46 \\ \footnotesize (.31)\end{tabular}  &  & \begin{tabular}[c]{@{}r@{}}.042 \\ \footnotesize (.0004)\end{tabular} & \begin{tabular}[c]{@{}r@{}}.067 \\ \footnotesize (.0002)\end{tabular} & \begin{tabular}[c]{@{}r@{}}.142 \\ \footnotesize  (.0005)\end{tabular} & \begin{tabular}[c]{@{}r@{}}.083 \\ \footnotesize (.0002)\end{tabular} & \begin{tabular}[c]{@{}r@{}}.045 \\ \footnotesize (.0003)\end{tabular} & \begin{tabular}[c]{@{}r@{}}.067 \\ \footnotesize (.0003)\end{tabular} \\[0.9em]
		AS J7                & \begin{tabular}[c]{@{}r@{}}319.65 \\ \footnotesize (15.36)\end{tabular} & \begin{tabular}[c]{@{}r@{}}628.51\\  \footnotesize (37.35)\end{tabular} &  & \begin{tabular}[c]{@{}r@{}}.083 \\ \footnotesize (.0030)\end{tabular} & -                                                         & -                                                          & \begin{tabular}[c]{@{}r@{}}.082 \\ (\footnotesize .0070)\end{tabular} & \begin{tabular}[c]{@{}r@{}}.057 \\ (\footnotesize .0010)\end{tabular} & -                                                          \\[0.9em]
		AS J15               & \begin{tabular}[c]{@{}r@{}}298.5\\ \footnotesize (1.21)\end{tabular}    & \begin{tabular}[c]{@{}r@{}}647.01\\  \footnotesize (.14)\end{tabular}  &  & \begin{tabular}[c]{@{}r@{}}.057 \\ \footnotesize (.0030)\end{tabular} & \begin{tabular}[c]{@{}r@{}}.141 \\ \footnotesize (.0020)\end{tabular} & -                                                          & \begin{tabular}[c]{@{}r@{}}.088 \\ \footnotesize (.0050)\end{tabular} & \begin{tabular}[c]{@{}r@{}}.056 \\ \footnotesize (.0030)\end{tabular} & -                                                          \\[0.9em]
		AP                   & 299                                                       & 648                                                       &  & .060                                                     & .140                                                     & -                                                          & .080                                                     & .055                                                     &  - \\[.2em]
		\hline
		\label{table:comparison_methods}                                                                                                                                                 
	\end{tabular}
\end{table}

It becomes clear that the detection of periodicities by
AdaptSPEC is affected by the specification of the number of spline basis functions  used for the smoothing, where increasing the number of basis function yields a better performance for AdaptSPEC. The example also shows that smoothing by splines may lead to  peaks in the periodogram to be over-smoothed and 
neighbouring close peaks to be merged. AutoPARM seems to also suffer from the latter problem.

When we increased the residual variance to the high 
levels set originally,  AdaptSPEC failed to detect any change-points for both $J=7$ and $J=15$, with posterior probability $\widehat{\pi} \, ( k = 0 \, | \, \bm{y})$ of 69\% and 93\%, respectively, while AutoPARM found 7 change-points and thus severely overestimates their number. Our conclusion from this comparison is that   
although   AdaptSPEC and AutoPARM may be well suited for time series 
processes with smooth time-varying spectra with few or no peaks, both methods are severely challenged in detecting changes in spectra that exhibit pronounced peakedness, possibly at nearby frequencies, as can be expected to occur in reality for the type of time series that we wish to analyze.

\subsection{Misspecified Model} \label{misspecified_model}

 We investigate the performance of our proposed method for identifying spectral peaks when the model is misspecified relative to the generating process.  In particular, we explored simulation studies under three different settings. In the first two scenarios we generated data from two types of autoregressive (AR) processes, namely a piecewise AR process and a slowly varying AR process. We compare the performance of our procedure with AutoPARM and AdaptSPEC.  In the third setting we assumed that the innovations are t-distributed, and therefore violate the Gaussianity assumption of $\varepsilon_t$ in Equation \eqref{eq:residual_error}. For all models, our estimation algorithm was run for 20,000 iterations, 5,000 of which were used as burn in, and the hyperparameters were chosen as $\phi_s = 40$, $\lambda_\omega = 0.05$ and $\lambda_s = 0.01$.

\subsubsection{Piecewise Autoregressive Process}

\begin{figure}[htbp]
    \centering
    \includegraphics[height = 4.2cm, width = 14cm]{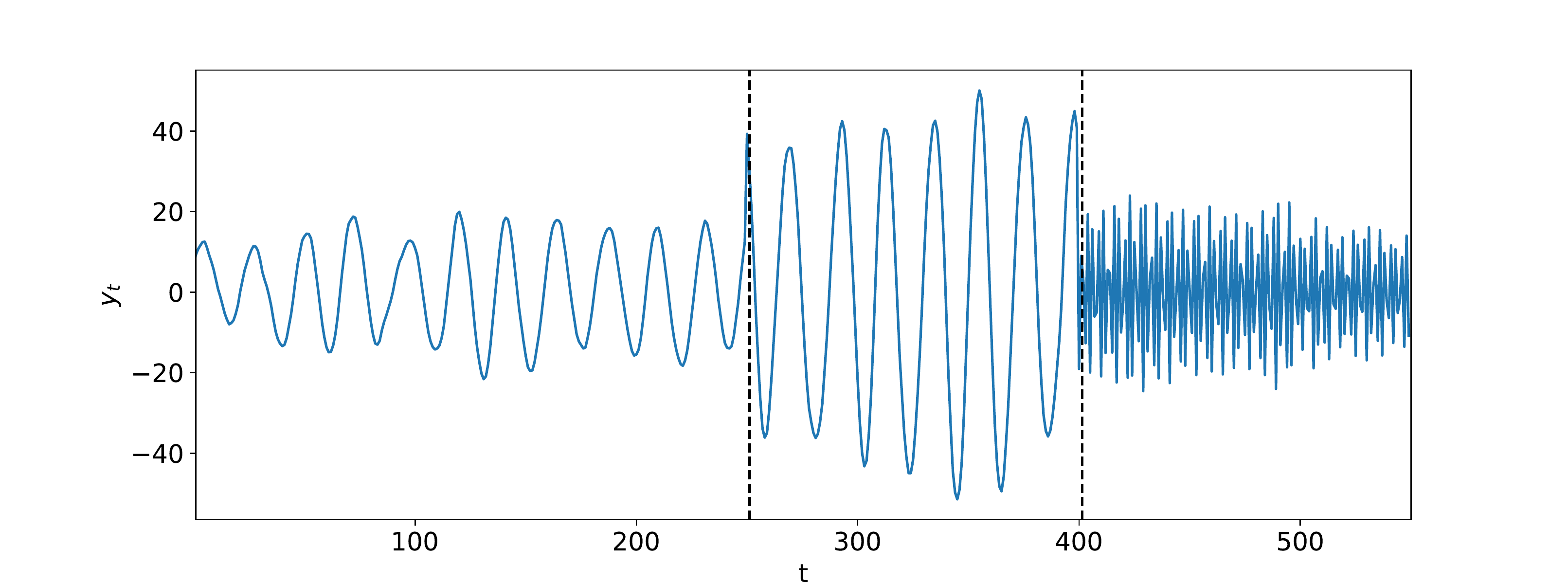}
    \vspace{0.1cm}
    \includegraphics[height = 4.2cm, width=.25\textwidth]{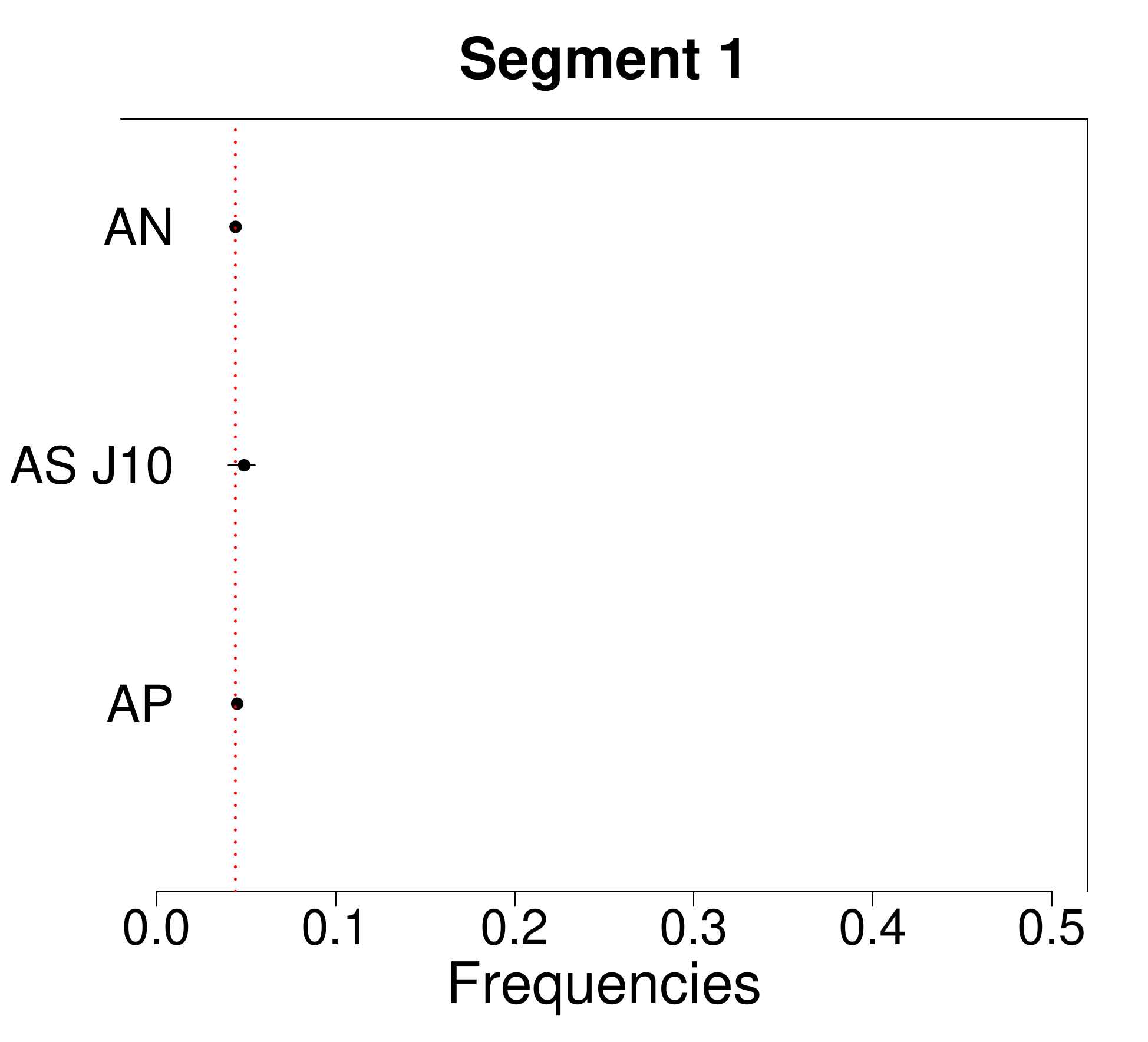}
	\includegraphics[height = 4.2cm,width=.25\textwidth]{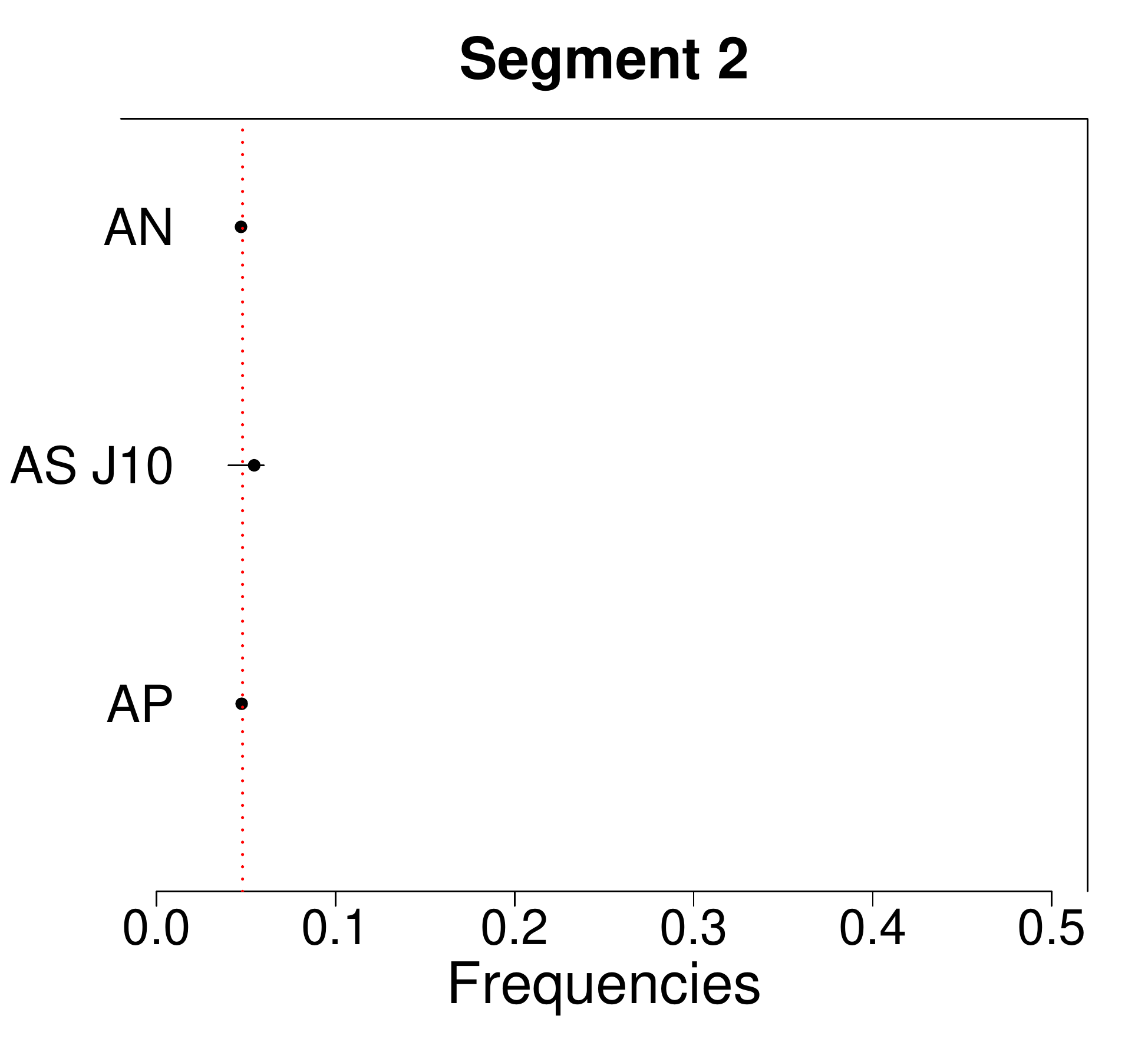}
	\includegraphics[height = 4.2cm,width=.25\textwidth]{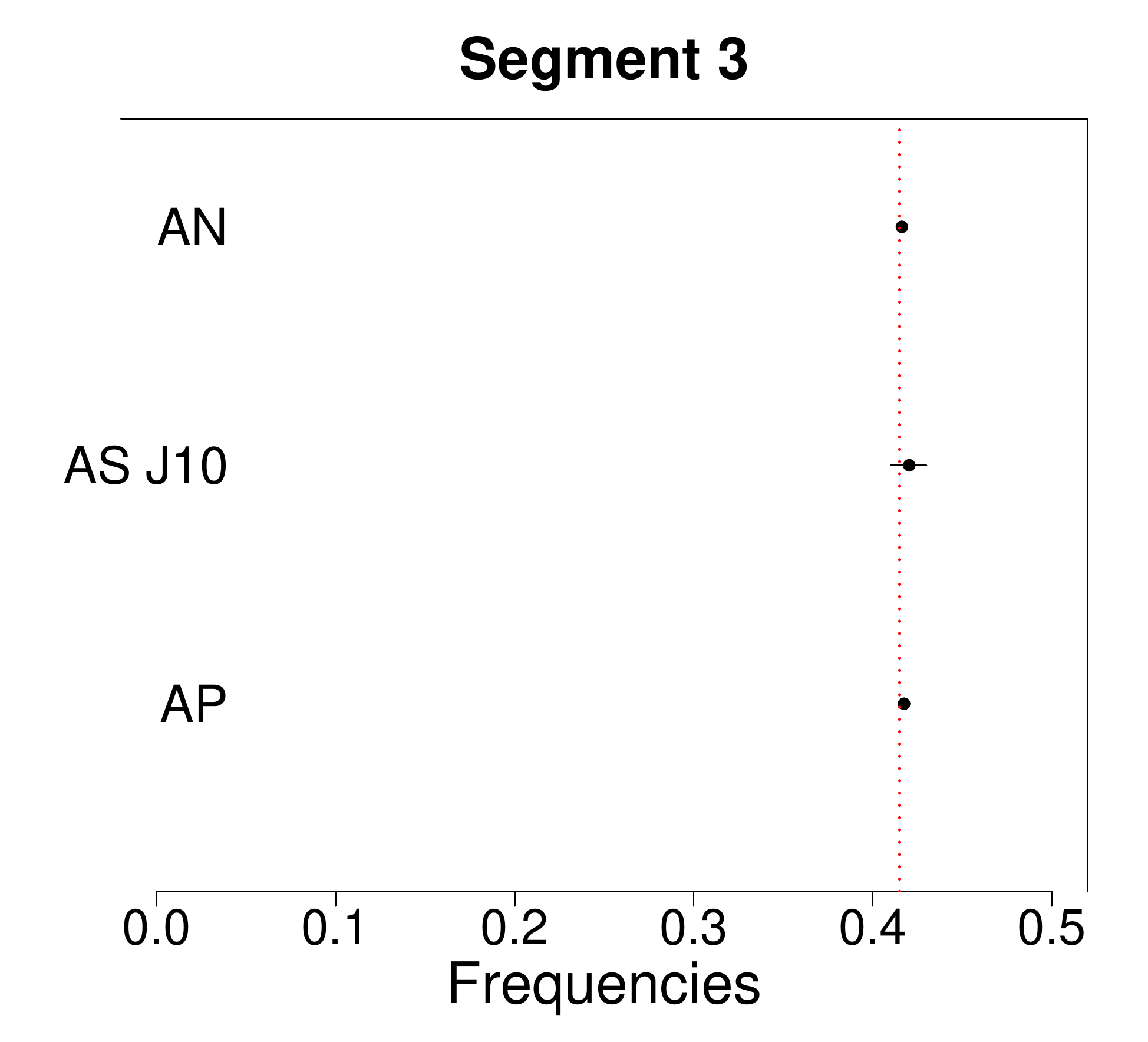}
	    \caption{Piecewise AR process. (Top) A realization from model \eqref{eq:piecewise_AR}. Vertical dotted lines are the estimated locations of the change-points. (Bottom) Estimated frequency peaks for AutoNOM (AN), AdaptSPEC (AS $J = 10$) and AutoPARM (AP); 95\% credible intervals (horizontal lines) are also reported for Bayesian methods. Dotted vertical lines are true locations of the frequency peaks.}
    \label{fig:piecewise_AR_data}
\end{figure}

As pointed out by a referee, although modeling a time series as a linear combination of a finite number of sinusoids plus noise is common in the signal processing literature, such line-spectrum based models are rare in the statistics literature.  In fact,  it is commonly assumed that the power spectrum is continuous across frequencies. We investigate the performance of the proposed procedure when analyzing data generated from a piecewise AR process whose local spectral density functions show sharp peaks. Specifically, a realization is simulated from

\begin{equation} \label{eq:piecewise_AR}
y_t = \begin{cases}
1.9 \, y_{t-1}  -.975 \, y_{t-2} + \varepsilon_{t}^{(1)} & \text{for } 1 \leq t \leq 250 \\ 
1.9 \,  y_{t-1}  -.991 \,  y_{t-2} + \varepsilon_{t}^{(2)} & \text{for } 251 \leq t \leq 400 \\
 -1.35 \, y_{t-1}  -.37  \, y_{t-2} + .36 \,y_{t-3} +  \varepsilon_{t}^{(3)} & \text{for } 401 \leq t \leq 550,
\end{cases}
\end{equation}
where $\varepsilon_t^{(1)} \stackrel{iid}{\sim} \mathcal{N}(0, 0.25)$ and $\varepsilon_t^{(i)} \stackrel{iid}{\sim} \mathcal{N}(0, 1)$ for $i = 2, 3$. Figure \ref{fig:piecewise_AR_data} (top panel) shows a realization from model \eqref{eq:piecewise_AR}. After applying our methodology AutoNOM, the posterior probability of two change-points is 97.93\% and the posterior means of the change-point locations are $\hat{\mathbb{E}} \, (s_1 \, | \,k = 2, \,  \bm{y}) = 251.19$ and $\hat{\mathbb{E}} \, (s_2 \, | \,k = 2, \,  \bm{y}) = 401.56$. The estimated location of the frequency peaks for our proposed procedure in comparison to AdaptSPEC and AutoPARM and the true values are shown in Figure \eqref{fig:piecewise_AR_data} (bottom panels). It is evident that the proposed and existing methodologies successfully identify the true location of the frequency peaks in each segment, with AdaptSPEC showing less precision.

\begin{figure}[htbp]
    \centering
    \begin{subfigure}[b]{0.495\textwidth}
    \centering
    \includegraphics[height = 4.5cm, width = 9cm]{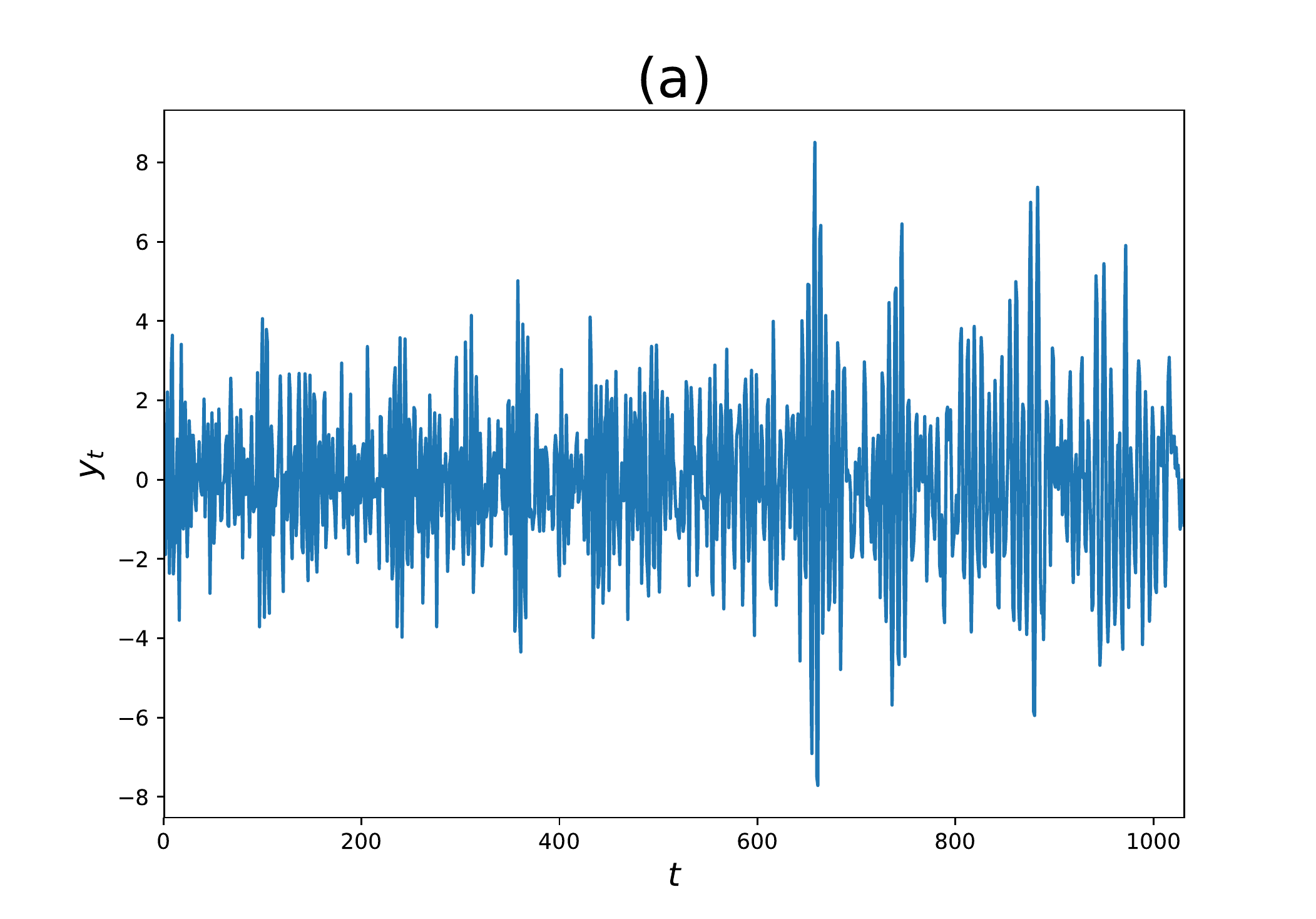}
    \end{subfigure}
    \begin{subfigure}[b]{0.495\textwidth}
    \centering
    \includegraphics[height = 4.5cm, width = 6.5cm]{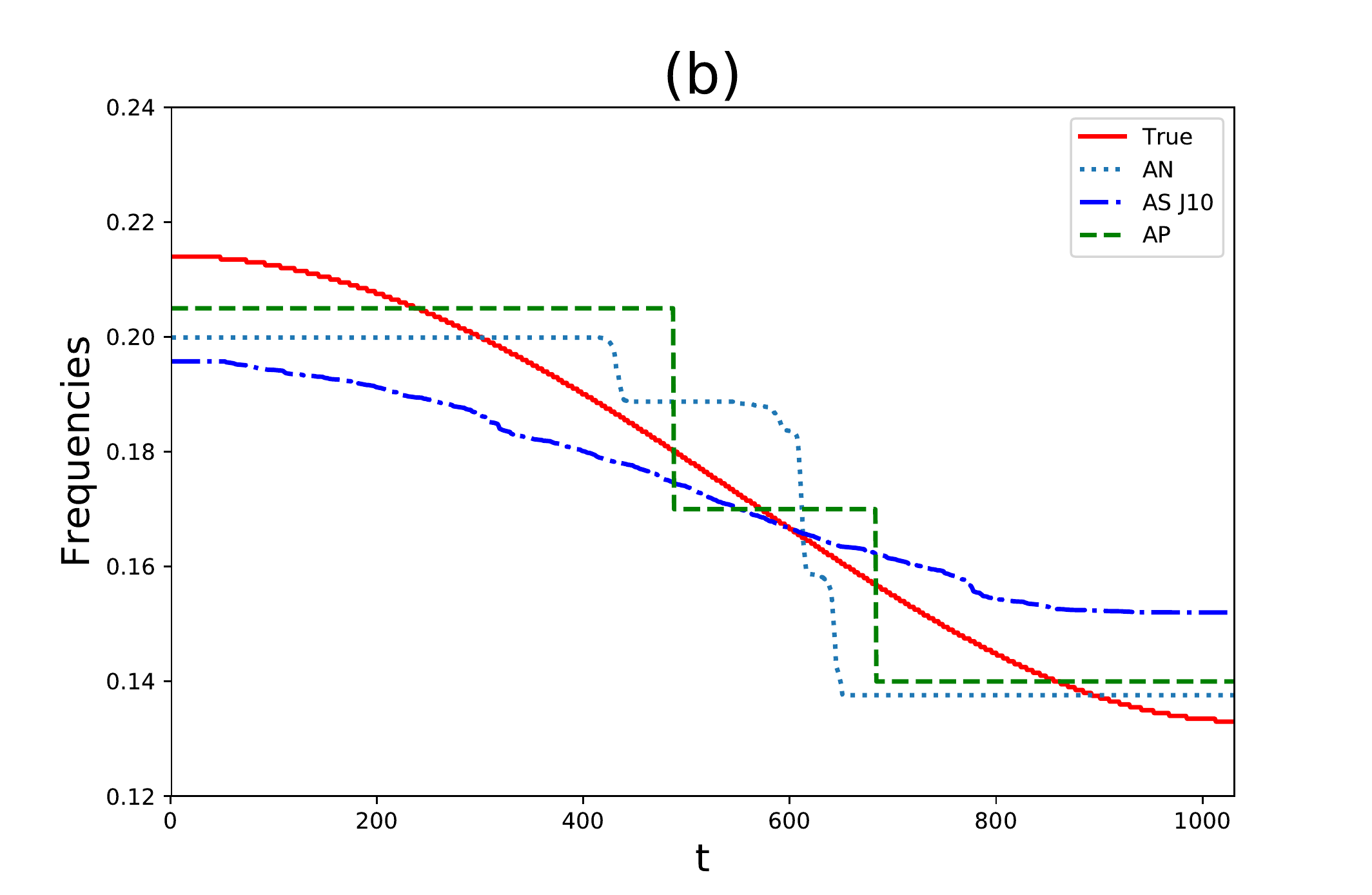}
    \end{subfigure}
    \caption{Slowly varying AR(2) process. (a) A realization from model  \eqref{eq:slowly_varying}. (b) True time varying frequency peak (solid line) and estimated time varying frequency peak for AutoNOM (AN), AdaptSPEC (AS $J = 10$) and AutoPARM (AP).}
    \label{fig:slowly_varying}
\end{figure}

\subsubsection{Slowly Varying Autoregressive Process}
In this section, we analyze an AR process whose continuous spectral density is changing slowly over time. We note though that this scenario is a large departure from the assumptions of our model. In particular, we consider the same slowly varying AR(2) process investigated by \citet{ombao2001automatic} and \citet{davis2006structural}, namely \begin{equation}
y_t = a_t \, y_{t-1} -.81 \,y_{t-2} + \varepsilon_t , \quad t = 1, \dots, 1031, 
\label{eq:slowly_varying}
\end{equation}
where $a_t = .8\,[1 - .5 \cos \,(\pi t/ 1031)]$ and $\varepsilon_t \stackrel{iid}{\sim} \mathcal{N}(0, 1)$. Notice that the parameter $a_t$ is changing gradually over time whereas the coefficient associated with the second lag remains constant.  A realization from model \eqref{eq:slowly_varying} is shown in Figure \ref{fig:slowly_varying} (a) and the corresponding time varying frequency peak is displayed in  Figure \ref{fig:slowly_varying} (b) as a solid line. Figure \ref{fig:slowly_varying} (b) also shows the estimated time varying frequency peak for AutoNOM, AdaptSPEC and AutoPARM. For AutoNOM and AdaptSpec, the time changing frequency peak has been averaged across the MCMC samples, giving a smoother estimate (especially for AdaptSPEC) than the one obtained by AutoPARM. For each method, we compute the residual sum of squares  $RSS = \sum_{t=1}^{1031}( \omega_t - \hat{\omega}_t )^2$ between the true time changing frequency peak $\omega_t$ and its estimate $\hat{\omega}_t$. The $RSS$ in this example was 0.111, 0.174, 0.085 for AutoNOM, AdaptSPEC and AutoPARM, respectively. It is clear that even in this scenario where the data generating model was very different from the underlying assumptions of our model, our approach seems to outperform AdaptSPEC and remains competitive with AutoPARM in estimating the time varying frequency peak.

\subsubsection{Non-Gaussian Time Series}
\begin{figure}[htbp]
     \centering
     \includegraphics[scale = 0.4]{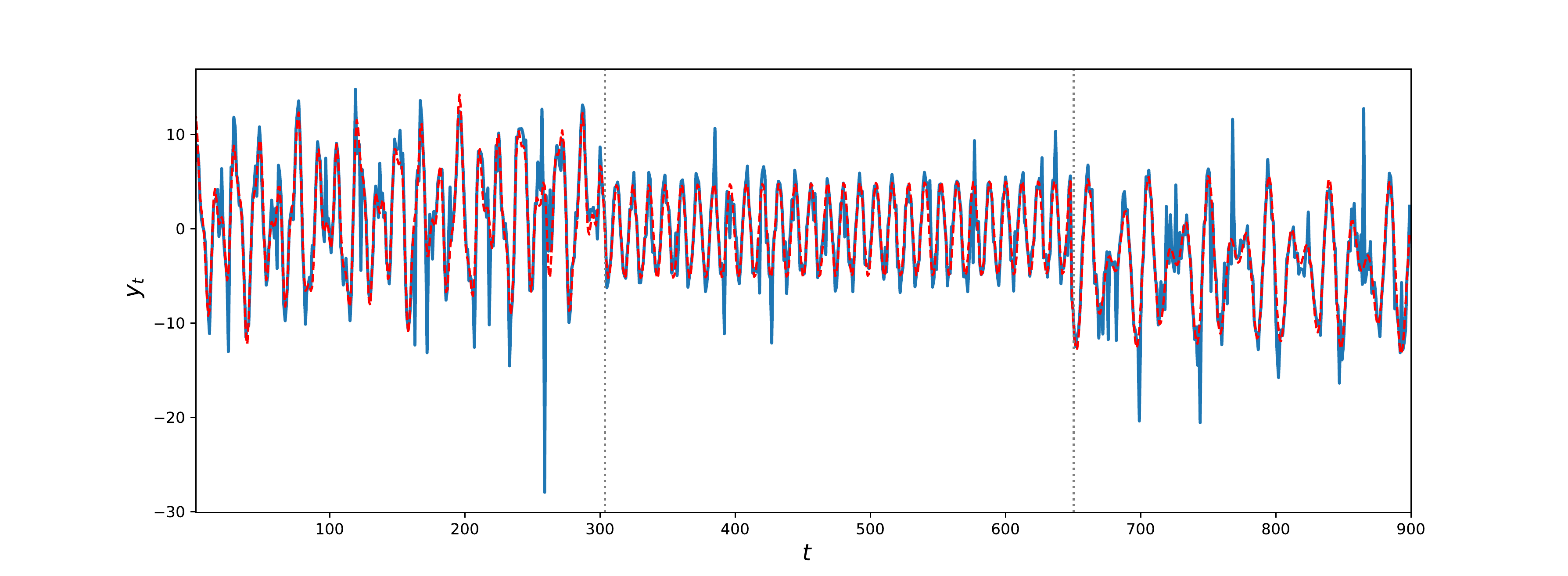}
     \caption{Illustrative example with t-distributed residual variances. Simulated time series (solid line) and estimated signal (dotted line). The dotted vertical lines represent the estimated location of the change-points.  }
     \label{fig:student_t_error}
 \end{figure}

We investigate the performance of our approach in the scenario when the innovations  are t-distributed. We simulate a time series from the same simulation model presented in  Section \ref{illustrative_example}, where errors were generated from a t-distribution with  2, 3, and 2 degrees of freedom for the sequence of three segments, respectively. The degrees of freedom were chosen low such that the corresponding distributions show heavy tails. A realization of this time series is shown in  Figure \ref{fig:student_t_error}. Our proposed methodology correctly identifies the 2 change-points, as the estimated posterior probability  $\hat{\pi} \, ( k = 2 \, | \, \bm{y})$ is 0.99. The posterior means of the change-point locations are $\hat{\mathbb{E}} \, (s_1 \, | \,k = 2, \,  \bm{y}) = 303.6$ and $\hat{\mathbb{E}} \, (s_2 \, | \,k = 2, \,  \bm{y}) = 650.5$, showing an excellent match to the true values $\bm{s}_{\,(2)} = (300, 650)$. Furthermore, the posterior mode of the number of frequencies in each segment is  ${\bm{\hat{m}}}_{\,(2)} = (3, 1, 2)$, which is a correct estimate of $\bm{m}_{\,(2)} = (3, 1, 2)$. We also display in Figure \ref{fig:student_t_error} the estimated signal (using Equation (1), Supplementary Material) as a dotted line. We can conclude that, although our model assumes Gaussianity, AutoNOM seems to perform well even in the case where the oscillatory underlying process is t-distributed with heavy tails.

%\begin{table}[htbp]
%	\centering
%	\caption{AdaptSpec simulation; (left panel) posterior probabilities of number of change-points, for different number of basis functions; (right panel) posterior means of change-point locations, conditioned on three segments, for different number of basis functions $J$.}
%	\begin{tabular}{lcccccc}
%		\hline \\[-0.9em]
%		\hline \\[-1.1em]
%		&  $\hat{\pi} \, (\, k = 0 \, | \, \bm{y})$  & $\hat{\pi} \, (\, k = 1 \, | \, \bm{y})$ & $\hat{\pi} \, (\, k = 2 \, | \, \bm{y})$ & &$\hat{\mathbb{E}} \, (s_1 \, | \,k = 2, \,  \bm{y})$ & $\hat{\mathbb{E}} \, (s_2 \, | \,k = 2, \,  \bm{y})$  \\ \cmidrule{2-4} \cmidrule{6-7} 
%		$J$ = 7 & 0.02 & 0.01 &  0.97 & & 304.8 & 646.1\\
%		$J$ = 10 & 0.75 & 0.16 & 0.09 & &313.9 & 540.1\\ 
%		$J$ = 15 & 0.23 & 0.01 & 0.76  & & 305.5 & 643.9\\ \hline 
%	\end{tabular}
%	\label{table:AdaptSpec_posterior_k_and_locations}
%\end{table}

\section{Case Studies} \label{applications}
The development of our methodology was motivated by the following two case studies where
dense physiological signals were observed which exhibit unknown periodicities  whose role may 
change over time in a more or less abrupt manner and where their
detection is of relevance to the health and well-being of the subject.

\subsection{Analysis of Human Skin Temperature}

%The development of information and communication technologies, in particular widespread internet access and availability of mobile phones and tablets, allows considering new developments in the health care system. Non-invasive measurement of skin surface temperature and
%rest-activity data of a healthy subject were recorded every 5
%minutes over 4 days, through a new mobile chest sensor platform, called PiCADo.
%These data were collected by F. Lévi’s lab and Cancer
%Chronotherapy Team, University of Warwick, UK and INSERM,
%Paris. 

\begin{figure}[htbp]
	\begin{subfigure}{0.48\textwidth}
		\includegraphics[height =5.7cm, width = 17.1cm]{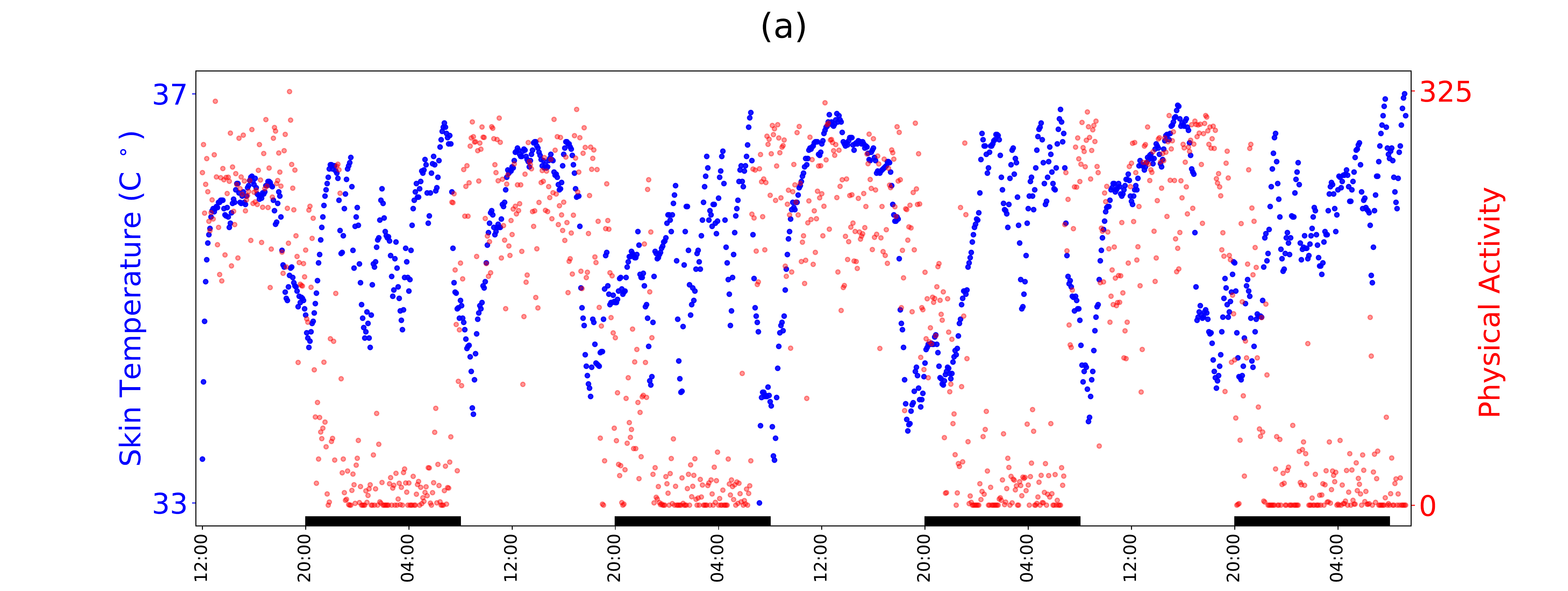}
	\end{subfigure}
	
	\begin{subfigure}{0.48\textwidth}
		\includegraphics[height =5.7cm, width = 17.1 cm]{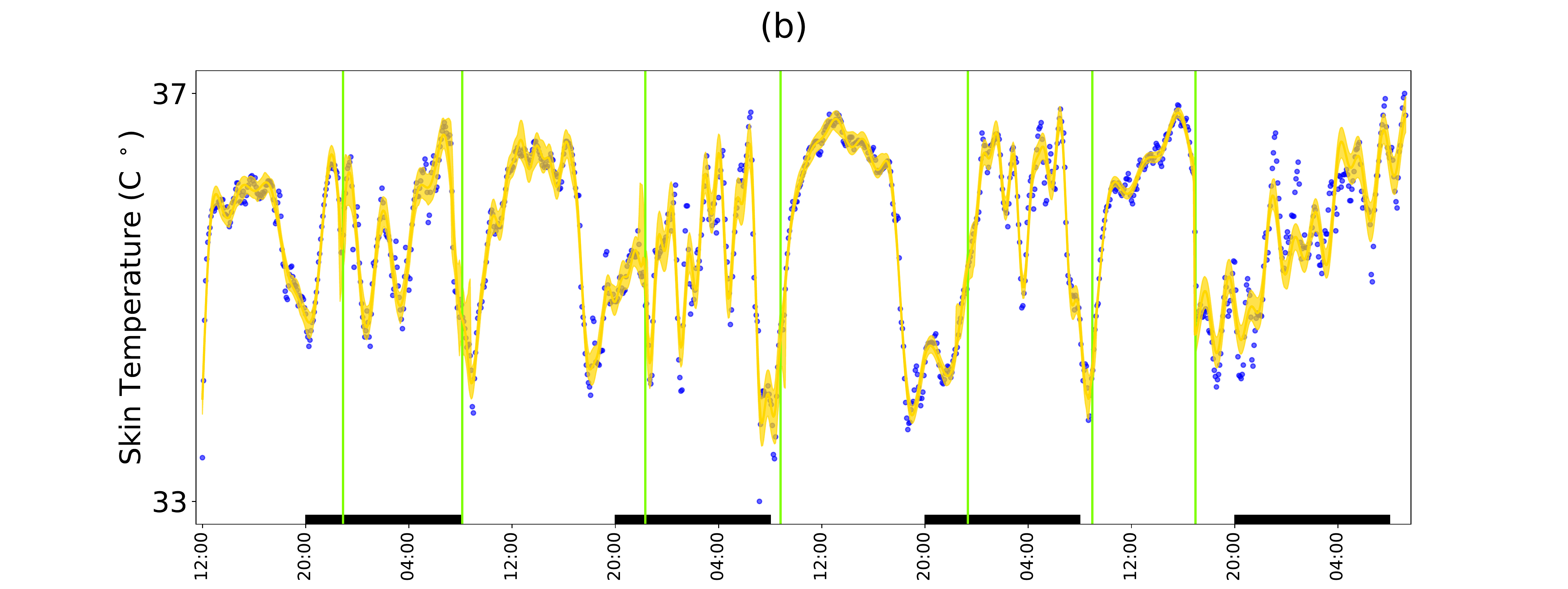}
	\end{subfigure}
	
	\begin{subfigure}{0.48\textwidth}
		\includegraphics[height = 3.5cm, width = 17.1 cm]{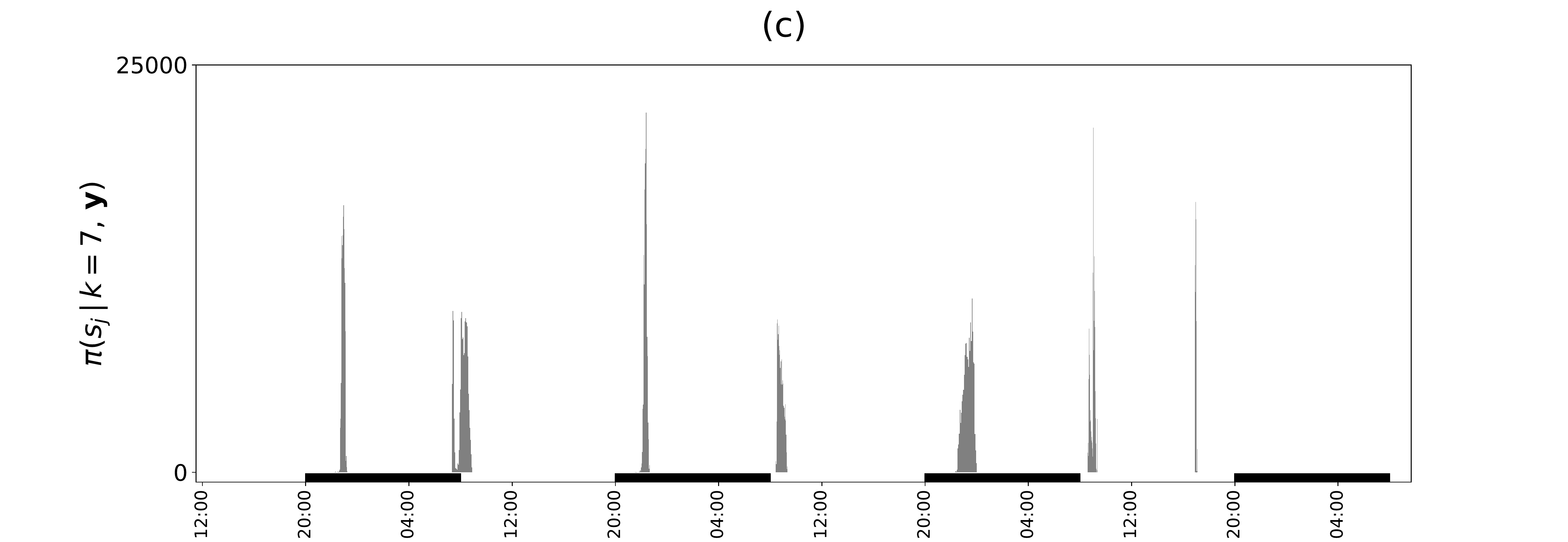}
	\end{subfigure}
	
	%and so on
	\caption{Analysis of skin temperature and upper chest activity of a healthy subject. Panel (a) are the time series of skin temperature and corresponding physical activity. Panel (b) is the estimated signal (solid line) along with its 95\% credible interval; vertical lines are the estimated locations of the change-points. Panel(c) is the estimated posterior density histogram of the locations of the changes, conditioned on $ k = 7$ change-points. Rectangles on the time axis of each plot correspond to periods from 20.00 to 8.00. The variation in skin temperature finds analogies with the rest-activity pattern that alternates between day activity and night rest.}
	\label{fig:rest_activity_temperature_summary}
\end{figure}

The development of information and communication technologies, in particular widespread internet access and availability of mobile phones and tablets, allows considering new developments in the health care system. To address the issue of personalized medical treatment according to the circadian timing system of the patient, referred to as {\it chronotherapy} \citep{levi2007circadian}, a
novel and validated non-invasive mobile e-Health platform pioneered by the French project PiCADo (\citealt{komarzynski2018}) is used to record and teletransmit
skin surface temperature as well as physical activity data \citep{huang2018hidden}
from an upper chest e-sensor. 
Figure \ref{fig:rest_activity_temperature_summary} (a) shows an example of 4 days 
of 5-minutes temperature recording for a healthy individual.
The circadian rhythms in core and skin surface temperature are usually 8-12 hours out of phase, with respective maxima occurring near 16:00 at day time, and near 2:00 at night  \citep{krauchi1994circadian}. The early night drop in core body temperature, which is critical for triggering the onset of sleep  \citep{van2006mechanisms}, results from the vasodilatation of the skin vessels and associated rise in skin surface temperature  \citep{krauchi2002circadian}.
 Under the assumption of stationarity \citet{komarzynski2018} analyzed the skin temperature time series identifying both strong 12 hours (circahemidian) and 24 hours (circadian)
rhythms. 

\begin{figure}[htbp]
	\centering
	\begin{subfigure}[b]{0.475\textwidth}
		\centering
		\includegraphics[height = 4.0cm, width = 8cm]{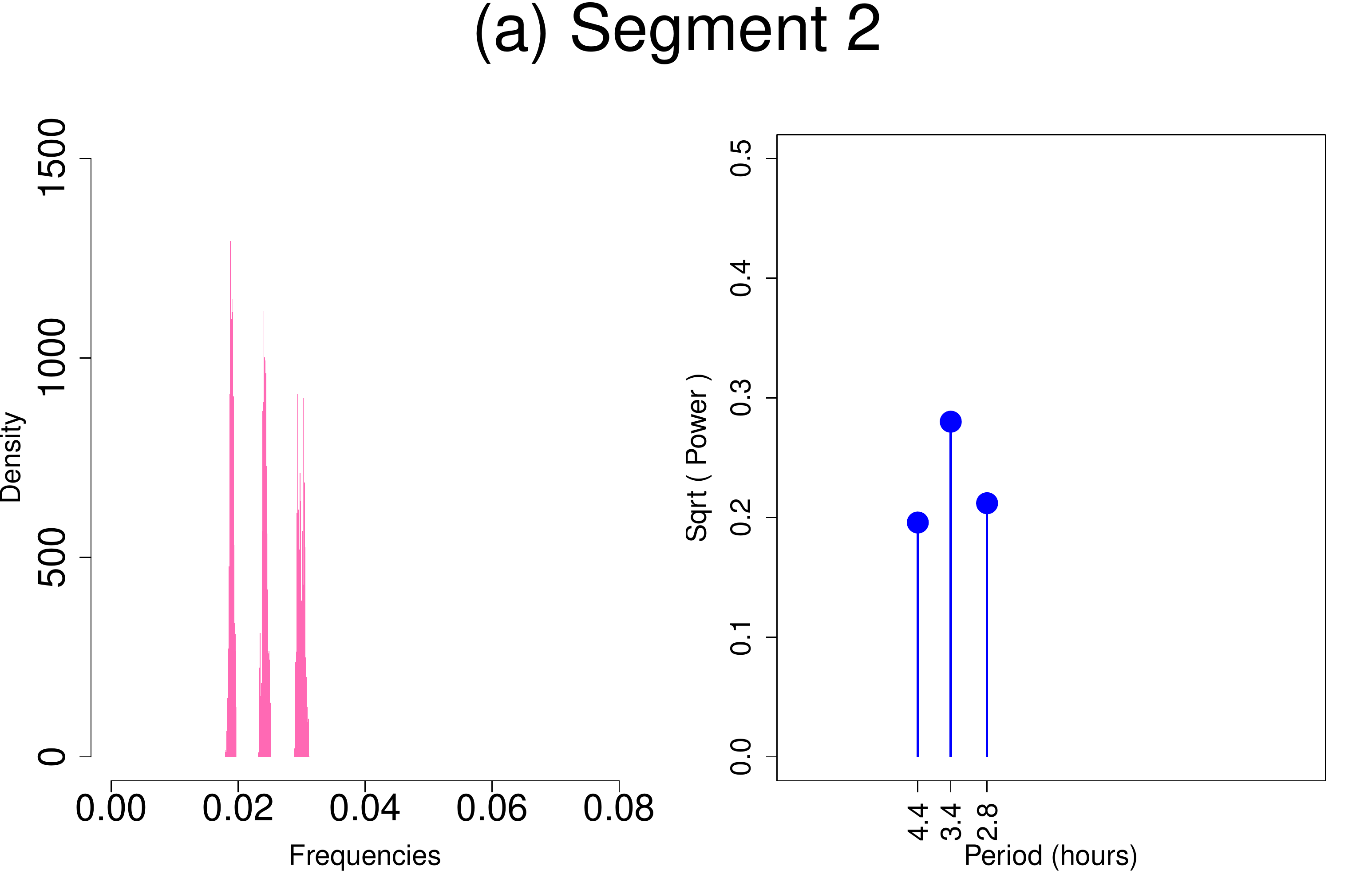}
	\end{subfigure}
	\hfill
	\begin{subfigure}[b]{0.475\textwidth}  
		\centering 
		\includegraphics[height = 4.0cm, width = 8cm]{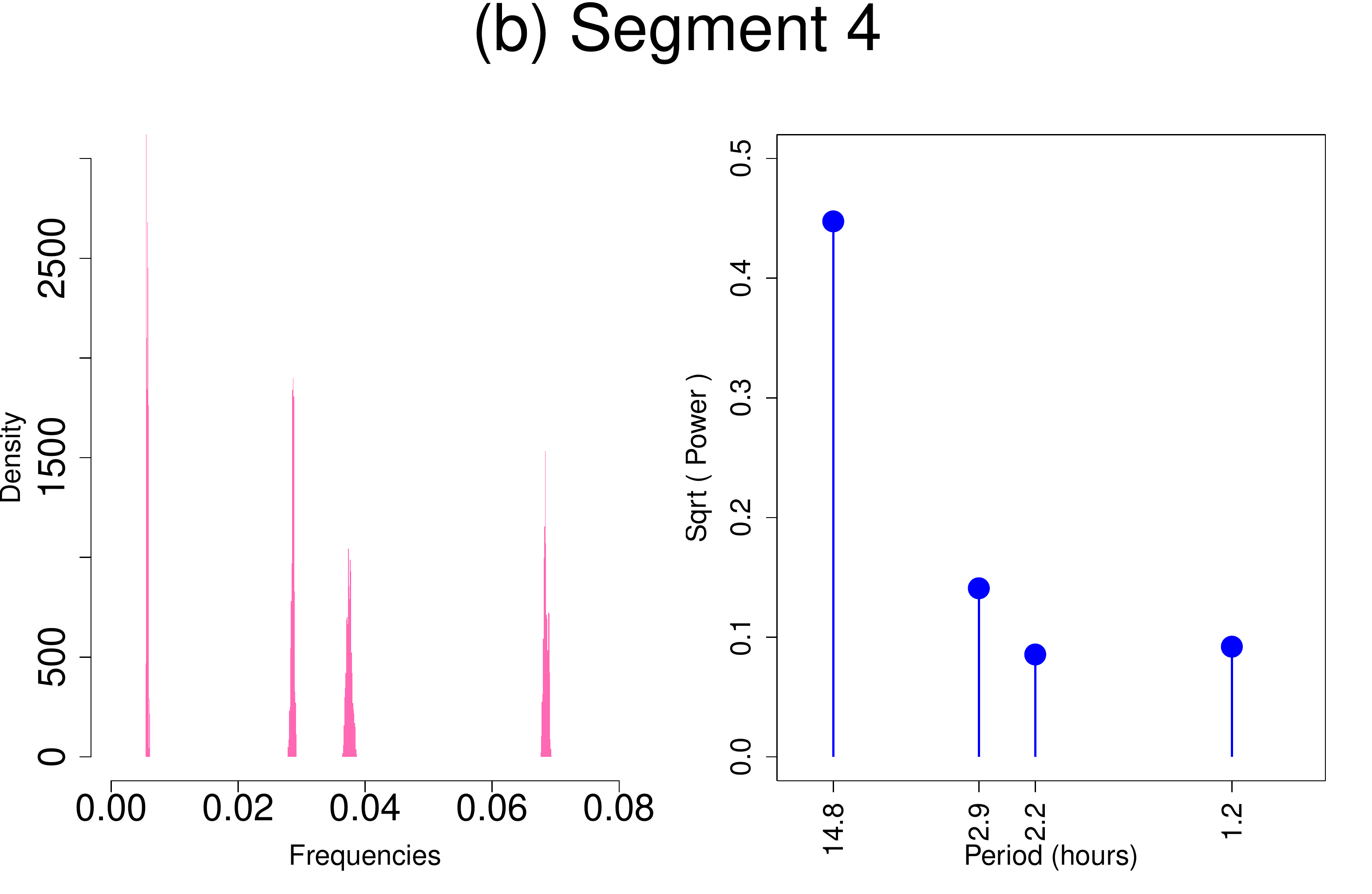}
	\end{subfigure}
	\vskip\baselineskip
	\begin{subfigure}[b]{0.475\textwidth}   
		\centering 
		\includegraphics[height = 4.0cm, width = 8cm]{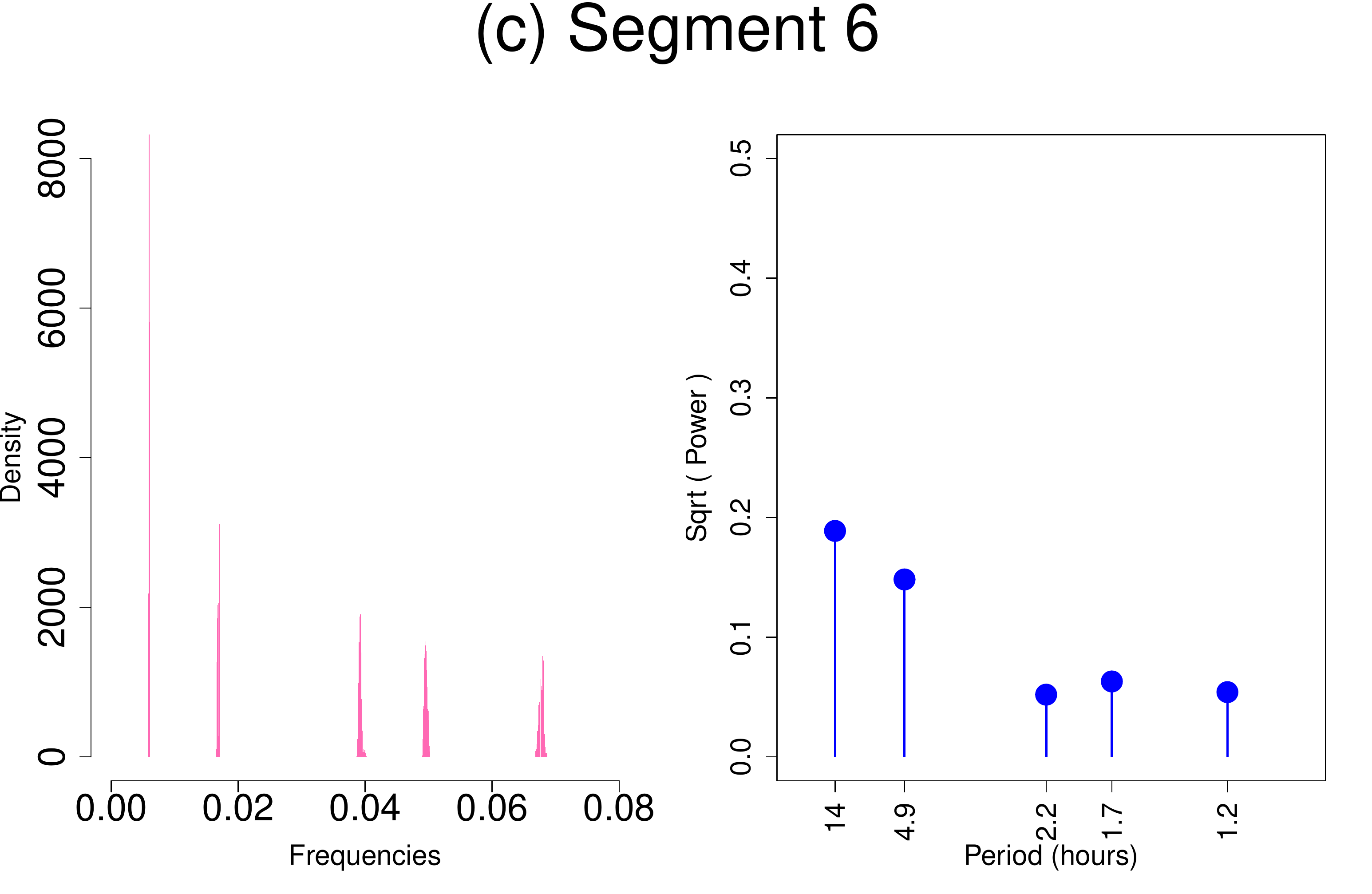}
	\end{subfigure}
	\quad
	\begin{subfigure}[b]{0.475\textwidth}   
		\centering 
		\includegraphics[height = 4.0cm, width = 8cm]{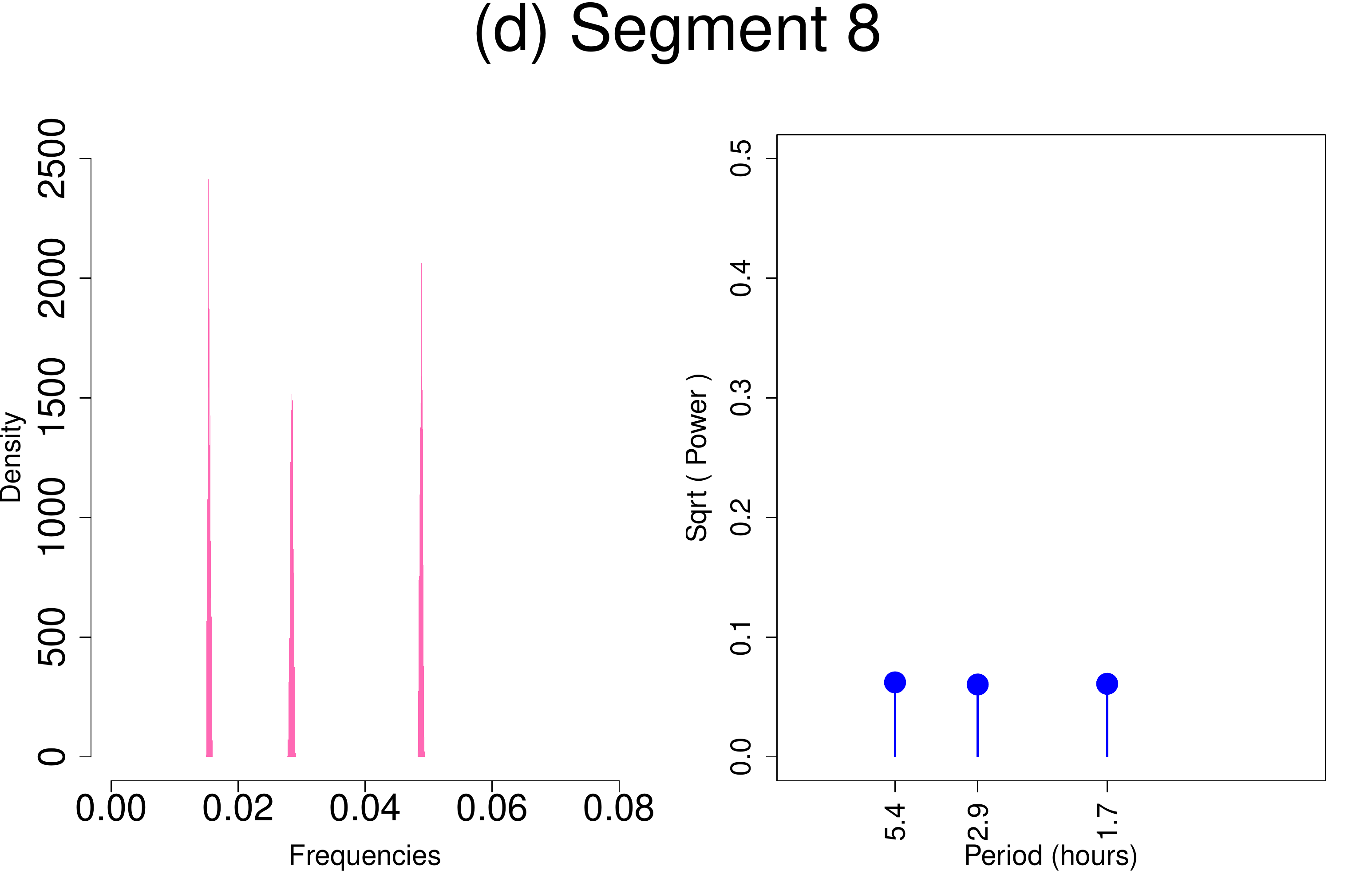}
	\end{subfigure}
	\caption{Spectral properties for segments corresponding to night rest. Estimated posterior distribution of the frequencies along with square root of the estimated power of the corresponding frequencies. The results are conditional on the modal number of frequencies per segment.} 
	\label{fig:spectral_properties_segments}
\end{figure}

Here, we applied our methodology to the skin-temperature time series shown in Figure \ref{fig:rest_activity_temperature_summary} (a) for 300,000 iterations, discarding the first 100,000 updates as burn-in. 
The maximum number of change-points $k_\text{max}$ was set to 10, whereas the maximum number of frequencies per regime $m_\text{max}$ was set to 5. The estimated number of change-points had a mode at 7, with $\hat{\pi} \, (\, k = 7 \, | \, \bm{y}) = 0.97$ and their estimated posterior  distributions are shown in Figure  \ref{fig:rest_activity_temperature_summary} (c). Inspecting them alongside the physical activity data we can see that the change points  mainly correspond to the start and end points of the prolonged rest periods at nights showing that skin temperature  alternates between day activity and night rest including sleep. 
Figure  \ref{fig:spectral_properties_segments} shows the estimated posterior distribution of the frequencies for the sleep segments  (2, 4, 6, 8)
along with the square root of the estimated power of the corresponding frequencies, where the power of each is  frequency $\omega_{j,l}$ is summarized by the sum of squares of the corresponding linear coefficients, i.e.  $ I \, (\omega_{j,l} ) \, = \beta_{j,l}^{\,(1)^{\,2}} + \beta_{j,l}^{\,(2)^{\,2}} $ \citep{Shumway:2005:TSA:1088844}. Figure \ref{fig:rest_activity_temperature_summary} (b)  shows the piecewise fitted signal, along with a 95 \% credible interval  obtained  from the 2.5 and 97.5 empirical percentiles of the posterior sample using Equation (1), Supplementary Material. Cycles of approximately 3 hours appear in segments 2, 4 and 6; cycles that range approximately  1-1.5 hour appear in segments 2, 4, 8 and  cycles of around 2 hours appear in segments 4 and 6 while some longer periods identified in  segments 4, 6, 8 indicate the presence of a trend. 

Stages of sleep are characterized by ultradian oscillations between  rapid eye movements (REM) and non-rapid eye movements (non-REM). The biological functionality that regulates the alternations between these two types of sleep is not yet much understood \citep{altevogt2006sleep}. However, several physiological changes that occur over night differ between REM and non-REM phases, such as heart rate, brain activity, muscle tone and body temperature (\citealp{berlad1993power, pace2002neurobiology}). The body cycles between REM and non-REM sleep stages with an average length that ranges approximately between 70 to 120 minutes, and there are usually four to six of these sleep cycles each night  (\citealp{carskadon2005normal, shneerson2009sleep}). Our analysis was able to use skin temperature data alone to detect periods of sleep throughout the day and identify oscillatory behaviour during the night, whose frequencies are compatible with ultradian oscillations between REM and different non-REM sleep stages.

% A comparison with the current state-of-the-art methods, AutoPARM and AdaptSPEC, is provided in the Supplementary Material where we show that the
% existing methods  fail to detect both circadian and ultradian rhythmicity which are instead elicited by our method and are to be expected as body temperature is known to be a circadian biomarker \citep{krauchi1994circadian}. 

A comparison with the current state-of-the-art methods, AutoPARM and AdaptSPEC, is provided in the Supplementary Material, Section 4.1. Circadian and ultradian rhythmicity are expected because body temperature is known to be a circadian biomarker \citep{krauchi1994circadian}, but these existing methods fail. % where we show that these existing methods  fail to detect both circadian and ultradian rhythmicity which are instead elicited by our method and are to be expected as body temperature is known to be a circadian biomarker \citep{krauchi1994circadian}. 
Furthermore, we notice that in the framework of analyzing circadian biomarker data, such as body temperature, a change in acrophase may be of interest to the clinician as this may be indicative of a disruption of the bodyclock. The methodology can indeed be used to investigate phase which can be computed from the sinusoidal function that characterizes the j$^{th}$ segment (see Supplementary Material, Section 3).

\vspace{1cm}

\newpage
\subsection{Characterizing Instances of Sleep Apnea in Rodents}
Sleep apnea is the temporary ($\geq$ 2 breaths) interruption of breathing during sleep. Moderate or severe ($\geq 15$  events per hour) sleep apnea, occurs in about 50 \% of men and 25 \% of women over the age of 40 \citep{heinzer2015prevalence}, with 91\% of people with sleep apnea being undiagnosed \citep{tan2016prevalence}. Sleep apnea is linked to many diseases. Patients with sleep apnea are at increased risk of: cardiovascular events \citep{lanfranchi1999prognostic}, cancer \citep{nieto2012sleep}, liver disease \citep{sundaram2016nocturnal}, diabetes \citep{harsch2004effect}, metabolic syndrome \citep{parish2007relationship}, cognitive decline \citep{osorio2016orexin}, and increased risk of dementia 
in the elderly \citep{lal2012neurocognitive}. The motivation of this research is to provide a statistical methodology that can be applied to analyze large breathing data sets resulting from \textit{in vivo} plethysmograph studies in rats to characterize the occurrence of sleep apnea under different experimental conditions. If this could be attained, a concrete aid to the understanding of the pathological implications of this status could be provided to clinicians and experimental biologists.

An unrestrained whole-body plethysmograph is used to produce a breathing trace from freely behaving rats for periods of up to 3 hours. Plethysmographs were made using an 2L air-tight box connected to a pressure transducer, with an air pump and outlet valve producing a flow rate of 2L/min. Airflow pressure signals were amplified using Neurolog system (Digitimer) connected to a 1401 interface and acquired on a computer using {\it Spike2} software (CED). 

\begin{figure}[htbp] % "[t!]" placement specifier just for this example
	\begin{subfigure}{0.48\textwidth}
		\includegraphics[height = 3.2cm, width = 10cm]{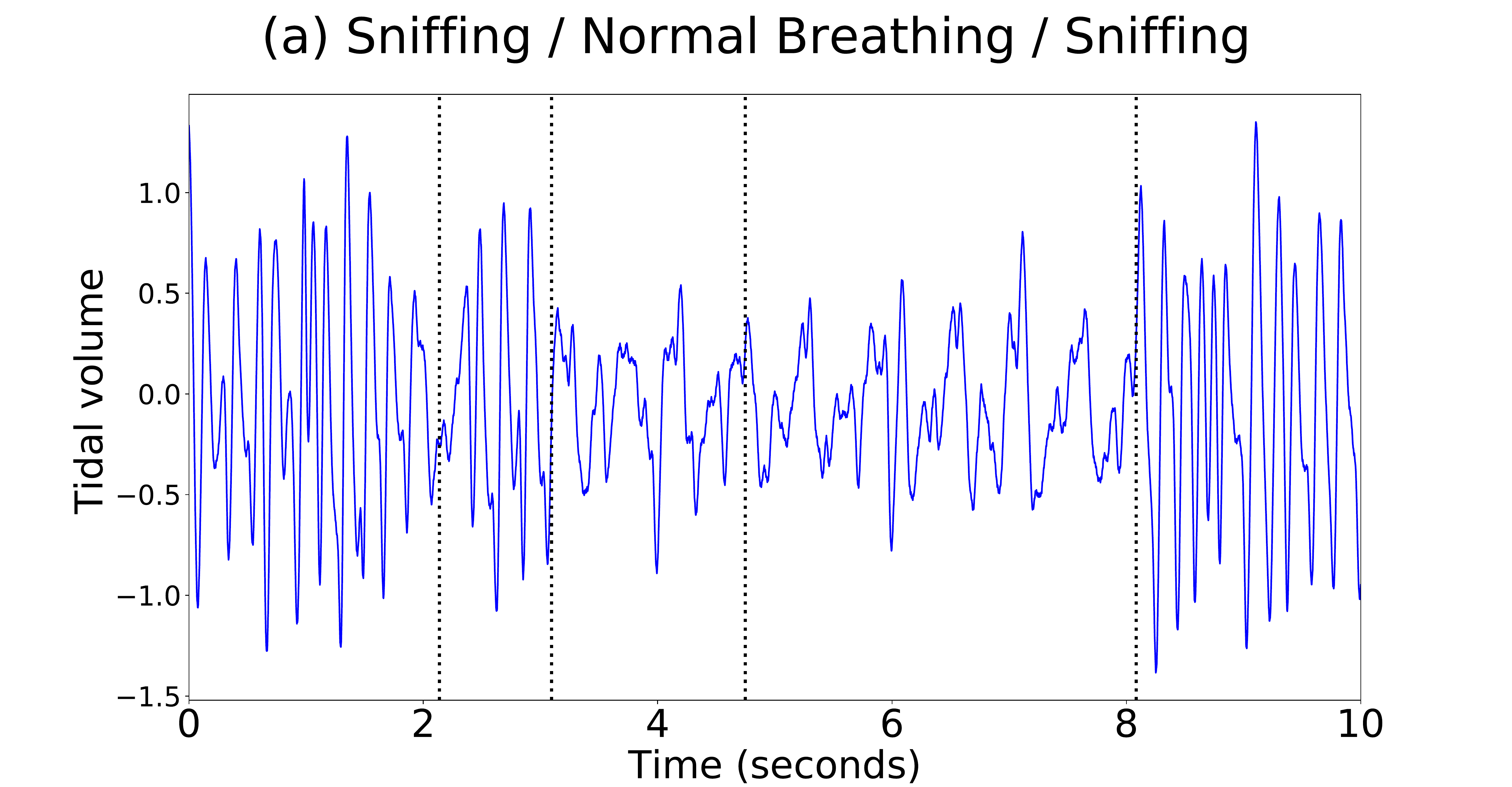}
	\end{subfigure}\hspace*{\fill} \qquad \qquad 
	\begin{subfigure}{0.48\textwidth}
		\includegraphics[height = 3.3cm, width = 6cm]{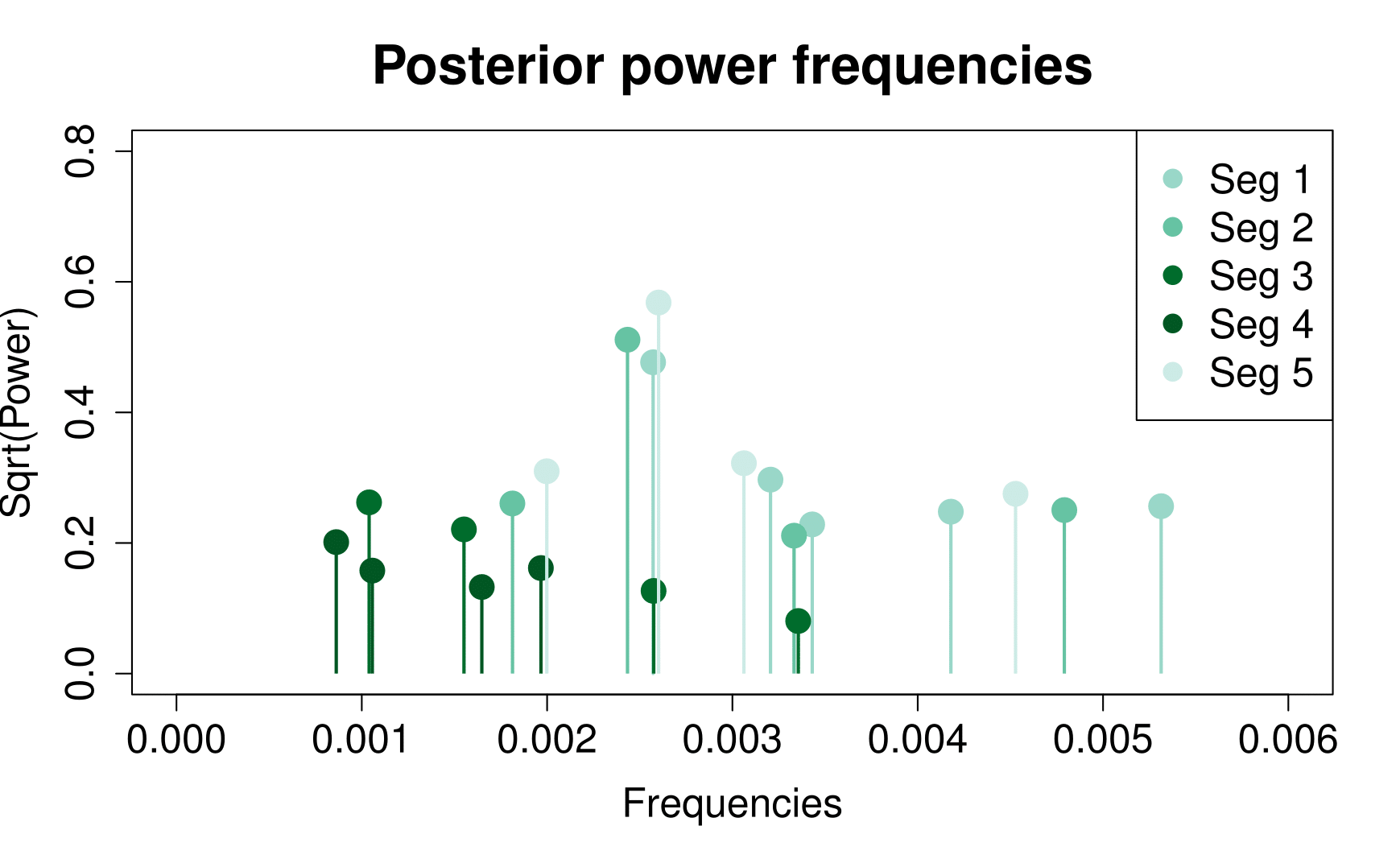}
	\end{subfigure}
	
	\begin{subfigure}{0.48\textwidth}
		\includegraphics[height = 3.2cm, width = 10cm]{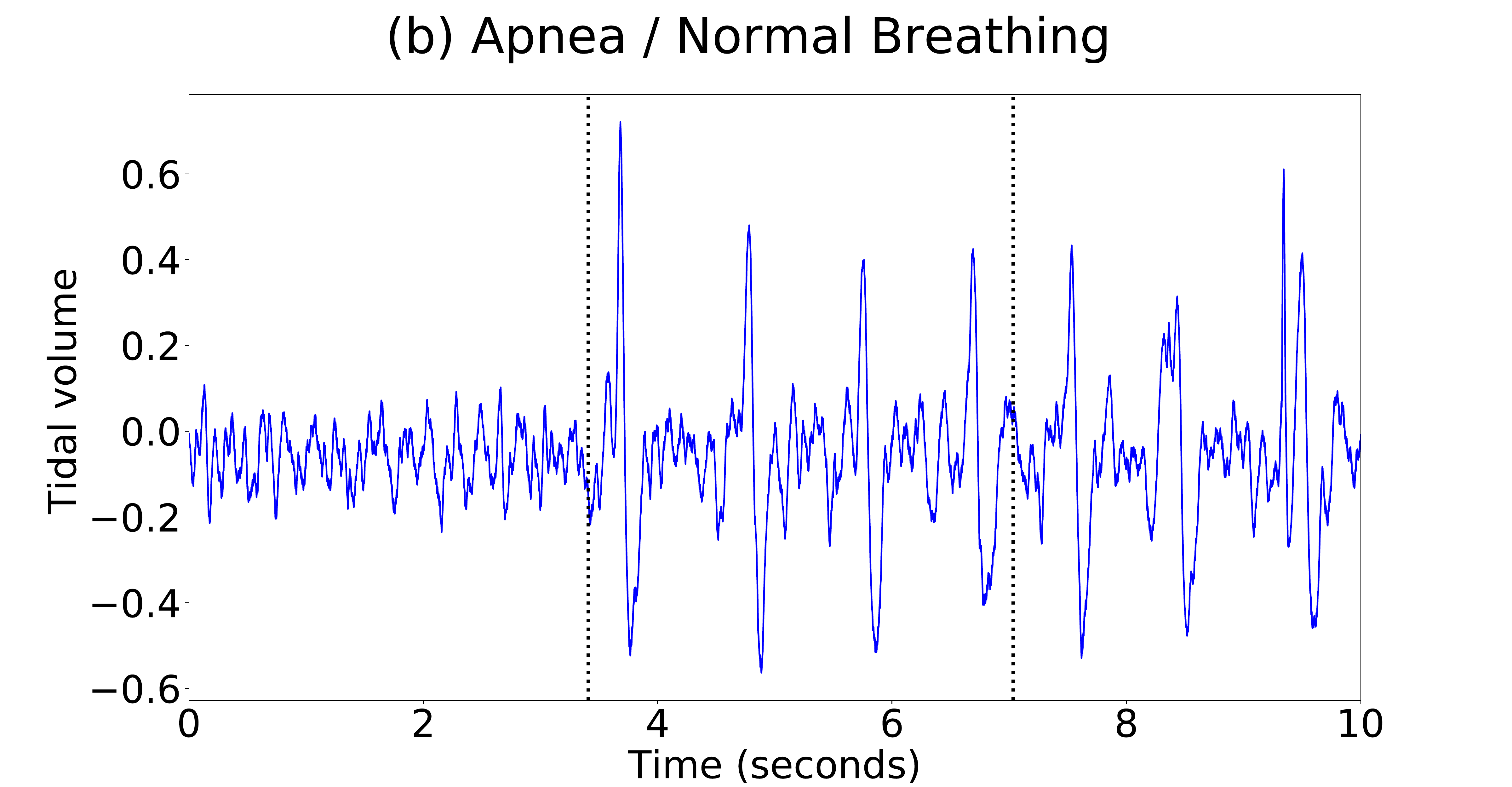}
	\end{subfigure}\hspace*{\fill}\qquad \qquad 
	\begin{subfigure}{0.48\textwidth}
		\includegraphics[height = 3.3cm, width = 6cm]{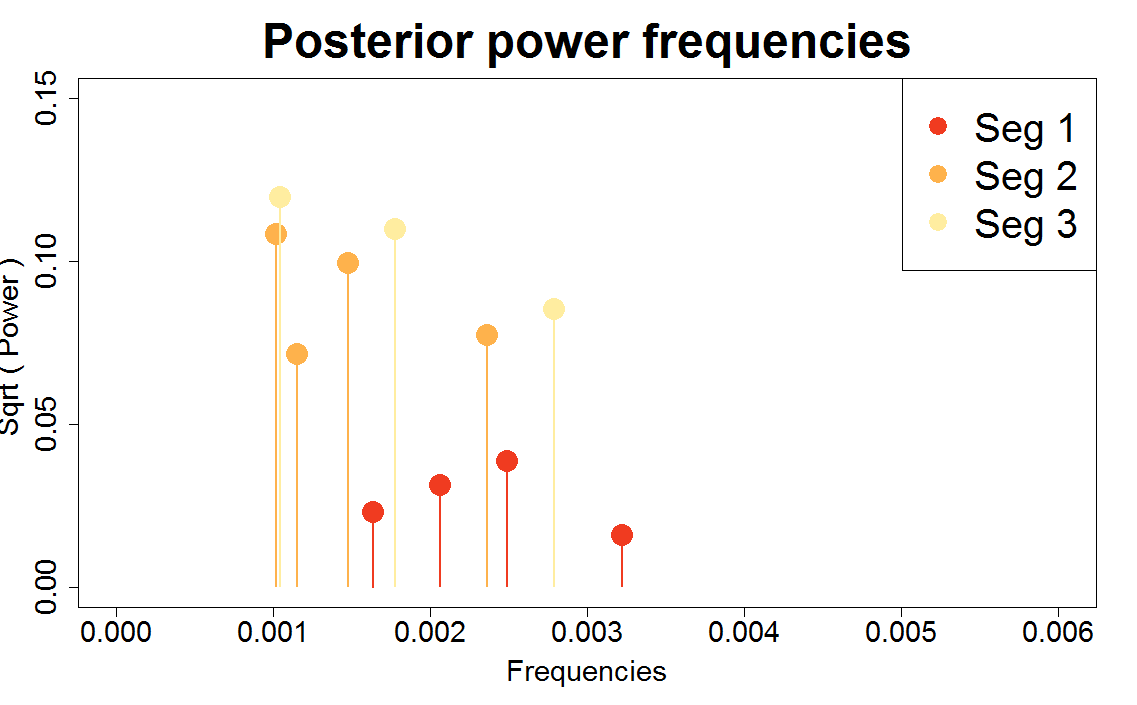}
	\end{subfigure}
	
	\begin{subfigure}{0.48\textwidth}
		\includegraphics[height = 3.2cm, width = 10cm]{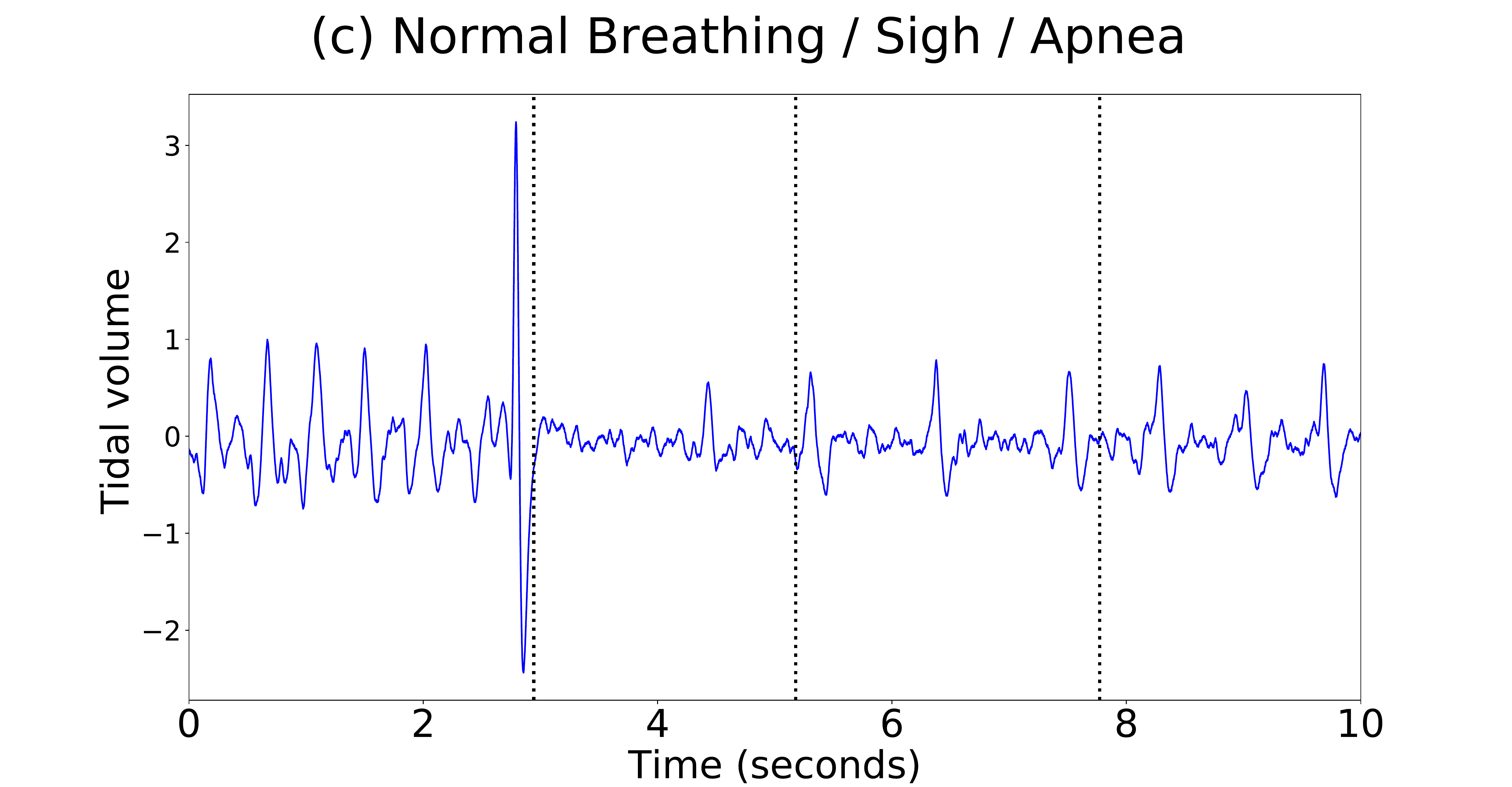}
	\end{subfigure}\hspace*{\fill} \qquad \qquad 
	\begin{subfigure}{0.48\textwidth}
		\includegraphics[height = 3.3cm, width = 6cm]{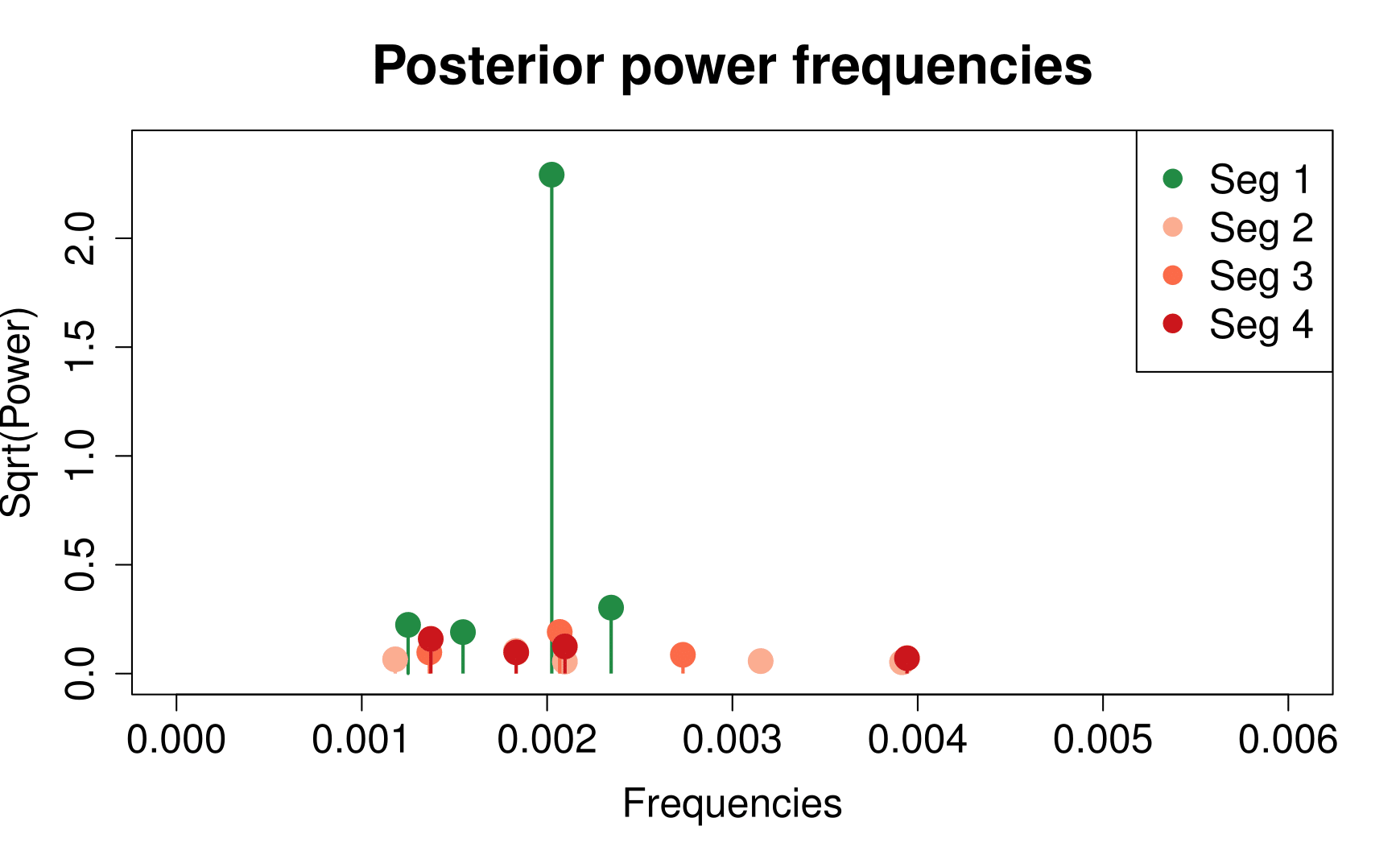}
	\end{subfigure}
	\caption{Plots of the respiratory traces of a rat (left panels) and corresponding estimated posterior power (right panels).  Panel (a) is characterised by an alternation of sniffing and normal breathing. Panel (b) is a plot of the trace of a spontaneous apnea, followed by normal breathing. Panel (c) shows normal breathing followed by a sigh, and a post-sigh apnea.   Dotted vertical lines correspond to the estimated locations of the change-points. } \label{fig:data_rats}
\end{figure}

Apneas are subclassified as post-sigh apneas, if the preceding breath was at least 25\% above the average amplitude of prior breaths, or spontaneous apneas, if there was no manifestation of a previous sigh \citep{davis2013rodent}. Airflow traces from the plethysmograph are shown in Figure \ref{fig:data_rats}  (left panels) and consist of three time series, which will be referred to as (a), (b) and (c). They correspond to different actions for this rat: (a) an alternation of sniffing and normal breathing; (b) spontaneous apnea followed by normal breathing; (c) normal breathing followed by a sigh, and a post-sigh apnea. We note that these actions were classified by eye by an experienced experimental researcher. Each time series contains 20,000 observations where the signal was sampled at 2000 Hz so that we have 2000 observations per second. 

Our procedure allows us to set an upper bound, $\phi_{\omega}$, (Appendix \ref{appendix_segment_between}) for the uniform interval where the new frequencies are sampled.  As the periodogram ordinates for these data were approximately zero for all frequencies larger than 0.01, we decided accordingly to set   $\phi_{\omega} = 0.01$. The locations of the changes (vertical lines) are displayed in Figure \ref{fig:data_rats} (left panels). The posterior power of the frequencies, for each time series, is shown in Figure \ref{fig:data_rats} (right panels). These results are conditional on the modal number of change-points and the modal number of frequencies per segment. For each data set, we summarise in Table \ref{tab:spectral_properties_rat} the spectral properties of each partition by displaying the periodicities corresponding to the first two largest values of the estimated power.  
  When the rat is sniffing, (a), the air flow trace oscillates with a dominant period of approximately 0.2 seconds, namely 5 cycles per second. Normal breathing, (a) and (b), is characterised by lower frequencies and lower magnitude than sniffing, by oscillating with a dominant period of around 0.5 seconds, namely around 2 cycles every second. Apneas, (b) and (c), appear to be characterised by higher frequencies than normal breathing but with a lower power, with dominant periods of around 0.25 and 0.35 seconds. Notice that in the first partition of (c), the highest value of the power corresponds to the frequency responsible for a sigh before apnea. Moreover, our methodology identifies different frequencies that explain the variation between the third and fourth partition of (c), leading to the hypothesis that there might be a time changing spectrum during the occurrence of an apnea instance.  A comparison of our results with the results from AutoPARM and AdaptSPEC is provided in the Supplementary Material, Section 4.2.

\begin{table}[htbp]
	\centering
	\caption{Spectral properties of respiratory traces of a rat. Periodicities (in seconds) corresponding to the first two largest values of the estimated power, for each time series (a), (b) and (c). }
	\label{tab:spectral_properties_rat}
	\begin{tabular}{llcclcclccl}
		\hline \\[-0.9em]
		\hline \\[-1.1em]
		
		{\color[HTML]{000000} }      & \multicolumn{1}{c}{{\color[HTML]{000000} }} & \multicolumn{1}{r}{{\color[HTML]{000000} $\qquad \qquad \, $ \textbf{(a)}}}                             & {\color[HTML]{000000} }                                                        & {\color[HTML]{000000} } & \multicolumn{1}{r}{{\color[HTML]{000000}  $\qquad \qquad \, $ \textbf{(b)}}}                             & \multicolumn{1}{l}{{\color[HTML]{000000} }}                                    &  & \multicolumn{1}{r}{ $\qquad \qquad \, $ \textbf{(c)}}                             & \multicolumn{1}{l}{}                                    &  \\
		{\color[HTML]{000000} }      & \multicolumn{1}{c}{{\color[HTML]{000000} }} & {\color[HTML]{000000} Period}                                              & {\color[HTML]{000000} Power}                                                   & {\color[HTML]{000000} } & {\color[HTML]{000000} Period}                                              & {\color[HTML]{000000} Power}                                                   &  & Period                                              & Power                                                   &  \\ \cmidrule{3-4} \cmidrule{6-7} \cmidrule{9-10}
		{\color[HTML]{000000} Segment 1} & \multicolumn{1}{c}{{\color[HTML]{000000} }} & {\color[HTML]{000000} \begin{tabular}[c]{@{}c@{}}.19\\ .15\end{tabular}} & {\color[HTML]{000000} \begin{tabular}[c]{@{}c@{}}.2272\\ .0883\end{tabular}} & {\color[HTML]{000000} } & {\color[HTML]{000000} \begin{tabular}[c]{@{}c@{}}.20\\ .25\end{tabular}} & {\color[HTML]{000000} \begin{tabular}[c]{@{}c@{}}.0015\\ .0010\end{tabular}} &  & \begin{tabular}[c]{@{}c@{}}.25\\ .21\end{tabular} & \begin{tabular}[c]{@{}c@{}}5.2540\\ .0915\end{tabular} &  \\[0.7em]
		{\color[HTML]{000000} Segment 2} & {\color[HTML]{000000} }                     & {\color[HTML]{000000} \begin{tabular}[c]{@{}c@{}}.20\\ .28\end{tabular}} & {\color[HTML]{000000} \begin{tabular}[c]{@{}c@{}}.2613\\ .0686\end{tabular}} & {\color[HTML]{000000} } & {\color[HTML]{000000} \begin{tabular}[c]{@{}c@{}}.49\\ .34\end{tabular}} & {\color[HTML]{000000} \begin{tabular}[c]{@{}c@{}}.0117\\ .0099\end{tabular}} &  & \begin{tabular}[c]{@{}c@{}}.27\\ .42\end{tabular} & \begin{tabular}[c]{@{}c@{}}.0107\\ .0043\end{tabular} &  \\[0.7em]
		{\color[HTML]{000000} Segment 3} & {\color[HTML]{000000} }                     & {\color[HTML]{000000} \begin{tabular}[c]{@{}c@{}}.48\\ .32\end{tabular}} & {\color[HTML]{000000} \begin{tabular}[c]{@{}c@{}}.0692\\ .0491\end{tabular}} & {\color[HTML]{000000} } & {\color[HTML]{000000} \begin{tabular}[c]{@{}c@{}}.48\\ .28\end{tabular}} & {\color[HTML]{000000} \begin{tabular}[c]{@{}c@{}}.0145\\ .0122\end{tabular}} &  & \begin{tabular}[c]{@{}c@{}}.24\\ .37\end{tabular} & \begin{tabular}[c]{@{}c@{}}.0365\\ .0095\end{tabular} &  \\[0.7em]
		{\color[HTML]{000000} Segment 4} & {\color[HTML]{000000} }                     & {\color[HTML]{000000} \begin{tabular}[c]{@{}c@{}}.58\\ .47\end{tabular}} & {\color[HTML]{000000} \begin{tabular}[c]{@{}c@{}}.0400\\ .0251\end{tabular}} & {\color[HTML]{000000} } & {\color[HTML]{000000} \_}                                                  & {\color[HTML]{000000} \_}                                                      &  & \begin{tabular}[c]{@{}c@{}}.36\\ .24\end{tabular} & \begin{tabular}[c]{@{}c@{}}.0253\\ .0155\end{tabular} &  \\[0.7em]
		{\color[HTML]{000000} Segment 5} & {\color[HTML]{000000} }                     & {\color[HTML]{000000} \begin{tabular}[c]{@{}c@{}}.19\\ .16\end{tabular}} & {\color[HTML]{000000} \begin{tabular}[c]{@{}c@{}}.3231\\ .1044\end{tabular}} & {\color[HTML]{000000} } & {\color[HTML]{000000} \_}                                                  & {\color[HTML]{000000} \_}                                                      &  & \_                                                  & \_                                                      &  \\[0.4em] \hline
	\end{tabular}
\end{table}


\section{Summary and Discussion} \label{conclusion}
We developed a novel Bayesian methodology for analyzing nonstationary time series that exhibit oscillatory behaviour. %, as an extension of the stationary model presented in \citet{andrieu1999joint}. % In a similar way to \citet{ombao2001automatic} and \citet{rosen2012adaptspec}, 
Our approach is based  on the assumption that, conditional on the position and number of change-points, 
the time series can be approximated by a piecewise changing sinusoidal regression model. The timing and number of changes
are unknown, along with the number and values of relevant periodicities in each regime. Bayesian inference is performed via a reversible jump MCMC  algorithm that can simultaneously estimate both the number and location of change-points, as well as the number, frequency and magnitude of sinusoids within each segment. Our methodology can be seen as a novel and relevant extension of the work in \citet{andrieu1999joint} to the nonstationary setting.

We illustrated the utility of our methodology in two case studies. First, we analyzed human skin temperature time series data obtained from a wearable device, which exhibited unknown periodicities that changed over time in an abrupt manner. %The detection and identification of these change points and the frequencies that drive the oscillatory signal has implications about the health and well-being of the subject. 
Our proposed methodology identified interesting oscillations whose frequencies are consistent with ultradian oscillations between REM and non-REM sleep stages. Second, we characterized the occurrence of sleep apnea in large breathing data sets resulting from \textit{in vivo} plethysmograph studies on rodents.
Our spectral investigation was able to distinguish very sharp peaks, corresponding to different nearby frequencies, that are responsible for the different actions of the rodent. 

Although we have not discussed this in detail here, several diagnostics for monitoring convergence were carried out in both simulation and case studies. In particular, we verified that the target posterior distribution reached a stable regime by analyzing the trace plot of the log likelihood across MCMC iterations \citep{marin2007bayesian}. We are aware that assessing convergence only based on this simple tool may sometimes be misleading since stable values of the log likelihood could simply mean that the Markov chain is stuck in some local mode of the posterior distribution. Additionally, conditioned on the modal number of change-points and modal number of frequencies per regime, we have also monitored (within-model) convergence by analyzing the traces and running averages plots for all parameters across MCMC iterations, with satisfactory results.  Comparable results were also obtained when running several chains starting at over-dispersed initial values. We notice that the diagnostic tool used by \citet{bruce2018conditional} and \citet{li2018adaptive} to assess convergence for reversible jump MCMC samplers  appears relevant. In the context of adaptive spectral analysis of nonstationary time series, they point out that although the number of partitions change across models, a power spectrum is defined at each time point. The power spectrum is modeled with a fixed number of splines, yielding a vector of summary measures of  parameters that maintain the same interpretation across models in their samplers.  However, our proposed sampler has a further layer of variable dimensionality, as not only the number and locations of the change-points may change from one iteration to the next, but also the number of frequencies in each segment are not fixed throughout the simulation.

 We conclude this article by noticing that, although a Gaussian distribution is assumed, it is conceivable that our model can be extended to allow for other error distributions in Equation \eqref{model_RJMCMC_1}. For example, a generalized linear model \citep{mccullagh1989generalized} may be used to model periodic count data by assuming that the observed data follows a Poisson distribution, i.e. $y_t \sim \text{Poisson} \, (\mu_t)$.  The logarithmic link function of the expected value $\mu_t$ of the response variable  $y_t$ may be expressed as $\log \,(\, \mu_t\, ) =  \sum_{j=1}^{k+1} f \, \big(t, \, \bm{\beta}_{ \, j}, \,  \bm{\omega}_{\, j} \, \big)\mathbbm{1}_{[ \, t \, \in \,  I_j \, ]}$, where the definitions of the variables are the same as for Equation \eqref{model_RJMCMC_1}. Bayesian inference can, in principle, be achieved in a similar way as described in the paper, namely by iterating between segment and change-point model moves, where the formulation of the acceptance probabilities and some proposal distributions need to be modified accordingly. We believe that such an extension would find use in several ranges of applications, for example in studying population cycles in ecology and epidemiology, where the abundance of species are measured as count variables \citep{white1996analysis, bhaskaran2013time, bramness2015analyzing}.

\section*{Acknowledgements} 

We wish to thank the Referees, the Associate Editor, Dr. Paul Jenkins, Zeda Li and Jack Jewson for their insightful and valuable comments. The work presented in this article was developed as part of the first author's PhD thesis at the University of Warwick. B. Hadj-Amar was supported by the Oxford-Warwick Statistics Programme (OxWaSP) and the Engineering and Physical Sciences Research Council (EPSRC) under Grant Number EP/L016710/1. B. Finkenst{\"a}dt  and F. L{\'e}vi   were supported by the Medical Research Council (MRC), Grant reference: MR/M013170/1.  F. L{\'e}vi was partly supported by the Conseil R{\'e}gional d'{\^I}le de France, the Conseil R{\'e}gional de Champagne-Ardenne, Mairie de Paris and the Banque Publique d'Investissement (BPI France) through the Fonds Unique Interminist{\'e}riel 12 (PiCADo, Contract 11017951), and the Institut de Recherches en Sant{\'e} Publique from France (CLOCK-DOM1, grant 2014-BDCR-EC). R. Huckstepp was supported by the MRC, Grant Reference: MC/PC/15070.

\section*{Supplementary Material}
Supplementary materials are available and include further details about simulation studies, acceptance rates, phase investigation, and performances of existing methods in the case studies. Code that implements the methodology and the data used in the case studies are available as online supplemental material and can be also found at \hyperlink{https://github.com/Beniamino92/AutoNOM}{https://github.com/Beniamino92/AutoNOM}.

\appendix
\section{Appendix - Details of the Sampling Scheme} \label{appendix_1}

\subsection{Updating the Segment Model} 

\subsubsection{Within-Model Move} \label{appendix_segment_within}

{\it \bf{Sampling} $\bm{\omega}$}: 
To obtain samples from the conditional posterior distribution $\pi \, ( \bm{\omega} \, | \,  \bm{\beta}, \, \sigma^2, \, m, \, \bm{y})$ (see Equation \eqref{posterior_omega}), we draw the frequencies one-at-time using a mixture of M-H steps. In order to explore the parameter space efficiently, we design a mixture distribution $q \, (\, \omega_l^{\, p}  \, | \, \omega_l^{\, c} \, )$, so that 	\begin{equation}
\label{mixture_proposal_freq}
q \, ( \,  \omega_l^{\, p}  \, | \, \omega_l^{\, c} \,) = \delta_{\omega} \, q_1 \, (\, \omega_l^{\, p}  \, | \, \omega_l^{\, c} \,) + (1 - \delta_{\omega} ) \, q_2 \, (\, \omega_l^{\, p}  \, | \, \omega_l^{\, c} \,), \qquad l = 1, \dots, m,
\end{equation} where $q_1$ is defined in Equation \eqref{q1_freq} below, $q_2$ is the density of a univariate Normal $\mathcal{N}\, (\omega_l^{\, c}, \sigma^2_{\omega})$,  $\delta_\omega$ is a positive real number such that $ 0 \leq \delta_\omega \leq 1$, and $c$ and $p$ refer to current and proposed values, respectively.  According to Equation \eqref{mixture_proposal_freq}  we carry out with probability $\delta_{\omega} $ a M-H step with proposal distribution $q_1 \, (\, \omega_l^{\, p}  \, | \, \omega_l^{\, c} \,)$, \begin{equation} \label{q1_freq}
q_1 \, (\, \omega_l^{\, p}  \, | \, \omega_l^{\, c} \,) \propto \sum_{h \, = \, 0}^{\tilde{n} - 1} I_h \, \mathbbm{1}_{\big[ \,  h/n \, \, \leq \, \, \omega_l^{\, p} \, < \, \,  (h+1)/n  \, \big]  \, },
\end{equation}
where $\tilde{n} = \floor{n/2}$ and $I_h$ is the value of the squared modulus of the Discrete Fourier transform of the observations $\bm{y}$ evaluated at frequency $h/n$

$$ I_h = \Big| \, \sum_{j \,=\, a}^{b} y_j \, \exp{\Big(-i \,  2 \pi \,  \frac{h}{n} \, \Big)}\, \Big|^{\, 2}.$$

In this way  frequencies are proposed from regions in parameter space  with high posterior density, yielding a Markov chain which converges quickly to its invariant distribution. The proposal distribution $q_1 \, (\, \omega_l^{\, p}  \, | \, \omega_l^{\, c} \,)$ is independent of the current state $\omega_l^{\, c}$. The acceptance probability for this move is 

\begin{equation*}
\alpha  = \min \Bigg\{1, \dfrac{\pi \, (\bm{\omega}^{\, p} \, | \, \bm{\beta},  \, \sigma^2, \, m, \, \bm{y}) }{\pi \, (\bm{\omega}^{\,c} \, | \, \bm{\beta},  \, \sigma^2, \, m, \, \bm{y}) } \times  \dfrac{q_1 \, (\, \omega_l^{\, c}  \, )}{q_1 \, (\, \omega_l^{\, p}  \, )} \Bigg\},
\end{equation*}
where $\bm{\omega}^{\,p} = (\omega_1^{\, c}, \dots, \omega_{l-1}^{\,c}, \omega_{l}^{\,p}, \omega_{l+1}^{\,c}, \dots, \omega_{m}^{\,c})^{'}$.
On the other hand, with probability 1 - $\delta_{\omega} $, we perform a random walk M-H step with proposal distribution $q_2 \, (\, \omega_l^{\, p}  \, | \, \omega_l^{\, c} \,)$, whose density is a univariate Normal distribution with mean $\omega_l^{\, c}$ and variance $\sigma^2_{{\omega} }$, i.e.
$\omega_l^{\, p} \, | \, \omega_l^{\, c} \, \sim \mathcal{N}(\,\omega_l^{\, c}, \, \sigma^2_{_{\omega} }\,)$.
This perturbation around the current value $\omega_l^{\, c}$ allows a local exploration of the conditional  posterior distribution. The acceptance probability for this move is 
\begin{equation*}
\alpha  = \min \Bigg\{1, \dfrac{\pi \, (\bm{\omega}^{\,p} \, | \, \bm{\beta},  \, \sigma^2, \, m, \, \bm{y}) }{\pi \, (\bm{\omega}^{\, c} \, | \, \bm{\beta},  \, \sigma^2, \, m, \, \bm{y}) } \Bigg\}.
\end{equation*}
Setting $\delta_{\omega} $ to a relative low value integrates a fairly high acceptance rate with a quick exploration of the parameter space.
For our experiments, we set  $\sigma^2_{\omega} = (1/50n)^2 $ and $\delta_{\omega} = 0.2$. 

\vspace{0.1cm} 

 {\it \bf {Sampling} $\bm{\beta}$:} 
Given values of $\bm{\omega}$ and $\sigma^2$, the vector of linear parameters $\bm{\beta}$ can be sampled via
a M-H step from the target posterior conditional distribution $\pi\, ( \, \bm{\beta} \, | \,  \bm{\omega}, \, \sigma^2, \,  m, \, \bm{y})$ (see Equation \eqref{posterior_beta}). The proposed vector of coefficients $\bm{\beta}^{\,p}$ can be drawn from normal approximations 
to their posterior conditional distribution (\citealp{gilks1995markov, rosen2012adaptspec}), e.g
\begin{equation} 
\label{multivariate_proposal} 
\bm{\beta}^{\,p}  \sim \bm{\mathcal{N}}_{2m+2} \, (\, \hat{\bm{\beta}}^{\,p}, \, \hat{\bm{\Sigma}}^{\, p}), 
\end{equation} 
where 
$$   \hat{\bm{\beta}}^{\,p} = \argmax_{\bm{\beta}^{\,p}} \,  \pi\, ( \, \bm{\beta}^{\,p} \, | \, \bm{\omega}, \, \sigma^2, \, m, \,  \bm{y}), $$
and
$$ \hat{\bm{\Sigma}}^{\, p} = \Bigg\{ -\dfrac{\partial^2 \log \, \pi\, ( \,\bm{\beta}^{\,p} \, | \, \bm{\omega}, \, \sigma^2, \, m, \,  \bm{y})}{\partial \bm{\beta}^{\,p} \, \partial \bm{\beta}^{\,p\, '}} \Big|_{\bm{\beta}^{\,p} \, = \, \hat{\bm{\beta}}^{\,p}} \Bigg\}^{-1}.$$

The proposal  for  $ \bm{\beta}^{\, p}$ is independent of the current values $\bm{\beta}^{\, c}$, and the acceptance probability for this move is 
$$ \alpha = \min\Bigg\{1, \dfrac{ \pi\, ( \, \bm{\beta}^{\,p} \, | \,  \bm{\omega}, \, \sigma^2, \,  m, \, \bm{y})}{ \pi\, ( \, \bm{\beta}^{\,c} \, | \,  \bm{\omega}, \, \sigma^2, \,  m, \, \bm{y})} \times \dfrac{q \, ( \, \bm{\beta}^{\,c} \, )}{q \, ( \, \bm{\beta}^{\,p}  \, )} \Bigg\}, $$
where  $ \, q \, ( \, \bm{\beta}^{\,c}  \, )$ and $q \, ( \, \bm{\beta}^{\,p}  \, )$ denote the Normal proposal densities $\bm{\mathcal{N}}_{2m+2} \, (\, \hat{\bm{\beta}}^{\,c}, \, \hat{\bm{\Sigma}}^{\, c})$ $   $ and $\bm{\mathcal{N}}_{2m+2} \, (\, \hat{\bm{\beta}}^{\,p}, \, \hat{\bm{\Sigma}}^{\, p})$, respectively. 

\subsubsection{Between-Model Moves} \label{appendix_segment_between}
The number of frequencies on a segment is  proposed to either increase  or decrease  by one. Let $\bm{\theta}^{\, c}  = ( \, \bm{\beta}^{\, c \,'}, \, \bm{\omega}^{\, c \,'}, \, \sigma^{\, 2c \,'} \,)^{\,'}$  and assume the Markov chain is currently at $( \, m^{\,c}, \, \bm{\theta}^{\, c} \,)$. We propose a move to $( \, m^{\,p}, \, \bm{\theta}^{\, p} \,)$ by drawing from a proposal density $q \, (\, m^{\,p}, \, \bm{\theta}^{\, p} \, | \, m^{\,c}, \, \bm{\theta}^{\, c} \,) $
and accepting this update with probability  \begin{equation}
\label{general_acceptance}
\alpha = \min \Bigg\{ 1, \dfrac{\mathscr{L}(\, m^{\,p}, \, \bm{\theta}^{\, p} \, | \, \bm{y})}{\mathscr{L}(\,  m^{\,c}, \,\bm{\theta}^{ \, c} \, | \, \bm{y})} \times \dfrac{\pi(m^{\,p}) \, \pi(\bm{\theta}^{ \, p} \, | \, m^{\,p})}{\pi(m^{\,c}) \, \pi(\bm{\theta}^{ \, c} \, | \, m^{\,c})} \times \dfrac{q \, ( m^{\,c}, \, \bm{\theta}^{ \, c} \, | \,  m^{\,p}, \, \bm{\theta}^{\, p} \,)}{q \, ( m^{\,p}, \, \bm{\theta}^{ \, p} \, | \,  m^{\,c}, \, \bm{\theta}^{ \, c}  \,)}\Bigg\}.
\end{equation}  The proposed state  $( \, m^{\,p}, \, \bm{\theta}^{ \, p}  \,)$ is drawn by first drawing $m^{\, p}$, followed by $\bm{\omega}^{\, p}$, $\bm{\beta}^{\, p}$ and $\sigma^{\,2p}$.
In fact, we can rewrite the proposal density as \begin{equation*}
\begin{split}
q \, ( m^{\,p}, \, \bm{\theta}^{\, p} \, | \,  m^{\,c}, \, \bm{\theta}^{ \, c}  \,) &= q \, ( m^{\, p} \, | \, m^{\, c}  ) \times q \, (\bm{\theta}^{\, p} \, | \, m^{\,p}, \,  m^{\,c}, \, \bm{\theta}^{\,c}) \\
&= q \, ( m^{\, p} \, | \, m^{\, c}  ) \times q \, (\bm{\omega}^{\, p} \, | \, m^{\,p}, \, m^{\,c}, \, \bm{\theta}^{ \, c} )  \\
& \qquad \qquad \quad  \, \, \, \,   \times q \, (\, \bm{\beta}^{\, p} \, | \, \bm{\omega}^{\, p}, \,  m^{\,p}, \, m^{\,c}, \, \bm{\theta}^{ \, c}) \\
& \qquad \qquad \quad  \, \, \, \,   \times q \, (\,  \sigma^{\,2p} \, \, | \, \, \bm{\beta}^{\, p}, \, \bm{\omega}^{\,p}, \,  m^{\,p}, \, m^{\, c}, \, \bm{\theta}^{ \, c} ).
\end{split}
\end{equation*}

 {\it \bf Birth move: } If a birth move is proposed, we have that $m^{\,p} = m^{\,c} + 1$.  The proposed frequency vector $\bm{\omega}^{\, p}$ is constructed as $$ \bm{\omega}^{\, p} = (\omega_1^{c}, \dots, \, \omega_{m^c}^{c}, \, \omega_{m^p}^{\,p})^{'}, $$
namely by keeping the current vector of frequencies and proposing an additional frequency $\omega_{m^p}^{\,p}$. Alternatively to \citet{andrieu1999joint}, we choose to sample  $\omega_{m^p}^{\,p}$ uniformly on the interval $(0, \, \phi_{\omega})$, where $0 < \phi_{\omega} < 0.5$. The value of $\phi_\omega$ can be chosen to reflect prior information about the significant frequencies that drive the variation in the data, for example by choosing $\phi_\omega$ in the low frequencies range ( $ 0 < \phi_\omega < 0.1 $).  Additionally, for computational and/or modelling reasons, we would like not to sample frequencies that are too close to each other.  Hence, we choose to draw a candidate value $\omega_{m^p}^{p}$ uniformly from the union of intervals of the form $[\omega_l^{c} + \psi_\omega, \, \omega_{l+1}^{c} - \psi_\omega]$, for $l = 0, \dots, m_{c}$ and denoting $\omega_0^{\, c} = 0$ and $\omega_{\, m^c +1}^c = \phi_\omega$.  Here, $\psi_\omega$ is a fixed minimum distance between frequencies, which is chosen larger than $\frac{1}{n}$; in fact, when the separation of two frequencies is less than the \textit{Nyquist step} \citep{priestley1981spectral}, i.e. $| \,  \omega_{\,l} - \omega_{\, l+1} \,  | < \frac{1}{n}$, the two frequencies are indistinguishable \citep{dou1995bayesian}. Moreover, we sort the proposed vector of frequencies $\bm{\omega}^{\, p}$ to ensure identifiability and perform practical estimation, as suggested in \citet{andrieu1999joint}. For proposed $\bm{\omega}^{\, p}$ and given $\sigma^{\,2c}$, the proposed vector of linear coefficients $\bm{\beta}^{\, p}$ is sampled following the same procedure of Section \ref{appendix_segment_within}, namely we draw $\bm{\beta}^{\, p}$  from a normal approximation to their posterior conditional distribution  $\pi\, ( \, \bm{\beta}^{\, p} \, | \, \bm{\omega}^{\, p}, \, \sigma^{\,2c},  \, m^{\, p}, \, \bm{y} \,  )$. Finally, the residual variance $\sigma^{\, 2p}$ is sampled directly from its  posterior conditional distribution $\pi \, ( \sigma^{\, 2p} \, | \, \bm{\omega}^{\, p},  \,  \bm{\beta}^{\, p}, \, m^{\, p}, \,  \bm{y})$ (see Equation \eqref{inverse_gamma}). The proposed state $(m^{p}, \, \bm{\theta}^{\, p} \,)$ is accepted or reject in a Metropolis-Hastings step with probability 
\begin{equation*}
\label{acceptance_birth}
\alpha = \min \Bigg\{ 1, \,  \dfrac{\mathscr{L}(\, \bm{\theta}^{\, p} , \, m^{\, p} \, | \, \bm{y})}{\mathscr{L}( \, \bm{\theta}^{\, c}, \, m^{\, c} \, | \, \bm{y})} \times \dfrac{\pi \,(m^{\, p}) \, \pi\,(\bm{\theta}^{\, p} \, | \, m^{\,p})}{\pi \,(m^{\, c}) \, \pi\,(\bm{\theta}^{\, c} \, | \, m^{\,c})} \times \dfrac{d_{m^{\, p}} \cdot \big(\frac{1}{m^{\,p}}\big) \cdot q \, ( \, \bm{\beta}^{\, c}\,) \cdot q \, ( \, \sigma^{\, 2c} \,)}{b_{m^{\, c}} \cdot q \, (\omega_{m^{\,p}}^{\,p}) \cdot  q \, ( \, \bm{\beta}^{\, p} \, ) \cdot q \, ( \, \sigma^{\,2p} \,) } \Bigg\},
\end{equation*}

where the likelihood function is given in Equation \eqref{likelik_segment}, $\pi(\, m \, ) $ is the density of the Poisson distribution truncated at $m_{\text{max}}$,  $b_{m^{\, c}}$ and $ d_{m^{\, p}}$ are defined in Equation \eqref{transition_prob}, $ q \, (\omega_{m^p}^{p})$ is the density of the uniform proposal for sampling the additional frequency, $q \, ( \bm{\beta}^{\, c} )$ and  $q \, ( \bm{\beta}^{\,p} )$ are the multivariate Normal proposal densities $\bm{\mathcal{N}}_{2m^{\,c}+2} \, (\, \hat{\bm{\beta}}^{\,c}, \, \hat{\bm{\Sigma}}^{\, c})$ and $\bm{\mathcal{N}}_{2m^{\,p}+2} \, (\, \hat{\bm{\beta}}^{\,p}, \, \hat{\bm{\Sigma}}^{\, p})$, respectively; $ q \, ( \, \sigma^2_p \, )$ and   $q \, ( \,\sigma^2_c \, )$ are the Inverse-Gamma proposal densities defined in Equation \eqref{inverse_gamma}.

\vspace{0.3cm}

 {\it \bf Death move: }If a death move is proposed, then $m^{\, p} = m^{\, c} - 1$. A vector of frequencies $\bm{\omega}^{\, p}$ is constructed by randomly selecting with probability $\frac{1}{m^{\,  c}}$ one of the current frequencies as the candidate frequency for removal. Given $\bm{\omega}^{\, p}$ and $\sigma^{\,2c}$, a vector of linear coefficients $\bm{\beta}^{\, p}$ is sampled from a normal approximation to its posterior conditional distribution. Finally, conditioned on $\bm{\omega}^{\, p} $ and  $\bm{\beta}^{\, p}$, the residual variance $\sigma^{\,2p}$ is drawn from its posterior Inverse-Gamma distribution. It is straightforward to see that the acceptance probability for the death step has the same form as the birth step, with the proper change of labelling of the variables, and the ratio terms inverted.
 
 \subsection{Updating the Change-Point Model}  
 
 \subsubsection{Within-Model Move}
 Let $\bm{s}_{\,(k)} ^{\, c} = (s_1^{\,c}, \dots, s_k^{\, c})^{\,'}$ be the current vector of change-points 
 locations, $\bm{m}_{\,(k)}^{\, c} = (m_1^{\, c}, \dots, m_{k+1}^{\,c})^{\,'}$ be the current vector of 
 number of frequencies, $\bm{\omega}_{\,(k)}^{\, c} = (\bm{\omega}_{1}^{\, c \,'}, \dots, \bm{\omega}_{k+1}^{\, c \,'})^{\,'}$
 be the current vector of frequencies.  
 Let $\bm{\beta}_{\,(k)}^{\, c } = (\,\bm{\beta}_{1}^{\, c \,'}, \dots, \bm{\beta}_{k+1}^{\, c \,'})^{\,'}$ 
 and $\bm{\sigma}^{\, 2}_{ (k)} = (\,\sigma^{\,2c}_{1}, \dots, \sigma^{\,2c}_{\, k+1} )^{\,'}$ 
 be the current vectors of linear coefficients and residual variances, respectively. 
 
 Let us also define $\bm{\theta}_{\,(k)}^{\, c } = (\bm{\beta}_{\,(k)}^{\, c \,'}, \, \bm{\omega}_{\,(k)}^{\, c \,'}, \,  \bm{\sigma}^{\, 2c \,'}_{(k)})^{\,'}. $
 Following \citet{green1995reversible}, a  change-point,  $s_j^{\,c}$ say, is randomly selected with probability $\frac{1}{k}$ from the 
 existing set of  change-points. In order to explore the parameter space in an efficient way and similar to above we 
 construct a mixture distribution $q \, ( \, s_j^{\,p} \, | \, s_j^{\,c} \, )$, as
 \begin{equation}
 \label{proposal_s_within}
 q \, ( \, s_j^{\,p} \, | \, s_j^{\,c} \, ) =   \delta_s \, q_1 \, ( \, s_j^{\,p} \, | \, s_j^{\,c} \, ) + ( 1- \delta_s) \, q_2 \, ( \, s_j^{\,p} \, | \, s_j^{\,c} \, )
 \end{equation}
 where $q_1$ is the density of a Uniform  $[  s_{j-1}^{\, c} + \psi, s_{j+1}^{\, c} - \psi ] $, $q_2$ is the density 
 of a univariate Normal $\mathcal{N}\, (s_j^{\,c}, \sigma^2_{s})$ and 
 $\delta_s$ is a positive real number such that $ 0 \leq \delta_s \leq 1$.  We  propose with probability $\delta_s$  
 a candidate value $s_j^{\, p}$  from the above uniform distribution 
 where $\psi$ is a fixed minimum time between change-points avoiding  change-points being too close to each other.
 On the other hand, with probability $(1 - \delta_s)$, $s_j^{\, p}$  arises as a Normal random walk proposal centered at  
 the current change-point $s_j^{\, c}$.  
 The proposed vector of change-points locations is denoted by 
 $$\bm{s}_{\,(k)}^{\, p} = (s_1^{\, c}, \dots, s_{j-1}^{\, c}, s_j^{\, p}, s_{j+1}^{\, c}, \dots, s_{k}^{\, c})^{\,'}, $$ 
 and hence the  proposed value $s_j^{\, p}$ induces a new proposed data partition on $[s_{j-1}^{\, c}, s_{j+1}^{\, c}]$ 
 corresponding to $[s_{j-1}^{\, c}, s_{j}^{\, p})$ and $[s_{j}^{\, p}, s_{j+1}^{\, c})$.
 We denote the vectors of observations belonging to these two proposed segments as 
 $\bm{y}_j^{\,p}$ and $\bm{y}_{j+1}^{\,p}$, which include $n_j^{\, p}$ and $n_{j+1}^{\, p}$ observations, respectively. 
 Given  $\bm{s}_{\,(k)}^{\, p}$, the proposed number of frequencies $m_j^{\, p}$, $m_{j+1}^{\, p}$ are set equal 
 to the current ones $m_j^{\, c}$, $m_{j+1}^{\, c}$, so that $\bm{m}_{\,(k)}^{\, p} = \bm{m}_{\,(k)}^{\, c}$. 
 Similarly, the proposed pair of frequency vectors $\bm{\omega}_{j}^{\, p}$, $\bm{\omega}_{j+1}^{\,p}$ is 
 chosen equal to the current pair $\bm{\omega}_{j}^{\, c}$, $\bm{\omega}_{j+1}^{\, c}$ in the corresponding segments, 
 i.e. $\bm{\omega}_{\,(k)}^{\, p} = \bm{\omega}_{\,(k)}^{\, c}$. 
 The proposed  vectors  $\bm{\beta}_{j}^{\, p}$, $\bm{\beta}_{j+1}^{\, p}$ are sampled 
 from normal approximations to their posterior conditional distributions 
 $\pi \, (\, \bm{\beta}_{j}^{\, p} \, | \, \bm{\omega}_{j}^{\,p}, \,  \sigma^{\,2c}_{\, j}, \,  m_j^{\, p}, \, \bm{y}_j^{\,p} \, )$ 
 and $\pi \, (\, \bm{\beta}_{j+1}^{\, p} \, | \, \bm{\omega}_{j+1}^{\,p}, \,  \sigma^{\,2c}_{\, j+1}, \,  m_{j+1}^{\, p}, \, \bm{y}_{j+1}^{\,p} \, )$ 
 and are accepted  in a M-H step with probability 
 \begin{equation*}
 \alpha = 
 \min \Bigg\{ \, 1, \dfrac{\mathscr{L} ( \,  k, \, \bm{m}_{\,(k)}^{\, p}, \, \bm{s}_{\,(k)}^{\, p}, \, \bm{\theta}_{\,(k)}^{\, p} \, | \,  \bm{y} \,)}{\mathscr{L} ( \,  k, \, \bm{m}_{\,(k)}^{\, c}, \, \bm{s}_{\,(k)}^{\, c}, \, \bm{\theta}_{\,(k)}^{\, c} \, | \,  \bm{y} \,)} \times \dfrac{\pi(\, \bm{s}_{\,(k)}^{\, p} \, | \, k) \, \pi ( \, \bm{\theta}_{\,(k)}^{\, p} \, | \, \bm{m}_{\,(k)}^{\, p}, \,  k) }{\pi(\, \bm{s}_{\,(k)}^{\, c} \, | \, k) \,\pi ( \, \bm{\theta}_{\,(k)}^{\, c} \, | \, \bm{m}_{\,(k)}^{\, c}, \,  k \, )} \times \dfrac{ \prod_{h=j}^{j+1} q \, ( \, \, \bm{\beta}^{\, c}_{h} \, ) \, }{\prod_{h=j}^{j+1} q \, ( \, \, \bm{\beta}^{\, p}_{h} \, ) \, } \, \Bigg\}, 
 \end{equation*}
 where the likelihood  is specified in Equation \eqref{log_likelik_RJMCM} and $q \, ( \,  \bm{\beta}^{\, c}_{h} \, )$ and $q \, ( \,  \bm{\beta}^{\, p}_{h} \, )$ 
 are multivariate Gaussian as in Equation \eqref{multivariate_proposal}. 
 Note that the likelihood ratio and the prior ratio differ from one only for the two  segments affected by the move of the change-points. 
 Finally, the residual variances $\sigma^{\,2p}_{j}$, $\sigma^{\,2p}_{j+1}$ are drawn   from their posterior conditional distributions $\pi \, ( \sigma^{\,2p}_{j} \, | \, \bm{\omega}^{\,p}_{j},  \,  \bm{\beta}^{\, p}_{j}, \, m_j^{\, p}, \,  \bm{y}_j^{\, p})$, $\pi \, ( \sigma^{\,2p}_{j+1} \, | \, \bm{\omega}^{\, p}_{j+1},  \,  \bm{\beta}^{\, p}_{j+1}, \, m_{j+1}^{\, p}, \,  \bm{y}_{j+1}^{\, p})$ in a Gibbs step.
 
 \subsubsection{Between-Model Moves}
  Let $\bm{\xi}_{\,(k^{\, c})}^{\, c} = (\, \bm{s}_{\,(k^{\, c })}^{\, c \,'}, \, \bm{m}_{\,(k^{\, c})}^{\, c \,'}, \, \bm{\theta}_{\,(k^{\, c})}^{\, c \,'} \, )^{\, '}$ and assume the Markov chain is at $(\, k^{\, c}, \, \bm{\xi}_{\,(k^{\, c})}^{\, c} \, )$. We  propose  a move to $(\, k^{\, p}, \, \bm{\xi}_{\,(k^{\, p})}^{\, p} \, )$ by first  drawing  $k^{\, p}$,  followed by sampling the change-point locations $\bm{s}_{\,(k^{\, p})}^{\, p}$. The latter involves either merging  two segments (death)  or splitting a segment (birth). The number of frequencies and their values in the proposed segments are selected from the current state. We draw $\bm{\beta}_{\,(k^{\, p})}^{\, p}$ and jointly update the entire state $(\, k^{\, p}, \, \bm{\xi}_{\,(k^{\, p})}^{\, p} \, )$. Hence, we  propose  a move to $(\, k^{\, p}, \, \bm{\xi}_{\,(k^{\, p})}^{\, p} \, )$ by drawing from a proposal density of the form \begin{equation*}
  \begin{split}
  q \,(\, k^{\, p}, \, \bm{\xi}_{\,(k^{\, p})}^{\, p} \, | \, k^{\,c}, \, \bm{\xi}_{\,(k^{\, c})}^{\, c} \, ) &= q \, ( \, k^{\, p} \, | \, k^{\, c} \, ) \times q \, ( \, \bm{\xi}_{\,(k^{\, p})}^{\, p} \, | \, k^{\, p}, \, k^{\,c}, \, \bm{\xi}_{\,(k^{\, c})}^{\, c} \, ) \\ 
  &= q \, ( \, k^{\, p} \, | \, k^{\, c} \, ) \, \times q \, ( \,  \bm{s}_{\,(k^{\, p})}^{\, p} \, | \, k^{\, p}, \,  k^{\,c}, \, \bm{\xi}_{\,(k^{\, c})}^{\, c} \, )\\
  & \qquad \qquad \quad  \, \, \, \,   \times q \, ( \, \bm{m}_{\,(k^{\, p})}^{\, p}, \,  \bm{\omega}_{\,(k^{\, p})}^{\, p} \, | \, \bm{s}_{\,(k^{\, p})}^{\, p}, \, k^{\, p}, \,  k^{\,c}, \, \bm{\xi}_{\,(k^{\, c})}^{\, c} \, )\\
  & \qquad \qquad \quad  \, \, \, \,   \times q \, ( \bm{\sigma}_{\,(k^{\, p})}^{\, 2p} \, | \, \bm{m}_{\,(k^{\, p})}^{\, p}, \,  \bm{\omega}_{\,(k^{\, p})}^{\, p}, \,  \bm{s}_{\,(k^{\, p})}^{\, p}, \, k^{\, p}, \,  k^{\,c}, \, \bm{\xi}_{\,(k^{\, c})}^{\, c} \, )\\ 
  & \qquad \qquad \quad  \, \, \, \,   \times q \, ( \, \bm{\beta}_{\,(k^{\, p})}^{\, p} \, | \, \bm{\sigma}_{\,(k^{\, p})}^{\, 2p}, \, \bm{m}_{\,(k^{\, p})}^{\, p}, \,  \bm{\omega}_{\,(k^{\, p})}^{\, p}, \,  \bm{s}_{\,(k^{\, p})}^{\, p}, \, k^{\, p}, \,  k^{\,c}, \, \bm{\xi}_{\,(k^{\, c})}^{\, c} \, ).
  \end{split}
  \end{equation*}
  
  \vspace{0.1cm}
  
   {\it \bf Birth move: } If a birth move is proposed, we have that $k^{\, p} = k^{\, c} + 1$. We draw a new change-point uniformly on $f \, (\bm{s}_{\,(k^{\, c})}^{\, c}, \, \psi_s)$, the support of $\bm{s}_{\,(k^{\,c})}^{\, c}$ given the constraints imposed by $\psi_s$, i.e. $f \, (\bm{s}_{\,(k^{\, c})}^{\, c}, \, \psi_s) = [1 + \psi_s, \,  s_1^{\,c} - \psi_s] \,   \cup \,  [s_1^{\,c} + \psi_s, \,  s_2^{\,c} - \psi_s] \cup \,  \dots \, \cup \, [s_{k^{\,c}}^{\,c} + \psi_s, \,  n - \psi_s].$ Hence, the new proposed location $\tilde{s}_j$ is sampled from a $ \text{Uniform} \, \{ \, f \, (\bm{s}_{\,(k^{\, c})}, \, \psi_s) \}$, where the proposal density is given by  \begin{equation} \label{proposal_sampling_new_cp}
  q \, (\bm{s}_{\,(k^{\,p})}^{\,p} \, |\,  k^{\, p}, \, k^{\, c}, \,  \bm{\xi}_{\,(k^{\,c})}^{\,c}) = \frac{1}{(n - 2 \psi_s \,  (k^{\, c}+1) - 1)}.
  \end{equation} As the proposed location $\tilde{s}_j$ will lie within an existing interval $(s_j^{\, c}, \, s_{j+1}^{\, c})$ with probability one, we can define the proposed change-points location vector as $$\bm{s}_{\,(k^{\,p})}^{\,p} = (s_1^{\, c}, \dots, s_j^{\, c}, \tilde{s}_j, s_{j+1}^{\, c}, \dots, s_{k^{\, c}}^{\, c})^{\,'}. $$  The number of frequencies $m_{j}^{\, p}, m_{j+1}^{\, p} $ corresponding to the two newly proposed segments $[s_j^{\, c}, \tilde{s}_j)$ and $[\tilde{s}_j, s_{j+1}^{\, c})$ are set equal to the current number of frequencies on the whole segment $(s_j^{\, c}, \, s_{j+1}^{\, c})$. Therefore, we can construct the proposed vector of the number of frequencies $\bm{m}_{\, (k^{\, p})}^{\, p}$ and the proposed vector of frequencies $\bm{\omega}_{\,(k^{\, p})}^{\, p}$ as \begin{equation*}
  \begin{split}
  \bm{m}_{\,(k^{\, p})}^{\, p} &= (\, m_1^{\, c}, \dots, m_{j-1}^{\, c}, m_j^{\, c}, m_j^{\, c}, m_{j+1}^{\, c}, \dots, m_{k^{\, c} + 1}^{\, c})^{\,'}, \\
  \bm{\omega}_{\,(k^{\, p})}^{\, p}  &= ( \bm{\omega}_{1}^{\, c \, '}, \dots, \bm{\omega}_{j-1}^{\, c \, '}, \bm{\omega}_{j}^{\, c \, '}, \bm{\omega}_{j}^{\, c\, '}, \bm{\omega}_{j+1}^{\, c\, '}, \dots, \bm{\omega}_{k^{\, c}+1}^{\, c \, '})^{\,'}.
  \end{split}
  \end{equation*} The proposed vector of residual variances $\bm{\sigma}_{\,(k^{\, p})}^{\, 2p}$ is $$ \bm{\sigma}_{\,(k^{\, p})}^{\, 2p} = ( \, \sigma^{\, 2c}_{1}, \dots,\sigma^{\, 2c}_{j-1}, \sigma^{\,2p}_{j}, \sigma^{\,2p}_{j+1}, \sigma^{\, 2c}_{j+1}, \dots, \sigma^{\, 2c}_{k^{\, c} + 1} \, )^{\, '}, $$ where the residual variances $\sigma^{\, 2p}_j, \sigma^{\, 2p}_{j+1}$ for the split partition are constructed following \cite{green1995reversible}, namely as a perturbation of the current variance $\sigma^{\, 2c}_j$. Specifically, we draw $u \sim \text{Uniform}\,(0, 1)$ and let $\sigma^{\, 2p}_j, \sigma^{\, 2p}_{j+1}$ be deterministic transformations of $\sigma^{\, 2c}_j$, i.e \begin{equation} \label{transformation_variance}
  \sigma^{\, 2p}_j = \frac{u}{1-u} \sigma^{\, 2c}_j, \qquad \sigma^{\, 2p}_{j+1} = \frac{1-u}{u} \sigma^{\, 2c}_j.
  \end{equation} Finally, the proposed vector of linear coefficients $\bm{\beta}_{\, (k^{\, p})}^{\, p}$ is $$ \bm{\beta}_{\,(k^{\, p})}^{\, p} = (\, \bm{\beta}_{1}^{\, c \, '}, \dots,  \bm{\beta}_{j-1}^{\, c \, '}, \bm{\beta}_{j}^{\, p \, '}, \bm{\beta}_{j+1}^{\,p \, '}, \bm{\beta}_{j+1}^{\, c \, '}, \dots, \bm{\beta}_{k^{\, c } + 1}^{\, c\, '} \, )^{\, '},$$ 
  where the vectors $\bm{\beta}_{j}^{\, p}$, $\bm{\beta}_{j+1}^{\, p}$ are drawn from normal approximations of their posterior conditional distribution, as in Section \ref{segment_model_within}. The proposed move to the state $(k^{\, p}, \, \bm{\xi}_{\,(k^{\, p})}^{\, p})$ is accepted with probability  $$ \alpha = \min \Bigg\{ \, 1, \dfrac{\mathscr{L} ( \,  k^{\, p}, \, \bm{\xi}_{\,(k^{\, p})}^{\, p} \, | \,  \bm{y} \,)}{\mathscr{L} ( \,  k^{\, c}, \, \bm{\xi}_{\,(k^{\, c})}^{\, c} \, | \,  \bm{y} \,)}\times \dfrac{\pi\,(k^{\, p}) \, \pi \, (\bm{\xi}_{\,(k^{\, p})}^{\, p} \, | \, k^{\, p} ) }{\pi\,(k^{\, c}) \, \pi \, (\bm{\xi}_{\, (k^{\, c})}^{\, c} \, | \, k^{\, c} )}   \times \dfrac{d_{k^{\, p}} \cdot \frac{1}{k^{\, p}} \cdot \frac{1}{2} \cdot  q \, (\bm{\beta}_{j}^{\, c})}{b_{k^{\, c}} \cdot q \, (\bm{s}_{\,(k^{\,p})}^{\,p} )  \cdot \prod_{h=j}^{j+1} q \, (\bm{\beta}_{h}^{\, p}) }   \times \, \bm{J}_{\sigma^{\, 2}}   \, \Bigg\}, $$ where the likelihood function is provided in Equation \eqref{log_likelik_RJMCM}, $q \, (\bm{s}_{\,(k^{\,p})}^{\,p} )$ is the uniform density defined in Equation  \eqref{proposal_sampling_new_cp};  $q \, ( \bm{\beta}_{h}^{\, c})$,  $q \, ( \bm{\beta}_{h}^{\, p})$ are the multivariate Normal proposal densities, and  the Jacobian $ \bm{J}_{\sigma^{\, 2}} $ is  
  \begin{equation*}
  \bm{J}_{\sigma^{\, 2}} = \Bigg| \, \dfrac{\partial \, (\sigma^{\, 2p}_j,  \sigma^{\, 2p}_{j+1})}{\partial \, (\sigma^{\, 2c}_j, u) }  \Bigg| = 2\, \Big(\sqrt{\sigma^{\, 2p}_j} + \sqrt{\sigma^{\, 2p}_{j+1}} \,  \Big)^{\, 2}. 
  \end{equation*} The numerator of the proposal ratio is better understood by looking at the details of the death step, which are given below.

     {\it \bf Death Move: }If a death step is proposed, then $k^{\, p} = k^{\, c} - 1$. A candidate change-point $s^{\, c}_j$ to be removed is sampled uniformly from the vector of existing change-points; that is, we propose to remove $s^{\, c}_j$ with probability $\frac{1}{k^{\,c}}$. Then, the proposed vector of change-points locations $\bm{s}_{\,(k)}^{\,p}$ is defined as $$\bm{s}_{\,(k)}^{\,p} = (s_1^{\, c}, \dots, s_{j-1}^{\, c}, s_{j+1}^{\, c}, \dots, s_{k^{\, c}}^{\, c})^{\, '}.$$
  The number $m_j^{\, p}$ and the vector of relevant frequencies $\bm{\omega}_{j}^{\,p}$ of the newly merged segment $[s_{j-1}^{\, c}, s_{j+1}^{\,c})$ are selected by drawing an index at random from $\{j, j+1\}$, obtaining say $j^{\, *}$, and setting the proposed parameters equal to the current ones relative to the selected index. That is, we set $m_j^{\, p} = m_{j^{\, *}}^{\, c}$ and $\bm{\omega}_{j}^{\, p} = \bm{\omega}_{j^{\, *}}^{\, c}$. Hence, the proposed vectors of number of frequencies $\bm{m}_{\, (k^{\, p})}^{\, p}$ and their values $\bm{\omega}_{\, (k^{\, p})}^{\, p}$  are constructed as follows 
  \begin{equation*}
  \begin{split}
  \bm{m}_{\, (k^{\, p})}^{\, p} & = ( \, m_1^{\, c}, \dots, m_{j-1}^{\, c}, m_j^{\, p}, m_{j+2}^{\, c}, \dots, m_{k^{\, c}+1}^{\, c}\, )^{\, '},\\
  \bm{\omega}_{\, (k^{\, p)}}^{\, p} &= (\, \bm{\omega}_{1}^{\,c \, '}, \dots, \bm{\omega}_{j-1}^{\, c \, '}, \bm{\omega}_{j}^{\, p \, '}, \bm{\omega}_{j+2}^{\, c \, '}, \dots, \bm{\omega}_{k^{\, c}+1}^{\, c \, '})^{\, '}.
  \end{split}
  \end{equation*} 
  The residual variance  $\sigma^{\, 2p}_{j}$ of the newly merged segment is obtained by inverting the  transformation of Equation \eqref{transformation_variance}. Specifically, we construct $\sigma^{\, 2p}_{j} = \sqrt{\sigma^{\, 2c}_{j} \,  \sigma^{\, 2c}_{j+1}}$ and set the proposed vector of residual variances $\bm{\sigma}_{\, (k^{\, p})}^{\, 2p}$ as $$ \bm{\sigma}_{\, (k^{\, p})}^{\, 2p} = ( \, \sigma_{1}^{\, 2c}, \dots, \sigma_{j-1}^{\,2c}, \sigma_{j}^{\, 2p}, \sigma_{j+2}^{\, 2c}, \dots, \sigma_{k^{\, c}+1}^{\, 2c} \, )^{\, '}. $$  The proposed vector of linear coefficients $\bm{\beta}_{\, (k^{\, p})}^{\, p}$ is 
  $$\bm{\beta}_{\, (k^{\, p})}^{\, p} = (\, \bm{\beta}_{1}^{\, c \, '}, \dots, \bm{\beta}_{j-1}^{\, c \, '}, \bm{\beta}_{j}^{\,p \, '}, \bm{\beta}_{j+2}^{\,c \, '}, \dots, \bm{\beta}_{k^{\, c}+1}^{\, c \, '})^{\, '}, $$
  where the vector of coefficients $ \bm{\beta}_{j}^{\, p}$  is drawn from Normal approximation to its posterior conditional distribution.  The acceptance probability for the death step has the same form of the birth step, with the proper change of labelling of the variables, and the ratio terms inverted.

\bibliographystyle{agsm}
\bibliography{Biblio}

\newpage

\section{Supplementary Material}

\subsection{Further Details of Illustrative Example  } \label{sec:illustrative}
We provide more details about the illustrative example considered in Section 5.1 of the manuscript. In particular, we give the values of the parameterization of the study and we investigate the sensitivity of our methodology AutoNOM on the prior means $\lambda_{s}$ and $\lambda_{\omega}$. 

\vspace{0.5cm}

\subsubsection{Parameter Values}

\begin{table}[htbp]
	\centering
	\caption{Illustrative example. Parameter values for simulation from  model (1) of the manuscript.}
	\label{table:parameter_illustrative_1}
	\begin{tabular}{cccccc}
		%\multicolumn{15}{c}{}\\[-1.5em]
		\hline \\[-0.9em]
		\hline
		Frequencies     &      & Linear coefficients &            & Trends and variances &         \\[.2em] \hline
		$\omega_{1, 1}$ & 1/24 &    $\bm{\beta}_{1,1}$                 & (2.0, 3.0) &      $\mu_1$                & .010   \\
		$\omega_{1, 2}$ & 1/15  &   $\bm{\beta}_{1,2}$                    & (4.0, 5.0)  &    $\mu_2$                    & .000   \\
		$\omega_{1, 3}$ & 1/7  &    $\bm{\beta}_{1,3}$                   & (1.0, 2.5) &        $\mu_3$                & -.005  \\
		$\omega_{2, 1}$                & 1/12 &      $\bm{\beta}_{2,1}$                 & (4.0, 3.0) &         $\sigma^2_1$             & $4.0^2$ \\
		$\omega_{3, 1}$ & 1/22 &      $\bm{\beta}_{3,1}$                 & (2.5, 4.0) &        $\sigma^2_2$              & $3.5^2$ \\ 
		$\omega_{3,2}$ & 1/15 &     $\bm{\beta}_{3,2}$                  & (4.0, 2.0) &           $\sigma^2_3$           & $2.8^2$ \\[.2em]
		\hline
	\end{tabular}
\end{table}
The value of the intercept was set to zero for every segment. 

\vspace{0.5cm}

\subsubsection{Sensitivity Analysis} To investigate the influence of the  prior means $\lambda_{\omega}$ and $\lambda_{s}$
we simulate 10 realizations from the same model 
and run our estimation algorithm for  combinations of values for  $\lambda_\omega$ and $\lambda_s$, ranging from 0.1 to 10.0. 
Table \ref{table:average_prob_correct_model_illustrative} shows the average posterior probability of choosing the correct model, i.e. $\hat{\pi} \, ( \, k = 2, \, m_1 = 3, \, m_2 = 1, \, m_3 = 2 \, | \, \bm{y} \, ).$ Table \ref{table:average_MSE_illustrative}  displays the average mean squared error $$ \text{MSE} =  \dfrac{1}{n} \sum_{t=1}^{n} \Big\{ \hat{f}_t - f_t \Big\}^2, $$  to asses the distance between the true underlying signal $f_t$ and  the estimated signal $\hat{f}_t$. The latter is  obtained by averaging across models of differing number of components, in contrast to model selection. Specifically, if we run our procedure for $S$ iterations, then the estimated signal $\hat{f}_t$ is defined as 

\begin{equation} \label{estimated_signal}
\hat{f}_t  = \dfrac{1}{S} \sum_{s=1}^{S}  \sum_{j=1}^{k^{\,(s)}+1} f \, \big(t, \, \bm{\beta}_{ \, j}^{\, (s)}, \,  \bm{\omega}_{\, j}^{\, (s)} \, \big)\mathbbm{1}_{[ \, t \, \in \,  I_j^{\, (s)} \, ]}, \quad t = 1, \dots, n, 
\end{equation} where the superscript $(s)$ denotes the $s^{\,th}$ sample of the Markov chain.  Both analyses suggest that, for this example,  the choice of the prior means $\lambda_\omega$ and $\lambda_s$  has hardly noticeable  impact on the results. However,  our experience is  that small values for these hyper-parameters are preferable as they prevent the algorithm from overfitting and seems to  be more robust to model misspecification. 

\vspace{0.5cm}

\begin{table}[htbp]
	\centering
	\caption{Sensitivity analysis of illustrative example. Average probability of chosing the correct model from 10 replications with varying $\lambda_\omega$ and $\lambda_s$}
	\begin{tabular}{lccccccc}
		\hline \\[-0.9em]
		\hline
		\multicolumn{1}{c}{}    & $\lambda_s$ = .1 & $\lambda_s$ = .2 & $\lambda_s$ = 0.5 & $\lambda_s$ = 1.0 & $\lambda_s$ = 2.0 & \multicolumn{1}{l}{$\lambda_s$ = 5.0} & \multicolumn{1}{l}{$\lambda_s$ = 10.0} \\ \cmidrule{2-8}
		$\lambda_\omega$ = .1  & .98              & .97              & .97              & .99              & .93              & .91                                  & 1.0                                    \\[.05em]
		$\lambda_\omega$ = .2  & .99              & .99              & .99              & .99              & .99              & .99                                  & .98                                   \\[.05em]
		$\lambda_\omega$ = .5  & .99              & 1.0               & .98              & .95              & .98               & 1.0                                   & .99                                   \\[.05em]
		$\lambda_\omega$ = 1.0  & .99              & .94              & .93              & .99              & .99              & .99                                  & .99                                   \\[.05em]
		$\lambda_\omega$ = 2.0  & .99              & .99              & .99              & .99              & .99              & .99                                  & .99                                   \\[.05em]
		$\lambda_\omega$ = 5.0  & .97              & .96              & .96              & .97              & .97              & .93                                  & .98                                   \\[.05em]
		$\lambda_\omega$ = 10.0 & .94              & .95              & .95              & .91              & .93              & .95                                  & .85      \\[.2em] \hline                            
	\end{tabular}
	\label{table:average_prob_correct_model_illustrative}
\end{table}

\newpage
\begin{table}[htbp]
	\centering
	\caption{Sensitivity analysis of illustrative example. Average MSE from 10 replications with varying $\lambda_\omega$ and $\lambda_s$}
	\begin{tabular}{lccccccc}
		\hline \\[-0.9em]
		\hline
		\multicolumn{1}{c}{}    & $\lambda_s$ = .1 & $\lambda_s$ = .2 & $\lambda_s$ = .5 & $\lambda_s$ = 1.0 & $\lambda_s$ = 2.0 & \multicolumn{1}{l}{$\lambda_s$ = 5.0} & \multicolumn{1}{l}{$\lambda_s$ = 10.0} \\ \cmidrule{2-8}
		$\lambda_\omega$ = .1  & .349             & .435             & .370             & .416             & .446             & .470                                 & .319                                  \\[.05em]
		$\lambda_\omega$ = .2  & .354             & .360             & .378             & .361             & .370             & .404                                 & .394                                  \\[.05em]
		$\lambda_\omega$ = .5  & .400             & .347             & .378             & .447             & .430             & .346                                 & .364                                  \\[.05em]
		$\lambda_\omega$ = 1.0  & .302             & .392             & .400             & .382             & .321             & .400                                 & .337                                  \\[.05em]
		$\lambda_\omega$ = 2.0  & .307             & .369             & .404             & .340             & .407             & .329                                 & .391                                  \\[.05em]
		$\lambda_\omega$ = 5.0  & .360             & .387             & .362             & .396             & .346             & .350                                 & .324                                  \\[.05em]
		$\lambda_\omega$ = 10.0 & .355             & .386             & .340             & .387             & .420             & .393                                 & .428      \\[.2em] \hline                           
	\end{tabular}
	\label{table:average_MSE_illustrative}
\end{table}

\vspace{0.5cm}

\subsection{Acceptance Rates} \label{sec:acceptance_rates}
Here we report the acceptance rates in the illustrative example and in the case studies. The overall acceptance rate for the simulation study is 28.3\% and for the analysis of the skin temperature is 15.2\%. The overall acceptance rates for the three time series of airflow traces of a rat are 24.0\%, 33.8\% and 33.9\%, respectively. These rates are the proportion of samples accepted in all Metropolis-Hastings steps evaluated in the sampling scheme.  We also report the acceptance rates for segment model and change-point model moves, grouped by within-model, birth and death steps for both simulation and case studies (see Table \ref{table:illustrative_acceptance}, \ref{table:temperature_acceptance}, \ref{table:rats_a_acceptance}, \ref{table:rats_b_acceptance} and \ref{table:rats_c_acceptance}). 
    
    \vspace{0.5cm}
    
    \begin{table}[htbp]
	\centering
	\caption{Illustrative example. Acceptance rates for segment model and change-point model moves, grouped by within-model, birth and death steps. The overall acceptance rate is 28.3\%.}
\begin{tabular}{lcc}
		\hline \\[-0.9em]
		\hline
\multicolumn{1}{c}{} & Segment & \multicolumn{1}{c}{Change-Point}  \\ \cmidrule{2-3}
Within               & .367    & .144                             \\
Birth                & .003    & .001                             \\
Death                & .002    & .001  \\ \hline               
\end{tabular}
\label{table:illustrative_acceptance}
\end{table}

\newpage
\begin{table}[htbp]
	\centering
	\caption{Analysis of skin temperature of a healthy subject. Acceptance rates for segment model and change-point model moves, grouped by within-model, birth and death steps. The overall acceptance rate is 15.2\%.}
\begin{tabular}{lcc}
		\hline \\[-0.9em]
		\hline
\multicolumn{1}{c}{} & Segment & \multicolumn{1}{c}{Change-Point}  \\ \cmidrule{2-3}
Within               & .202    & .100                          \\
Birth                & .004    & .004                            \\
Death                & .002    & .001  \\ \hline               
\end{tabular}
\label{table:temperature_acceptance}
\end{table}

\vspace{1cm}

\begin{table}[htbp]
	\centering
	\caption{Characterizing instances of sleep apnea in rodents: time series (a). Acceptance rates for segment model and change-point model moves, grouped by within-model, birth and death steps. The overall acceptance rate is 24.0. \%}
\begin{tabular}{lcc}
		\hline \\[-0.9em]
		\hline
\multicolumn{1}{c}{} & Segment & \multicolumn{1}{c}{Change-Point}  \\ \cmidrule{2-3}
Within               & .280    & .143                          \\
Birth                & .002    & .001                            \\
Death                & .003    & .001  \\ \hline               
\end{tabular}
\label{table:rats_a_acceptance}
\end{table}

\vspace{1cm}

\begin{table}[htbp]
	\centering
	\caption{Characterizing instances of sleep apnea in rodents: time series (b). Acceptance rates for segment model and change-point model moves, grouped by within-model, birth and death steps. The overall acceptance rate is 33.8\%.}
\begin{tabular}{lcc}
		\hline \\[-0.9em]
		\hline
\multicolumn{1}{c}{} & Segment & \multicolumn{1}{c}{Change-Point}  \\ \cmidrule{2-3}
Within               & .415    & .151                   \\
Birth                & .001    & .002                           \\
Death                & .001    & .001  \\ \hline               
\end{tabular}
\label{table:rats_b_acceptance}
\end{table}

\newpage
\begin{table}[htbp]
	\centering
	\caption{Characterizing instances of sleep apnea in rodents: time series (c). Acceptance rates for segment model and change-point model moves, grouped by within-model, birth and death steps. The overall acceptance rate is 33.9\%.}
\begin{tabular}{lcc}
		\hline \\[-0.9em]
		\hline
\multicolumn{1}{c}{} & Segment & \multicolumn{1}{c}{Change-Point}  \\ \cmidrule{2-3}
Within               & .282    &  .146                    \\
Birth                & .001    &  .001                       \\
Death                & .001    &   .001 \\ \hline               
\end{tabular}
\label{table:rats_c_acceptance}
\end{table}

\subsection{Phase Shift} \label{sec:phase_shift}
Our proposed methodology can be used to investigate phase since the sinusoidal function $f \, \big(t, \, \bm{\beta}_{ \, j}, \,  \bm{\omega}_{\, j} \big)$ that characterizes the $j^{\,th}$ segment (see Equation (2) of the manuscript) can be re-written using trigonometric identities\footnote{\label{note1} $\cos(a \pm b) = \cos(a) \cos(b) \mp \sin(a) \sin(b) $} as 
\begin{equation*} \label{eq:signal_with_phase}
f \, \big(t, \, \bm{B}_{ \, j}, \,  \bm{\omega}_{\, j},  \, \bm{\tau}_{\, j} \, \big) = \alpha_j + \mu_j \, t + \sum_{l=1}^{m_j} \Bigg(  B_{j, \,l} \cos(2\pi\omega_{j,\,l} \, t + \tau_{j, \,l}) \Bigg),
\end{equation*}
where $\bm{B}_{\, j} = (B_{j, \,1}, \dots, B_{j, \,m_{\,j}})$ and $\bm{\tau}_{\, j} = (\tau_{j, \,1}, \dots, \tau_{j, \,m_{\,j}})$. With this notation, $B_{j, \,l}$ is the amplitude of the frequency $\omega_{j,\,l}$ and $\tau_{j, \,l}$ is the phase shift of the corresponding frequency.  The phase $\tau_{j, \,l}$  of a frequency of interest $\omega_{j,\,l}$ can be estimated in terms of the  coefficients $\beta_{j, \,l}^{\, (1)}$ and $\beta_{j, \,l}^{\, (2) }$ using the following equality
\begin{equation*}
        \tau_{j, \,l} = \arctan \Bigg( -\dfrac{\beta_{j, \,l}^{\, (2)}}{\beta_{j, \,l}^{\, (1)}} \, \Bigg), \qquad -\pi \leq \tau_{j, \,l} \leq \pi,
\end{equation*}
where credible intervals can be easily obtained from the empirical percentiles of the posterior sample.  In the framework of analyzing circadian biomarker data, such as body temperature, a change in acrophase may be of interest to the clinician as this may be indicative of a disruption of the bodyclock.

\subsection{Case Studies:  Comparison with Existing Methods} \label{sec:comparison}

In this section we investigate how AutoPARM \citep{davis2006structural} and AdaptSPEC \citep{rosen2012adaptspec}, the current state-of-the-art methods, perform in the case studies (Section 6 of the manuscript). AdaptSPEC was fitted with $J=12$ basis functions and the results shown below are conditioned on the modal number of segments, whereas AutoPARM is performed with default tuning parameters.

        \vspace{0.1cm}
        \noindent
        \subsubsection{Analysis of Human Skin Temperature} 
\begin{figure}[htbp]
    \centering
    \includegraphics[height = 4.0cm, width = 9cm] {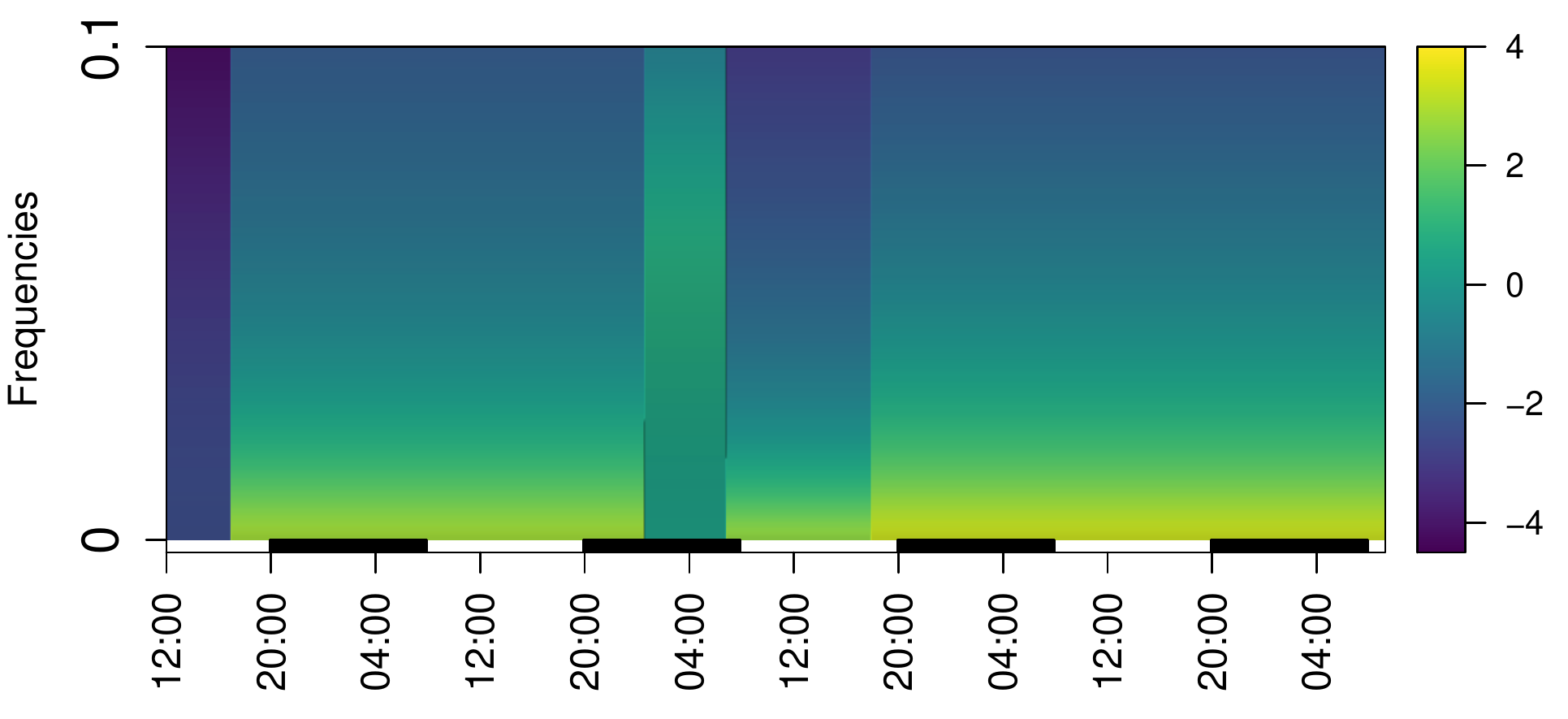}
    \includegraphics[height = 4.0cm, width = 9cm]{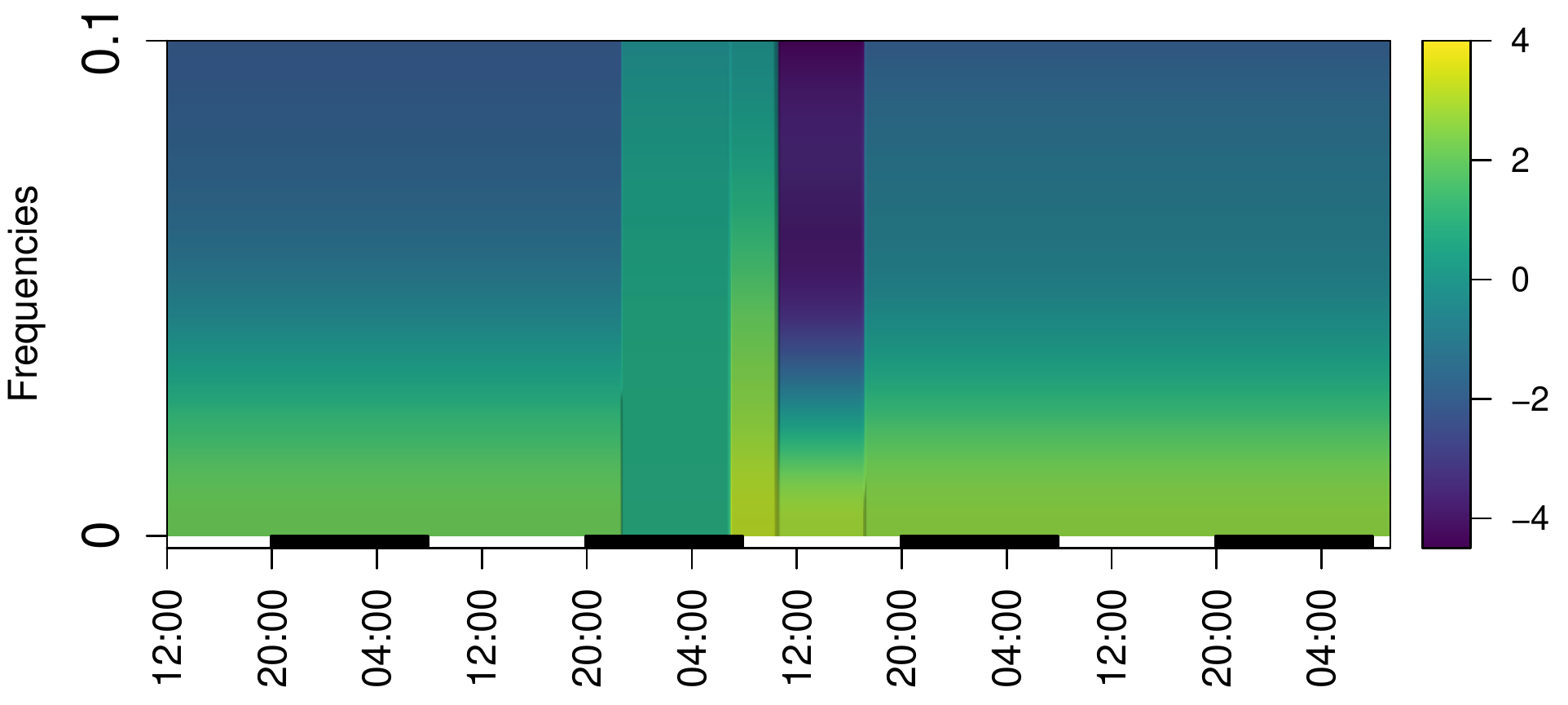}
    \caption{Estimated time-varying log spectrum for the skin temperature time series. (Top) AutoPARM. (Bottom) AdaptSPEC. Rectangles on the time axis of each plot correspond to periods from 20.00 to 8.00.}
    \label{fig:temperature_AP_AS}
\end{figure}

The estimated logarithm of the time-varying spectrum of the skin temperature time series is displayed in Figure \ref{fig:temperature_AP_AS}, for both AutoPARM (top panel) and AdaptSPEC (bottom panel). The elements of the time-varying spectrum are functions of frequency and time, and the locations of the change-points are  identified  visually  by inspecting  the abrupt changes in power over the time axis. Broadly speaking, both AutoPARM and AdaptSPEC identify five segments and show some general agreement with each other in estimating change-points and local spectra.  Both methodologies, which are based on continuous spectrum models, seem to smooth the local spectra at low frequencies in a considerable way.  The only frequency peak is estimated by AutoPARM in the third segment and  corresponds to a cycle of approximately 1.2 hour, which finds analogies with the spectral properties of Segment 4 estimated by our proposed approach. In particular, AutoPARM identifies the spectrum of an AR(2) process with autoregressive parameters (1.55, -0.69) in that segment. However, and most importantly, both existing methods clearly fail to detect either circadian or ultradian rhythmicity which were elicited by our method (see Figure 6 and Figure 7 of  the manuscript) and are to be expected as body temperature is known to be a circadian biomarker \citep{krauchi1994circadian}.

\vspace{0.3cm}

\newpage

\vspace{1.2cm}

\noindent
\subsubsection{Characterizing Instances of Sleep Apnea in Rodents } 
\begin{figure}[htbp]
\minipage{0.32\textwidth}
  \includegraphics[width=\linewidth, height = 4.0cm]{Plots/data_rats_a.pdf}
\endminipage\hfill
\minipage{0.32\textwidth}
  \includegraphics[width=\linewidth, height = 5cm]{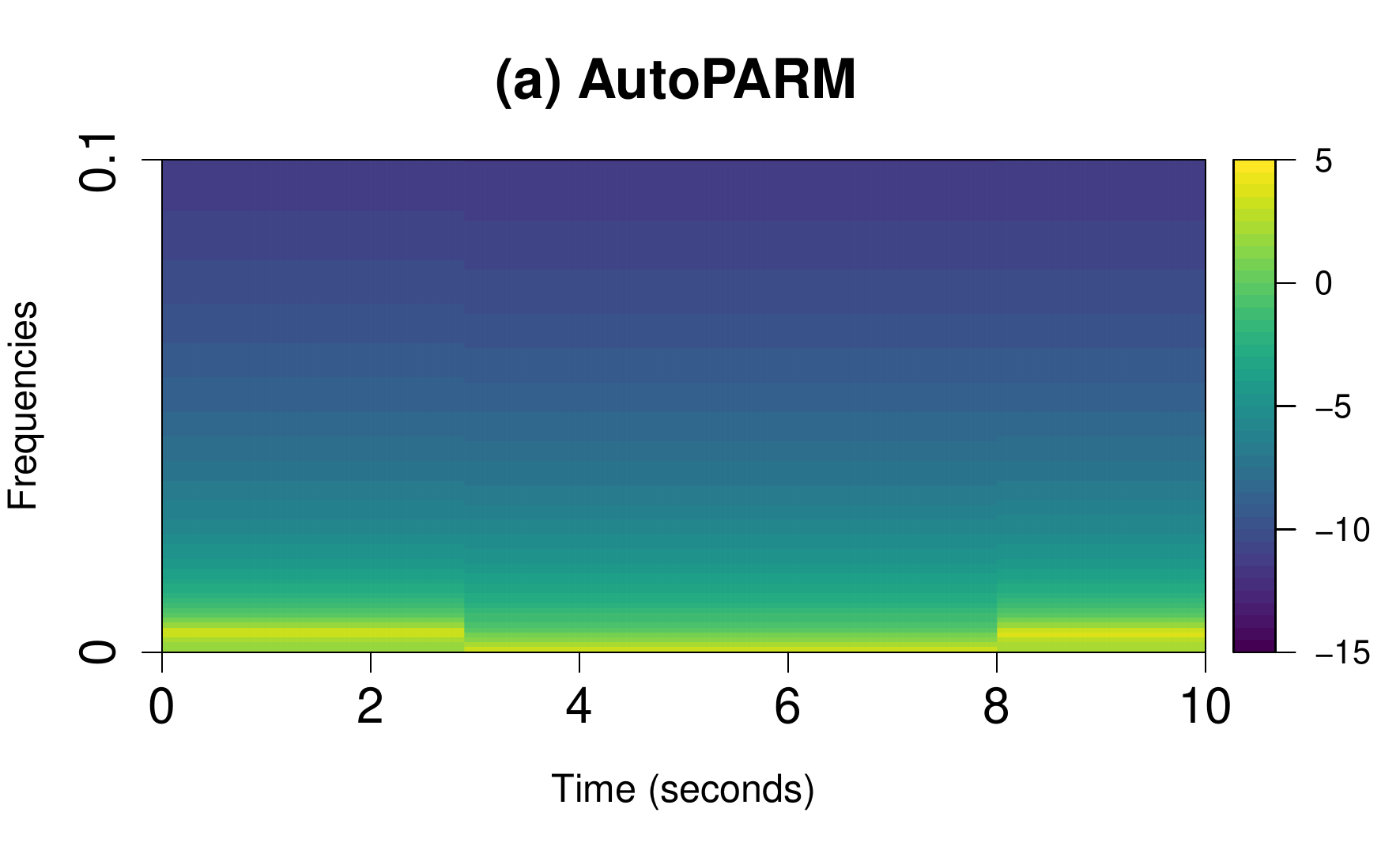}
\endminipage\hfill
\minipage{0.32\textwidth}%
  \includegraphics[width=\linewidth, height = 5cm]{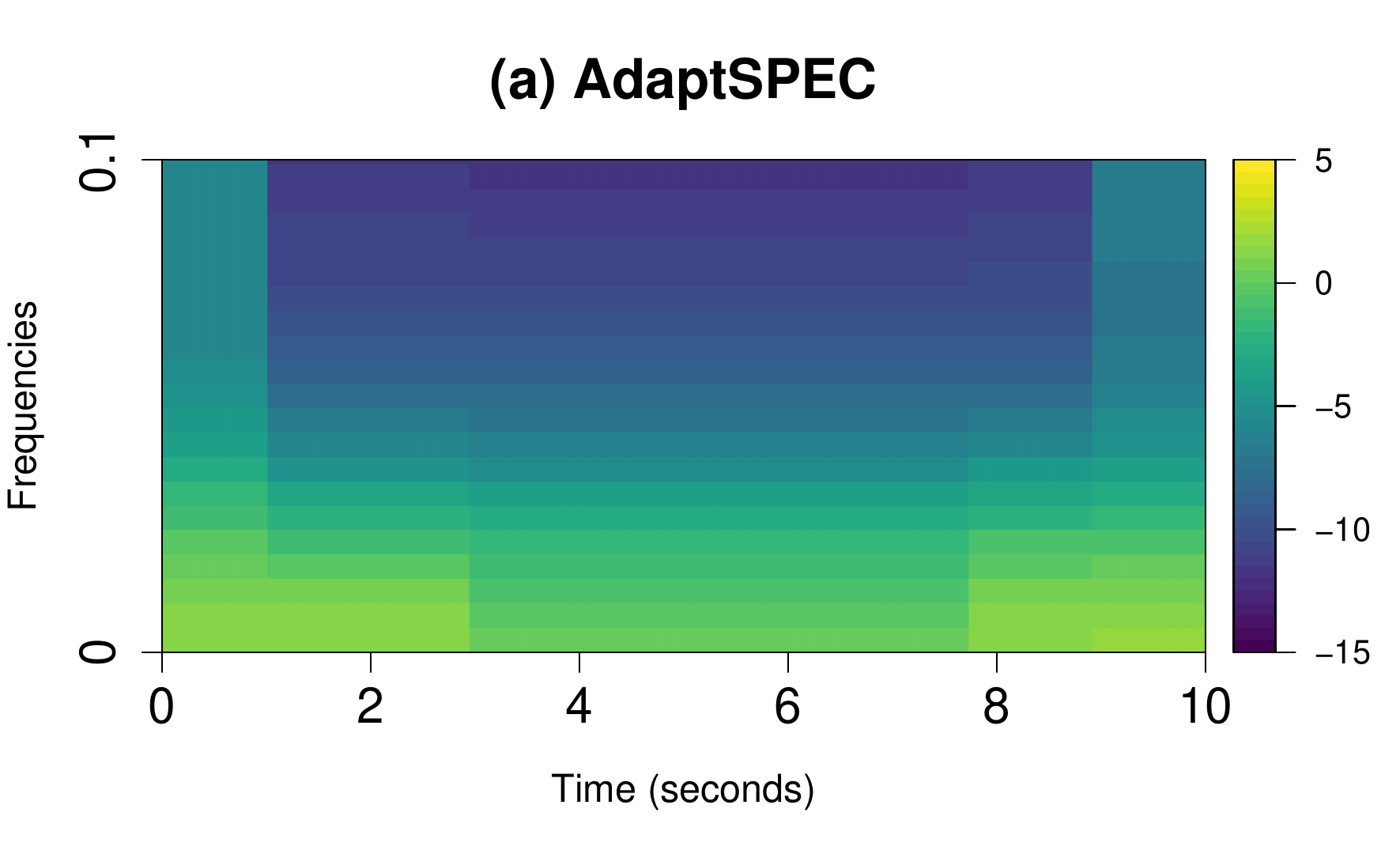}
\endminipage

\minipage{0.32\textwidth}
  \includegraphics[width=\linewidth, height = 4.0cm]{Plots/data_rats_b.pdf}
\endminipage\hfill
\minipage{0.32\textwidth}
  \includegraphics[width=\linewidth, height = 5cm]{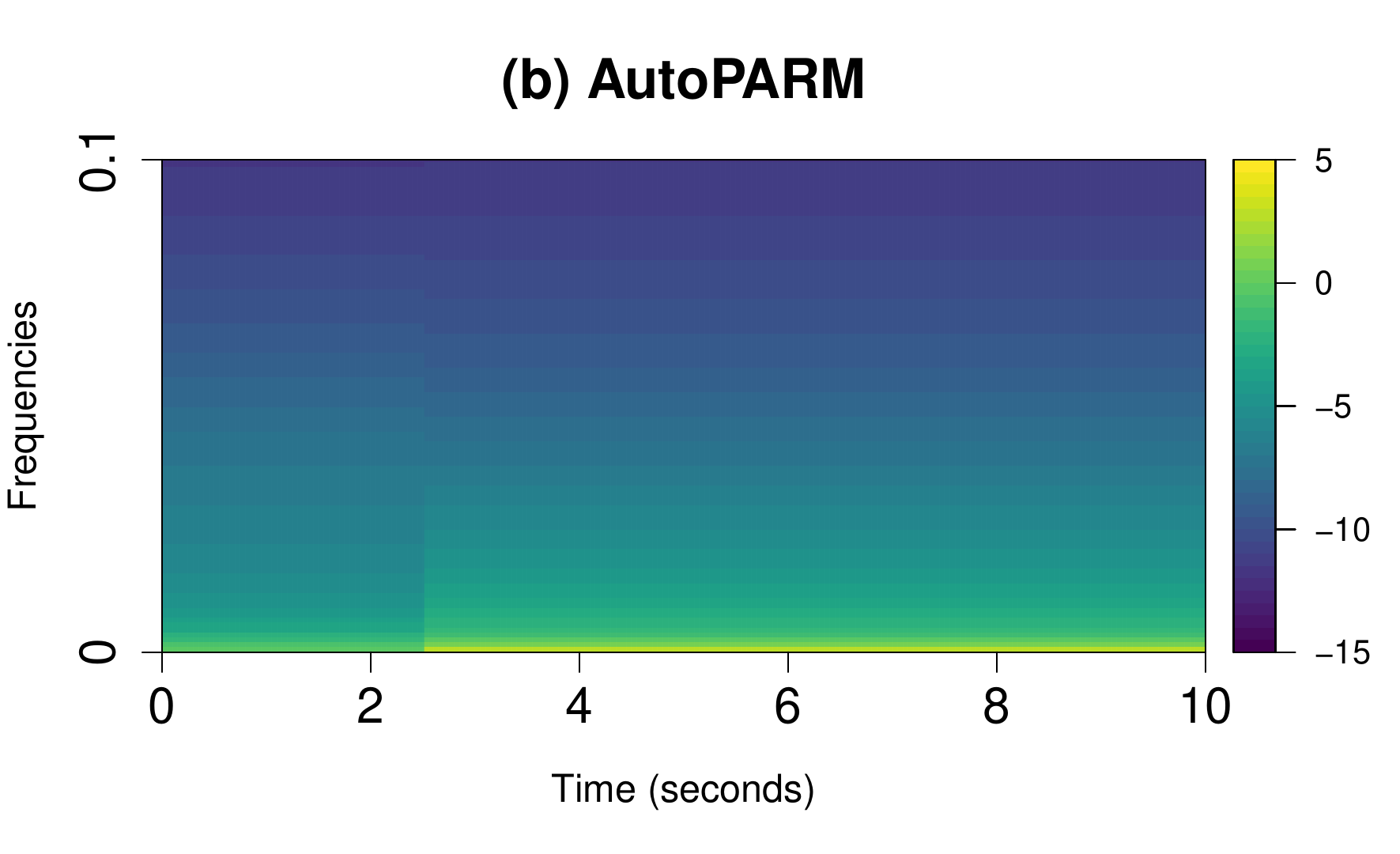}
\endminipage\hfill
\minipage{0.32\textwidth}%
  \includegraphics[width=\linewidth, height = 5cm]{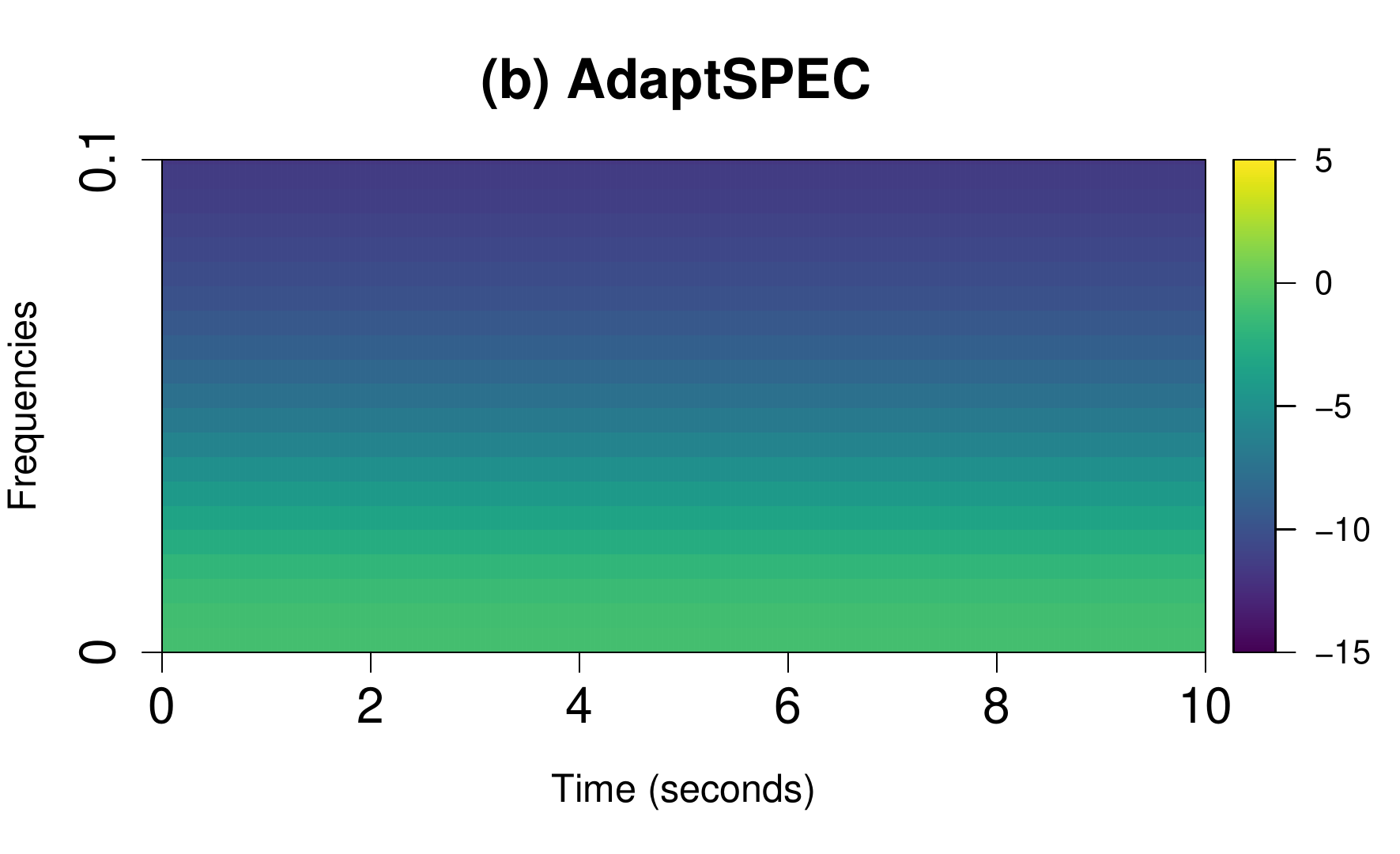}
\endminipage

\minipage{0.32\textwidth}
  \includegraphics[width=\linewidth, height = 4.0cm]{Plots/data_rats_c.pdf}
\endminipage\hfill
\minipage{0.32\textwidth}
  \includegraphics[width=\linewidth, height = 5cm]{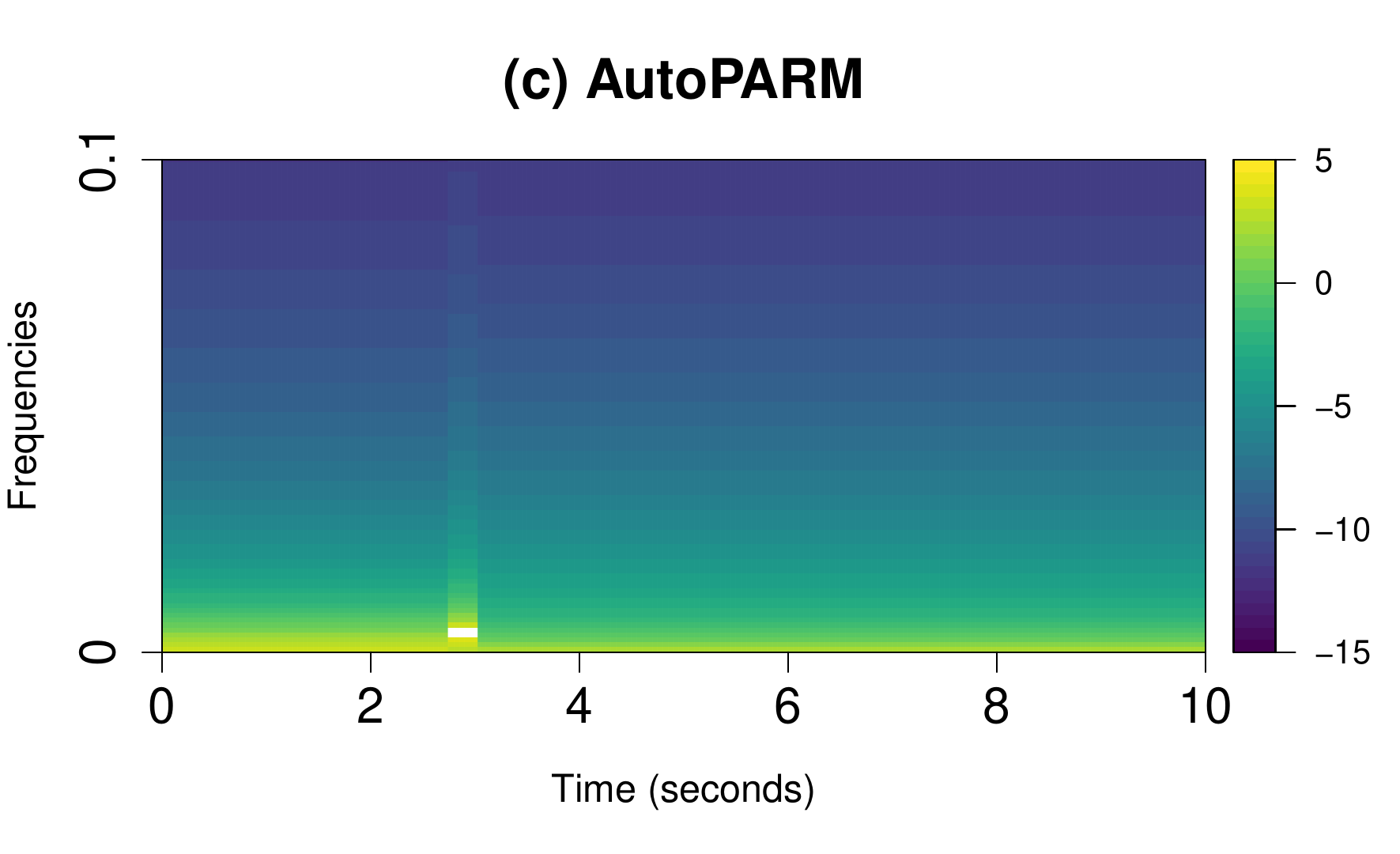}
\endminipage\hfill
\minipage{0.32\textwidth}%
  \includegraphics[width=\linewidth, height = 5cm]{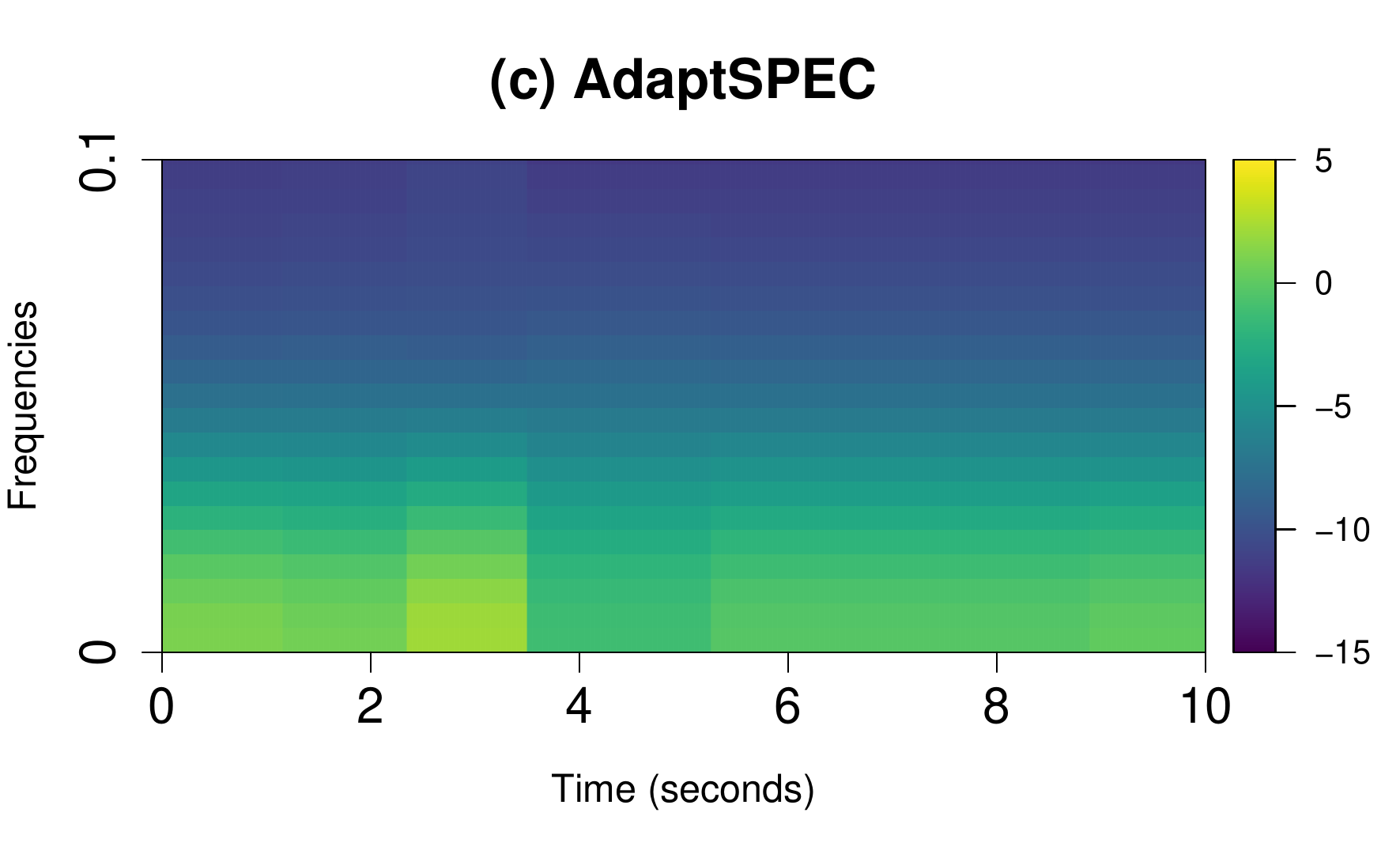}
\endminipage

	\caption{(Left) Plots of respiratory traces of a rat along with estimated change-points locations by AutoNOM. (a) is characterised by an alternation of sniffing and normal breathing. (b) is a plot of the trace of a spontaneous apnea, followed by normal breathing. (c) shows normal breathing followed by a sigh, and a post-sigh apnea. (Center) AutoPARM estimated time-varying log spectra for three different respiratory traces of a rat. (Right) AdaptSPEC estimated time-varying log spectra for three different respiratory traces of a rat. } 
	\label{fig:spectral_properties_rats}
\end{figure}

The estimated time-varying spectral properties for three  plethysmographic respiratory traces of the rat are displayed in Figure \ref{fig:spectral_properties_rats}, for both AutoPARM (center panels) and AdaptSPEC (right panels).  AutoPARM appears to identify fairly well changes in actions of this rat, such as (a) the alternation between sniffing and normal breathing, (b) the change from normal breathing to a spontaneous apnea,  and (c) different actions of apnea, sigh and post-sigh apnea. We note that these actions were classified by eye by an experienced experimental researcher. AdaptSPEC detects changes in a satisfactory way for (a) and (c)  but does not detect which distinct frequencies drive the oscillations in these data in particular as  all peaks corresponding to low frequencies are smoothed. Notice that the periodogram ordinates for these time series were approximately zero for all frequencies larger than 0.01. In addition, AdaptSPEC is not able to detect any changes from normal breathing to a spontaneous apnea since it identifies only one segment in (b). It seems that the AR building block of AutoPARM can better model peaked structures compared to the smoothing spline nature of AdaptSPEC.  Generally, our method find a larger number of change-points which are associated with changes in the spectrum, as seen in Figure 8 of the main paper (right panels). For example, in (c) AutoNOM identifies different frequencies that explain the variation between Segment 3 and Segment 4, leading to the hypothesis that there might be a time changing spectrum during the occurrence of an apnea instance.

\end{document}